\documentclass[aps,reprint,nofootinbib,superscriptaddress]{revtex4-2}

\usepackage{amsmath,amssymb,amsthm,mathtools, 
					  esint, 
				    hyperref, 
			      mathrsfs, 
		        graphicx} 

\newcommand{\cornell}{\affiliation{Cornell Center for Astrophysics and Planetary Science, Cornell University, Ithaca, New York 14853, USA}}

\newcommand{\caltech}{\affiliation{Theoretical Astrophysics 350-17, California Institute of Technology, Pasadena, CA 91125, USA}}

\newcommand{\tifr}{\affiliation{International Centre for Theoretical Sciences, Tata Institute of Fundamental Research, Bangalore 560089, India}}

\newcommand{\psu}{\affiliation{Institute for Gravitation and the Cosmos \& Physics Department, Penn State, University Park, Pennsylvania 16802, USA}}

\newcommand{\aei}{\affiliation{Max Planck Institute for Gravitational Physics (Albert Einstein Institute), Am M\"{u}hlenberg 1, Potsdam 14476, Germany}}

\begin{document}

\title{Multipole moments on the common horizon in a binary-black-hole simulation}

\author{Yitian Chen}
\email{yc2377@cornell.edu}
\cornell

\author{Prayush Kumar}
\tifr
\cornell

\author{Neev Khera}
\psu

\author{Nils Deppe}
\caltech

\author{Arnab Dhani}
\psu

\author{Michael Boyle}
\cornell

\author{Matthew Giesler}
\cornell

\author{Lawrence E. Kidder}
\cornell

\author{Harald P. Pfeiffer}
\aei

\author{Mark A. Scheel}
\caltech

\author{Saul A. Teukolsky}
\cornell
\caltech

\date{\today}

\begin{abstract}

We construct the covariantly defined
multipole moments on the common horizon of an equal-mass,
non-spinning, quasicircular binary-black-hole system. We see a strong correlation
between these multipole moments and the gravitational waveform. We
find that the multipole moments are well described by the fundamental
quasinormal modes at sufficiently late times. For each multipole
moment, at least two fundamental modes of different $\ell$
are detectable in the best model. These models provide faithful estimates
of the true mass and spin of the remnant black hole. We also show
that by including overtones, the $\ell=m=2$ mass multipole moment
admits an excellent quasinormal-mode description at all times after
the merger. This demonstrates the perhaps surprising
power of perturbation theory near the merger.

\end{abstract}

\maketitle

\section{Introduction} \label{sec:intro}

The black hole (BH) no-hair theorem \cite{Israel1_skip,
Carter1_skip} suggests that the final state of a charge-neutral
BH merger satisfies the Kerr solution, which is characterized by
only two parameters: mass and angular momentum (or equivalently,
spin). Numerical simulations of binary-black-hole (BBH) systems
have directly confirmed this theorem by comparing the quantities in
the final stage with the corresponding Kerr values \cite{0810.1767,
0811.3006, 1004.3768, 1711.00926}. The Kerr spacetime is axisymmetric and has
a simple geometry. In stark contrast, as brought out by numerical
simulations, the horizon of a merged BH is highly
distorted at its
formation, and undergoes large dynamical changes as it approaches
equilibrium. For a BH merger to lose its hair and settle down to
the final Kerr state, the horizon distortion must be washed away
by general relativity in the ringdown phase.

In numerical relativity, an event horizon is not a convenient notion of horizon, as it cannot be determined during the evolution of the spacetime. It is typically found
in post-processing, once the complete spacetime is known. Quasilocal objects like apparent horizons are more favored, because they can be computed on each time slice without the knowledge of the complete spacetime.
A recent topic in the study of quasilocal objects is seeking
a quantitative description of the horizon behavior of a BBH
merger.
One of the physical quantities used for such an investigation
is the gravitational flux falling into a horizon. It turns out that
the infalling energy flux is correlated with the outgoing flux
of gravitational waves \cite{1108.0060, 1801.07048}. This might
seem slightly surprising at first glance but is indeed reasonable,
because both the ingoing and outgoing flux are generated from the
same gravitational source. Besides the flux,
another quantity that can be used
in the analysis of BH horizons is the set of horizon multipole
moments. In the following discussion, we will discuss the multipole
moments only in the ringdown phase, though this concept is also
applicable in the inspiral phase (see, e.g., Ref.~\cite{2109.01193}).

Horizon multipole moments generalize the mass and spin of a
BH. It is fairly straightforward to define multipole moments on
the isolated horizon of a Kerr BH \cite{gr-qc/0401114}, or on
a dynamical horizon that is axisymmetric throughout the whole
ringdown phase \cite{gr-qc/0604015}. This is because in both
situations, the horizon possesses a rotational Killing vector,
which is associated with a natural choice of angular coordinates. In
a more general BBH configuration, however, choosing an appropriate
definition of multipole moments is a nontrivial task. One difficulty
comes from the nonaxisymmetry of the dynamical horizon. Moreover,
the coordinate system used to express the components of spacetime
quantities varies from simulations to simulations, which calls for
an invariant notion of multipole moments. Ashtekar \textit{et al.}
\cite{1306.5697} provide a definition of horizon multipole moments
that is appropriate for this task. They start with
the axisymmetry of
the final BH, construct weighting fields subject to this axisymmetry,
and transport these weighting fields backward along the dynamical
horizon. The resulting multipole moments are then spatially gauge
independent on a given dynamical horizon. This set of multipole
moments will be the subject of this paper, and we will explain the
construction process in greater detail
in later sections.

Regardless of different notions of multipole moments, an important goal in studying them is to
discover any universality in the horizon behavior of a remnant
BH. A natural avenue is to find inspiration from multipole
moments of the gravitational waveform in the ringdown phase.
BH perturbation theory shows
that the gravitational waves
radiated by a perturbed BH at late times
can be characterized by a superposition
of exponentially damped oscillations, called the \textit{quasinormal
modes} (QNMs) \cite{Teukolsky1_skip, Teukolsky2_skip, Press1_skip,
Teukolsky3_skip}. The frequency and the decay constant of each
mode are completely determined by the final mass and spin,
consistent with the no-hair theorem. The presence of quasinormal modes in
the late-time behavior of post-merger waveforms
has already been confirmed in numerical
simulations (e.g., \cite{gr-qc/0610122, gr-qc/0703053}). Recently,
Giesler \textit{et al.} \cite{1903.08284} discovered that including
overtones even allows a QNM model to describe the waveform
immediately after merger.

Although the waveform multipole moments are a superposition of QNMs
in the ringdown phase, we might not expect this behavior in multipole moments
of the dynamical horizon soon after the common horizon forms. After all,
this horizon is initially highly distorted compared to a Kerr horizon, so we
have no reason to expect perturbation theory to be valid. Moreover,
the time coordinate of the simulation is quite arbitrary compared
to the time coordinate of an observer at infinity, which is used to
define the frequency of QNMs. Nevertheless,
there is strong evidence
supporting the idea that horizon multipole
moments exhibit QNM behavior \cite{0907.0280, 1801.07048, 2006.03940,
2010.15186}. However, such evidence is based on either the special case of a
head-on collision of two BHs, or a definition of multipole moments
that does not refer to the connection among quasilocal horizons on
different time slices. A definition ignoring
the diffeomorphism content of a dynamical horizon is subject to
the arbitrariness of spatial coordinates.

In this paper, we calculate the horizon multipole moments that are
spatially gauge invariant on the common horizon of an equal-mass BBH
system, following the definition in Ref.~\cite{1306.5697}. To
investigate the dynamics of these multipole moments, we test their
balance laws, compare them with waveform multipole moments,
and model them as linear
combinations of QNMs. Regarding the QNM models, we use fundamental
tones to analyze the late-time behavior of multipole moments,
and then include overtones in the survey of their early-time
patterns. We will also consider
the effect of mode mixing, which
turns out to be significant in most of the multipole moments.

The rest of this paper is structured as follows. In
Sec.~\ref{sec:prelim}, we introduce the notions of horizons and
quasinormal modes. We also describe the construction
process of the horizon multipole moments proposed by Ashtekar
\textit{et al.} \cite{1306.5697}. In Sec.~\ref{sec:numeric}, we
describe the configuration of our BBH simulation and implement the
procedure to extract multipole moments on the common
horizon. In Sec.~\ref{sec:results}, we first look for potential
correlations between horizon and waveform behavior in the context of
their respective multipole moments. Then, we investigate the damped
sinusoidal patterns of multipole moments using QNM models, with or
without the inclusion of overtones. We finally summarize the results
and give remarks on possible future work in Sec.~\ref{sec:conclu}.

\section{Preliminaries} \label{sec:prelim}

\subsection{Dynamical horizons} \label{sec:dh}

A spacetime is a 4-dimensional Lorentzian manifold $\mathscr{M}$ equipped with a metric $g_{ab}$ of signature $(-,+,+,+)$. Here, we only consider a vacuum spacetime that is asymptotically flat.\footnote{The concepts in this section can be generalized in a non-vacuum spacetime.} Let $\nabla_a$ be the covariant derivative compatible with $g_{ab}$. Let $\mathcal{S} \subset \mathscr{M}$ be a smooth, orientable, spacelike 2-manifold with spherical topology $S^2$. Let $\tilde{q}_{ab}$ be the induced metric on $\mathcal{S}$. (All symbols with tilde in this paper represent quantities on or associated with $\mathcal{S}$.) The outgoing and ingoing future-directed null normals to $\mathcal{S}$, denoted as $l^a$ and $n^a$, are normalized subject to $l\cdot n = l^a n_a = -1$. The expansions of $l^a$ and $n^a$ are
\begin{align}
	\Theta_{(l)} &= \tilde{q}^{ab} \nabla_a l_b, \label{eqn:l_expan}\\
	\Theta_{(n)} &= \tilde{q}^{ab} \nabla_a n_b. \label{eqn:n_expan}
\end{align}
The shear of $l^a$ is 
\begin{align}
	\sigma_{ab} &= \tilde{q}_a^{\ c} \tilde{q}_b^{\ d} \nabla_c l_d - \frac{1}{2} \Theta_{(l)} \tilde{q}_{ab} \label{eqn:shear},
\end{align}
while the shear of $n^a$ is not used in this paper. Note that $\sigma_{ab}$ is related to but different from the shear spin coefficient $\sigma$, which is usually defined using a complex null tetrad. 

A \textit{marginally outer trapped surface (MOTS)} is a surface $\mathcal{S}$ satisfying $\Theta_{(l)} = 0$ (following the convention in Ref.~\cite{gr-qc/0503109}). A MOTS is called a \textit{future MOTS} if $\Theta_{(n)} < 0$, or a \textit{past MOTS} if $\Theta_{(n)} > 0$. The notion of a MOTS is quasilocal, which makes it very convenient because the calculation does not require the knowledge of a full spacetime. In numerical simulations of BHs, there are efficient algorithms \cite{gr-qc/9606010, gr-qc/9609059, gr-qc/9707050, gr-qc/0004062, gr-qc/0306056} that compute MOTSs to locate BHs on every Cauchy surface $\Sigma$.

A \textit{marginally trapped tube} is a smooth 3-manifold $\mathcal{H}$ foliated by future MOTSs \cite{gr-qc/0503109}. The 3-manifold $\mathcal{H}$ is said to be a \textit{dynamical horizon}\footnote{Other literature may use different definitions of a dynamical horizon. For example, Ref.~\cite{2006.03939} and the Appendix~B of Ref.~\cite{gr-qc/0308033} allow dynamical horizons to be timelike. We also note that the original definition of a dynamical horizon does not require $l^a$ and $n^a$ to be outgoing and ingoing \cite{gr-qc/0207080}.} \cite{gr-qc/0207080, gr-qc/0308033, gr-qc/0407042, gr-qc/0503109} if it is spacelike, or a \textit{timelike membrane} if it is timelike. We call $\mathcal{H}$ a \textit{non-expanding horizon} if it is null\footnote{The foliation in the definition of a non-expanding horizon only requires MOTSs, instead of future MOTSs. To define a non-expanding horizon in a non-vacuum spacetime, an additional condition is imposed on the stress-energy tensor $T_{ab}$: $-T^{ab} U_b$ is causal and future directed for any future-directed null normal $U^b$ to $\mathcal{H}$. This is an energy condition weaker than the dominant energy condition.} \cite{gr-qc/0005083, gr-qc/0006006, gr-qc/0111067}. A non-expanding horizon is called an \textit{isolated horizon}\footnote{In a non-vacuum spacetime, matter fields must be ``time'' independent on an isolated horizon as well, where ``time'' is understood as the parameter generated by $\mathring{l}^a$.} \cite{gr-qc/0005083, gr-qc/0006006, gr-qc/0111067} if there is a specific null normal $\mathring{l}^a$ to $\mathcal{H}$ such that
\begin{align}
	(\mathcal{L}_{\mathring{l}} D_a - D_a \mathcal{L}_{\mathring{l}}) W^a = 0, \label{eqn:IHdef}
 \end{align}
for any tangent vector $W^a$ on $\mathcal{H}$. Here $D_a$ is the covariant derivative compatible with the (degenerate) metric $q_{ab}$ induced on $\mathcal{H}$\footnote{Since $q_{ab}$ is degenerate, there exist infinitely many covariant derivatives compatible with it. The covariant derivative $D_a$ here is uniquely defined as the pullback of $\nabla_a$. This can be done, because the non-expanding horizon is shear free.} We are not interested in the specific form of $\mathring{l}^a$ on an isolated horizon, though it can be constructed from any null normal (see Sec.~IV~B in \cite{gr-qc/0111067}).

After the merger of a BBH, the outermost MOTSs (called the \textit{common horizons}) on Cauchy surfaces trace out a dynamical horizon.\footnote{Reference~\cite{2006.03939} shows that a tiny portion of early common horizons may admit $\Theta_{(n)}\ge0$, so the 3-manifold foliated by these early common horizons may not strictly obey the definition of a dynamical horizon used in this paper. However, a portion of $\Theta_{(n)}\ge0$ does not affect the conclusions of this paper.} As we expect the remnant BH to be Kerr, this dynamical horizon should asymptote to an axisymmetric isolated horizon \cite{gr-qc/0005083} as the BH settles down. We are only interested in this dynamical horizon (which is, the stack of common horizons) in the rest of this paper, so we reserve the symbol $\mathcal{H}$ to represent this dynamical horizon henceforth.

\begin{figure}[t]
\centering 
\includegraphics[width=\linewidth]{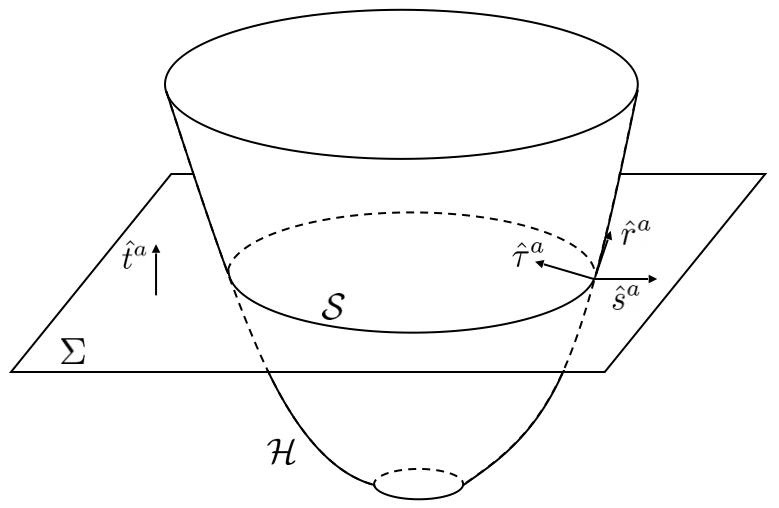}
\caption{Dynamical horizon in a numerical simulation. The
common horizon $\mathcal{S}$ is computed on the Cauchy
surface $\Sigma$ (the horizontal plane). The dynamical
horizon $\mathcal{H}$ (the paraboloid) consists of a stack
of $\mathcal{S}$. Note that these shapes do not reflect the
actual appearance of these quantities. See Sec.~\ref{sec:dh}
for the definitions of the vectors. This
figure is a modification of Fig.~1 of Ref.~\cite{1306.5697}.}
\label{fig:DH}
\end{figure}

We visualize the relation among $\mathcal{S}$, $\mathcal{H}$,
and $\Sigma$ in Fig.~\ref{fig:DH}. The figure is based
on Fig.~1 of Ref.~\cite{1306.5697}, with slightly different use of
symbols. This figure is merely illustrative:
the shapes of the
objects in this figure do not reflect their actual appearance in
a numerical simulation. The horizontal plane represents a Cauchy
surface $\Sigma$, and the circle on this plane represents the
common horizon $\mathcal{S}$. The common horizons on all Cauchy
surfaces constitute a dynamical horizon $\mathcal{H}$, shown as
the paraboloid. There are four vectors in this figure: $\hat{t}^a$
is the unit timelike normal to $\Sigma$, $\hat{\tau}^a$ the unit
timelike normal to $\mathcal{H}$ within the spacetime, $\hat{r}^a$
the unit spacelike normal to $\mathcal{S}$ within $\mathcal{H}$,
and $\hat{s}^a$ the unit spacelike normal to $\mathcal{S}$ within
$\Sigma$. Based on these unit vectors, we fix the scaling freedom
in $l\cdot n = -1$ by choosing
\begin{align}
	l^a = \hat{\tau}^a + \hat{r}^a, \qquad n^a = \frac{1}{2}(\hat{\tau}^a - \hat{r}^a). \label{eqn:ln_def}
\end{align}
We also define another set of null normals that satisfy the same normalization, \{$l',n'$\}, such that
\begin{align}
	l'^a = \hat{t}^a + \hat{s}^a, \qquad n'^a = \frac{1}{2}(\hat{t}^a - \hat{s}^a).
\end{align}

\subsection{Multipole moments} \label{sec:moment_def}

The notion of multipole moments on horizons was first introduced
for an isolated horizon \cite{gr-qc/0401114}. If an isolated
horizon is axisymmetric, multipole moments are defined as the
multipolar expansion of the Weyl scalar $\Psi_2$. Multipole moments were later generalized to a dynamical horizon in
Refs.~\cite{gr-qc/0604015, 0907.0280, 1306.5697}. As mentioned in the
previous section, we only consider a dynamical horizon $\mathcal{H}$
that asymptotes to an axisymmetric isolated horizon. In simulations, the late portion of $\mathcal{H}$ can be treated
as an axisymmetric isolated horizon to within numerical accuracy. We
construct multipole moments on such a dynamical horizon by following
Ref.~\cite{1306.5697}.

\subsubsection{Multipole moments on an axisymmetric $\mathcal{S}$} \label{sec:build_sph_coord_S}

We start by choosing a pair of angular coordinates $(\theta, \phi)$ on
$\mathcal{S}$. If $\mathcal{S}$ is axisymmetric (as in the late portion of
$\mathcal{H}$), there is a natural choice of $(\theta, \phi)$
\cite{gr-qc/0401114}. Let $\varphi^a$ on $\mathcal{S}$ be the rotational Killing
vector field, which generates closed integral curves and vanishes at exactly two
points (the poles). Let $\phi$ be the affine parameter of each closed integral curve with range $[0,2\pi)$. We then pick a new curve that connects the two poles and is orthogonal to $\varphi^a$ everywhere, and we set it to be the prime meridian $\phi=0$. We define a variable $\zeta$ that satisfies
\begin{align}
	&\tilde{D}_a \zeta = \frac{1}{R^2}\tilde{\epsilon}_{ba} \varphi^b, \label{eqn:zeta} \\
	&\oint_{\mathcal{S}} \zeta d^2V = 0,
\end{align}
where $\tilde{D}_a$ is the covariant derivative compatible with $\tilde{q}_{ab}$, $\tilde{\epsilon}_{ab}$ the area 2-form, $d^2V$ the corresponding area element, $R=\sqrt{A/4\pi}$ the areal radius, and $A$ the area. It is necessary that $\zeta$ has range $[-1,1]$. We obtain the angle $\theta$ via $\zeta=\cos\theta$. Note that there is a rotational degree of freedom in choosing the prime meridian, and we will fix this freedom in Sec.~\ref{sec:bbh_sim}. 

In the $(\theta, \phi)$ coordinates, the induced metric on $\mathcal{S}$ can be written as \cite{gr-qc/0401114}
\begin{align}
	\tilde{q}_{ab} &= \frac{R^4\sin^2\theta}{|\vec{\varphi}|^2} (d\theta)_a (d\theta)_b + |\vec{\varphi}|^2 (d\phi)_a (d\phi)_b,
\end{align}
where $|\vec{\varphi}|^2 = \varphi^a\varphi_a$. The compatible area element, $d^2V = R^2\sin\theta d\theta d\phi$, is the same as the area element of a fictitious round 2-sphere metric,
\begin{align}
	\mathring{q}_{ab} &= R^2 [(d\theta)_a (d\theta)_b +  \sin^2\theta (d\phi)_a (d\phi)_b].
\end{align}
Spherical harmonics\footnote{Spin-weighted spherical harmonics can be defined similarly, but we do not use them on a horizon in this paper.} are then defined as usual,
\begin{align}
	Y_{\ell m}(\theta, \phi) = \sqrt{\frac{2\ell +1}{4\pi} \frac{(\ell -m)!}{(\ell +m)!}} P_{\ell}^{m}(\cos\theta) e^{im\phi} \label{eqn:ylm},
\end{align}
where $P_{\ell}^{m}(x)$ are the associated Legendre polynomials (with the Condon–Shortley phase convention) \cite{Courant1_skip}. These $Y_{\ell m}$ are orthogonal on $\mathcal{S}$:
\begin{align}
	\oint_{\mathcal{S}} Y^*_{\ell m} Y_{\ell' m'} d^2V = R^2 \delta_{\ell \ell'} \delta_{mm'},
\end{align}
where * denotes complex conjugation, and the integration is with respect to the area 2-form of $\tilde{q}_{ab}$.

On an axisymmetric $\mathcal{S}$, we define \textit{mass multipole moments} (or simply \textit{mass moments}) $I_{\ell m}$ and \textit{spin multipole moments} (\textit{spin moments}) $L_{\ell m}$ as
\begin{align}
	I_{\ell m} &= \frac{1}{4} \oint_{\mathcal{S}} \tilde{\mathcal{R}} Y^*_{\ell m} d^2V,  \label{eqn:IHIlm} \\
	L_{\ell m} &= \frac{1}{2} \oint_{\mathcal{S}} \tilde{\epsilon}^{ab} \tilde{\omega}_b  \tilde{D}_a Y^*_{\ell m} d^2V \label{eqn:IHLlm_original}.
\end{align}
Here, $\tilde{\mathcal{R}}$ is the $\tilde{q}_{ab}$-compatible Ricci scalar\footnote{We use the following convention of the spacetime Riemann tensor $^{(4)}R_{abcd}$: $(\nabla_a\nabla_b - \nabla_b\nabla_a) v_c = {}^{(4)}{R_{abc}}^d v_d$ for any 4D 1-form $v_a$. The spacetime Ricci scalar is then defined as $^{(4)}R = {}^{(4)}{R^{ab}}_{ab}$. The Riemann tensor and Ricci scalar on a horizon follow similar conventions.} on $\mathcal{S}$, and $\tilde{\omega}_a$ is the rotational 1-form,
\begin{align}
	\tilde{\omega}_a = -\tilde{q}^{\ b}_a n^c \nabla_b l_c \label{eqn:omega1}.
\end{align}
These multipole moments are related to the Weyl scalar $\Psi_2$ by
\begin{align}
	I_{\ell m}+iL_{\ell m} &= -\oint_{\mathcal{S}} \Psi_2 Y^*_{\ell m} d^2V \label{eqn:psi2moments},
\end{align}
because $\Psi_2$ on an isolated horizon satisfies \cite{gr-qc/0401114}
\begin{align}
	\Psi_2 = -\frac{1}{4}\tilde{\mathcal{R}} + \frac{i}{2} \tilde{\epsilon}^{ab} \tilde{D}_a \tilde{\omega}_b \label{eqn:psi2_ricci_omega}.
\end{align}
Although the $m=0$ modes ($I_{\ell,0}$ and $L_{\ell,0}$) are the only nonvanishing modes because of the axisymmetry of $\mathcal{S}$, we keep $m$ arbitrary so that we can easily generalize multipole moments on any MOTS of $\mathcal{H}$ in the coming sections.

At the end of Sec.~\ref{sec:dh}, we fixed the scaling freedom in \{$l,n$\}, so there is no ambiguity in the definition of $\tilde{\omega}_a$. As the scaling freedom does not affect $\Psi_2$ and $L_{\ell m}$, we can replace the current pair \{$l,n$\} in Eq.~\eqref{eqn:omega1} by any other null \{$l,n$\} subject to $l\cdot n = -1$. For the purpose of this paper, it is more convenient and stable to use the pair \{$l',n'$\} in the definition of a rotational 1-form. We define
\begin{align}
	\omega_a = - \gamma^{\ b}_a n'^c \nabla_b l'_c = \gamma^{\ b}_a \hat{s}^c \nabla_b \hat{t}_c = \left(K^{\Sigma}\right)_{a b} \hat{s}^b, \label{eqn:omega2}
\end{align}
where $\gamma_{ab}$ is the spatial metric induced on $\Sigma$, and $\left(K^{\Sigma}\right)_{a b} = \gamma_a^{\ c} \nabla_c \hat{t}_b$ is the extrinsic curvature\footnote{We use a sign convention different from Ref.~\cite{Baumgarte1_skip}.} of $\Sigma$ within the spacetime. Replacing $\tilde{\omega}_a$ by $\omega_a$, we have an equivalent definition of spin moments,
\begin{align}
	L_{\ell m} &= \frac{1}{2} \oint_{\mathcal{S}} \tilde{\epsilon}^{ab} \omega_b  \tilde{D}_a Y^*_{\ell m} d^2V \label{eqn:IHLlm_stable}.
\end{align}
It is also useful to rewrite Eq.~\eqref{eqn:IHLlm_stable} as
\begin{align}
	L_{\ell m} &= -\frac{1}{2} \oint_{\mathcal{S}} \omega_a \varphi^a_{\ell m} d^2V \label{eqn:IHLlm_divfreevec}, \\
	\varphi^a_{\ell m} &= \tilde{\epsilon}^{ab} \tilde{D}_b Y^*_{\ell m}.
\end{align}
The vectors $\varphi^a_{\ell m}$ provide a complete basis for divergence-free vectors on $\mathcal{S}$ \cite{1306.5697}, and the vector $\varphi^a_{1,0}$ is parallel to the rotational Killing vector field $\varphi^a$ [see Eq.~\eqref{eqn:zeta}].

\subsubsection{2+1 decomposition of $\mathcal{H}$} \label{sec:2+1}

Except in special situations (e.g., head-on collisions of two BHs), an arbitrary MOTS $\mathcal{S}$ in $\mathcal{H}$ is not axisymmetric. It then becomes tricky to choose a suitable pair of angular coordinates $(\theta, \phi)$. We cannot simply apply the construction process in the previous section, since there is no longer a rotational Killing vector field on an arbitrary $\mathcal{S}$. However, we can still take advantage of the axisymmetry of those $\mathcal{S}$ in the late portion of $\mathcal{H}$. In particular, instead of defining $(\theta, \phi)$ separately and locally on every $\mathcal{S}$, we adopt the idea in Ref.~\cite{1306.5697} and build a vector $X^a$ on $\mathcal{H}$ that connects $(\theta, \phi)$ on all $\mathcal{S}$ in a canonical way. We call $X^a$ the \textit{stitching vector} and regard the coordinates $(\theta, \phi)$ as ``evolving'' along $X^a$ on $\mathcal{H}$.

A dynamical horizon is essentially a stack of MOTSs, so it naturally admits a 2+1 decomposition, similar to a 3+1 decomposition of spacetime (cf. \cite{Baumgarte1_skip} for an introduction of a 3+1 decomposition). Additionally, the foliation by MOTSs is unique for a dynamical horizon, in contrast to a non-expanding horizon \cite{gr-qc/0503109}. We treat $X^a$ as the time vector of the 2+1 decomposition, which has the form,
\begin{align}
	X^a = \tilde{\alpha} \hat{r}^a + \tilde{\beta}^a \label{eqn:X1},
\end{align}
where $\tilde{\beta}^a$ is a tangent vector to be specified on $\mathcal{S}$. The scalar $\tilde{\alpha}$ and the vector $\tilde{\beta}^a$ are the lapse and the shift in this 2+1 decomposition. We call $\tilde{\alpha}$ the 2-lapse and $\tilde{\beta}^a$ the 2-shift, to distinguish them from the usual lapse $\alpha$ and shift $\beta^a$ used in a 3+1 decomposition. 

The 2-lapse is required to preserve the foliation of MOTSs. Let the MOTSs be labeled by a parameter ${v}$ that is smooth on $\mathcal{H}$. In other words, each MOTS corresponds to a ${v}=\mathrm{constant}$ surface. (We will identify ${v}$ with simulation time $t$ in a numerical simulation, but we continue using ${v}$ here to keep the discussion general.) For $X^a$ being the time vector, we require ${v}$ to be the parameter of the integral curve generated by $X^a$, i.e., $X^a = (\partial_{v})^a$. This implies
\begin{align}
	\tilde{\alpha} = (q^{ab} D_a{v} D_b{v})^{-1/2},
\end{align}
where $q_{ab}$ is the induced metric on $\mathcal{H}$, and $D_a$ is the covariant derivative compatible with $q_{ab}$.\footnote{The definitions of $q_{ab}$ and $D_a$ on a dynamical horizon are consistent with the ones on an isolated horizon [Eq.~\eqref{eqn:IHdef}].} Note that $\tilde{\alpha}$ tends to 0 when $q_{ab}$ approaches a degenerate metric, as in the case when a merged BH approaches equilibrium. However, $X^a$ does not tend to 0, because the limiting behavior of $\hat{r}^a$ is nontrivial. This brings difficulties in the numerical calculation of $X^a$, and we will handle them in the next section.

Spin moments on an isolated horizon (or the late portion of $\mathcal{H}$) can be defined using a set of divergence-free vector fields [Eq.~\eqref{eqn:IHLlm_divfreevec}]. This inspires us to define spin moments on a general $\mathcal{S}$ that also uses divergence-free vector fields. We can obtain a canonical set of divergence-free vector fields on all $\mathcal{S}$ by imposing a mapping condition on $X^a$: $X^a$ maps divergence-free vector fields among different $\mathcal{S}$ isomorphically. Specifically, once $\varphi^a_{\ell m}$ (the divergence-free vectors on an axisymmetric $\mathcal{S}$) are known, we can Lie drag them along $X^a$ to all other MOTSs. In the mathematical language, we are looking for a vector $X^a$ on $\mathcal{H}$, that satisfies the following statement. Given a vector field $\xi^a$ that is divergence free on a particular MOTS $\mathcal{S}$, i.e., 
\begin{align}
\mathcal{L}_\xi \tilde{\epsilon}_{ab} \stackrel{\mathcal{S}}{=} 0 \label{eqn:divfreeS},
\end{align}
we can define $\xi^a$ on other MOTSs via $\mathcal{L}_X \xi^a = 0$, and the resultant vector field stays divergence free on all MOTSs, i.e.,
\begin{align}
\mathcal{L}_\xi \tilde{\epsilon}_{ab} \stackrel{\mathcal{H}}{=} 0 \label{eqn:divfreeH}.
\end{align}

The trivial choice $\tilde{\beta}^a = 0$ does not satisfy this mapping condition. To see this, we first note that Eq.~\eqref{eqn:divfreeH} implies $\mathcal{L}_X \mathcal{L}_\xi \tilde{\epsilon}_{ab} = 0$. Meanwhile, we know $\mathcal{L}_X \mathcal{L}_\xi \tilde{\epsilon}_{ab} = \mathcal{L}_\xi \mathcal{L}_X \tilde{\epsilon}_{ab} = \mathcal{L}_\xi (\tilde{\alpha}\tilde{K})$ because $\mathcal{L}_X \xi^a = 0$ and $\mathcal{L}_{\tilde{\alpha} \hat{r}} \tilde{\epsilon}_{ab}=\tilde{\alpha} \tilde{K} \tilde{\epsilon}_{ab}$. Here, $\tilde{K}_{ab} = \tilde{q}_{a}^{\ c} \tilde{q}_{b}^{\ d} \nabla_c \hat{r}_d$ is the extrinsic curvature of $\mathcal{S}$ within $\mathcal{H}$, and $\tilde{K} = \tilde{K}^a_{\ a}$ is its trace. The expression $\mathcal{L}_\xi (\tilde{\alpha}\tilde{K})$ is generally nonzero, which contradicts $\mathcal{L}_X \mathcal{L}_\xi \tilde{\epsilon}_{ab} = 0$. 

We can find a viable choice of $\tilde{\beta}^a$ by eliminating the inhomogeneity in $\tilde{\alpha} \tilde{K}$ from $\mathcal{L}_X \tilde{\epsilon}_{ab}$. In detail, the inhomogeneity is
\begin{align}
	\tilde{\alpha} \tilde{K} - \frac{1}{4\pi R^2}\oint_{\mathcal{S}} \tilde{\alpha} \tilde{K} d^2V = \tilde{\alpha} \tilde{K} - \frac{2\dot{R}}{R},
\end{align}
where $\dot{R} = dR/d{v}$. We choose $\tilde{\beta}^a$ such that
\begin{align}
	\tilde{D}_a \tilde{\beta}^a = -(\tilde{\alpha} \tilde{K} - 2\dot{R}/R) \label{eqn:beta},
\end{align}
which implies $\mathcal{L}_X \tilde{\epsilon}_{ab} = (2\dot{R}/R) \tilde{\epsilon}_{ab}$. Note that $2\dot{R}/R$ is only a function of ${v}$. The differential equation $\mathcal{L}_X \mathcal{L}_\xi \tilde{\epsilon}_{ab} = \mathcal{L}_\xi \mathcal{L}_X \tilde{\epsilon}_{ab} = (2\dot{R}/R) \mathcal{L}_\xi \tilde{\epsilon}_{ab}$, together   with the initial condition Eq.~\eqref{eqn:divfreeS}, admits the unique solution Eq.~\eqref{eqn:divfreeH}. In other words, this choice of $\tilde{\beta}^a$ [Eq.~\eqref{eqn:beta}] satisfies the mapping condition of $X^a$. In the numerical implementation of Eq.~\eqref{eqn:beta}, it is more convenient to define
\begin{align}
	\tilde{\beta}^a =  \tilde{q}^{ab} \tilde{D}_b g \label{eqn:beta_g}
\end{align}
and solve
\begin{align}
	\tilde{q}^{ab} \tilde{D}_a \tilde{D}_b g = -(\tilde{\alpha} \tilde{K} - 2\dot{R}/R) \label{eqn:gPDE1}
\end{align}
for $g$ on every $\mathcal{S}$. The integration constant in the solution of $g$ does not affect $\tilde{\beta}^a$ and can be selected arbitrarily. 

We have thus constructed the time vector $X^a$ that satisfies the following four properties:
\begin{enumerate}
	\item $X^a$ is constructed covariantly.
	\item $X^a$ preserves the foliation of $\mathcal{H}$.
	\item $X^a$ maps divergence-free vectors isomorphically among different $\mathcal{S}$.
	\item If $\mathcal{H}$ is axisymmetric, $X^a$ preserves the rotational Killing vector.
\end{enumerate}
Now, we are ready to define multipole moments on a general dynamical horizon $\mathcal{H}$ whose late portion is axisymmetric. We first construct $Y_{\ell m}(\theta,\phi)$ on an axisymmetric but otherwise arbitrary $\mathcal{S}$ as described in Sec.~\ref{sec:build_sph_coord_S}. We then extend $Y_{\ell m}(\theta,\phi)$ to the whole $\mathcal{H}$ by
\begin{align}
	\mathcal{L}_X Y_{\ell m} = 0. \label{eqn:liedrag}
\end{align}
We define mass (multipole) moments $I_{\ell m}$ and spin (multipole) moments $L_{\ell m}$ as functions of ${v}$ (or time $t$ in numerical simulations),\footnote{We define multipole moments using the complex conjugates of the spherical harmonics, instead of the spherical harmonics themselves. This is different from Ref.~\cite{1306.5697}.}
\begin{align}
	I_{\ell m} &= \frac{1}{4} \oint_{\mathcal{S}} \tilde{\mathcal{R}} Y^*_{\ell m} d^2V, \label{eqn:DHIlm} \\
	L_{\ell m} &= \frac{1}{2} \oint_{\mathcal{S}} \tilde{\epsilon}^{ab} \omega_b  \tilde{D}_a Y^*_{\ell m} d^2V, \label{eqn:DHLlm}
\end{align}
where $\tilde{\mathcal{R}}$ still represents the $\tilde{q}_{ab}$-compatible Ricci scalar and $\omega_a$ is still defined by Eq.~\eqref{eqn:omega2}. These multipole moments are dimensionless, so they are sometimes referred to as \textit{geometric multipole moments}. They extend Eqs.~\eqref{eqn:IHIlm} and \eqref{eqn:IHLlm_stable}, but the relation among $\Psi_2$, $\tilde{\mathcal{R}}$, and $\tilde{\epsilon}^{ab} \tilde{D}_a \tilde{\omega}_b$ is not as simple as Eq.~\eqref{eqn:psi2_ricci_omega}, so Eq.~\eqref{eqn:psi2moments} no longer holds on a general MOTS.\footnote{Penrose and Rindler studied the right hand side of Eq.~\eqref{eqn:psi2_ricci_omega} and called its additive inverse the complex curvature \cite{Penrose1_skip}
\begin{align}
	\mathcal{K} = \frac{1}{4}\tilde{\mathcal{R}} - \frac{i}{2} \tilde{\epsilon}^{ab} \tilde{D}_a \tilde{\omega}_b.
\end{align}
They also provide the relation between $\Psi_2$ and $\mathcal{K}$ in Ref.~\cite{Penrose1_skip}. The complex curvature is closely related to horizon's tendicity and vorticity, which are visualized in Ref.~\cite{1012.4869}.}
Also, see Refs.~\cite{0907.0280, 1306.5697} for other definitions of multipole moments on a dynamical horizon.

\subsubsection{Alternative calculation of $X^a$}

The 2+1 decomposition, Eq.~\eqref{eqn:X1}, nicely resembles the 3+1 decomposition of a spacetime, but there exist numerical difficulties in the implementation. For example, as $\mathcal{H}$ becomes null and $q_{ab}$ becomes degenerate, $\tilde{\alpha}$ tends to zero and the components of $\hat{r}^a$ diverge. References~\cite{gr-qc/0308033, 1306.5697} discuss these ill behaviors and provide an alternative solution to handle them. Using this alternative solution, we can compute $X^a$ stably on both dynamical and isolated horizons, as described below.

Let $V^a$ be a normal to $\mathcal{S}$ within $\mathcal{H}$ such that
\begin{align}
	V^a D_a{v} = 1. \label{eqn:V}
\end{align} 
The vector $V^a$ is unique and well defined on both dynamical and isolated horizons. It is null on an isolated horizon and reduces to the spacelike vector $\tilde{\alpha} \hat{r}^a$ on a dynamical horizon. Thus, it is more promising to use 
\begin{align}
	X^a = V^a + \tilde{\beta}^a \label{eqn:X2}
\end{align}
in numerical simulations. The 2-shift $\tilde{\beta}^a$ may also be
problematic because of its dependence on $\tilde{\alpha}$ and $\tilde{K}$
[Eq.~\eqref{eqn:gPDE1}]. As $\mathcal{H}$ becomes null, evaluating
$\tilde{\alpha}$ and $\tilde{K}$ may become unstable. A better way
to obtain $\tilde{\beta}^a$ is to use the following differential
equation for $g$, 
\begin{align}
	\tilde{q}^{ab} \tilde{D}_a \tilde{D}_b g = -(\frac{1}{2}\tilde{q}^{ab} \mathcal{L}_V \tilde{q}_{ab} - \frac{2\dot{R}}{R}). \label{eqn:gPDE2}
\end{align}
This generalizes Eq.~\eqref{eqn:gPDE1}, because $\tilde{q}^{ab} \mathcal{L}_V \tilde{q}_{ab}$ reduces to $2\tilde{\alpha} \tilde{K}$ on a dynamical horizon.

As a simple example, let us consider the event horizon of a Kerr BH in the Boyer-Lindquist coordinates \{$t_\mathrm{BL}, r_\mathrm{BL}, \theta_\mathrm{BL}, \phi_\mathrm{BL}$\}. This event horizon is automatically an isolated horizon \cite{gr-qc/0005083} and admits a foliation of MOTSs labeled by ${v}=t_\mathrm{BL}$. The 2-shift $\tilde{\beta}^a$ vanishes on the horizon, so $X^a$ coincides with the null Killing vector $V^a = (\partial_{t_\mathrm{BL}})^a + \Omega_H (\partial_{\phi_\mathrm{BL}})^a$, where $\Omega_H$ is the horizon angular velocity \cite{Wald1_skip}.

\subsubsection{Balance laws} \label{sec:ballaw}

Let $\Delta\mathcal{H}$ be the portion of a dynamical horizon $\mathcal{H}$ between any two MOTSs $\mathcal{S}_1$ and $\mathcal{S}_2$. The gravitational energy flux across $\Delta\mathcal{H}$ is defined as \cite{gr-qc/0207080, gr-qc/0308033, gr-qc/0407042}
\begin{align}
	\mathcal{F}_g (\Delta\mathcal{H}) = \mathcal{F}_{g,\sigma} (\Delta\mathcal{H}) + \mathcal{F}_{g,\zeta} (\Delta\mathcal{H}), \label{eqn:flux}
\end{align}
where the first term on the right hand side,
\begin{align}
	\mathcal{F}_{g,\sigma} (\Delta\mathcal{H}) &= \frac{1}{16\pi} \int_{\Delta\mathcal{H}} |dR| \sigma_{ab} \sigma^{ab} d^3V,
\end{align}
arises naturally at a perturbed event horizon \cite{Hawking1_skip}, and the second term,
\begin{align}
	\mathcal{F}_{g,\zeta} (\Delta\mathcal{H}) &= \frac{1}{8\pi} \int_{\Delta\mathcal{H}} |dR| \zeta_a\zeta^a d^3V,
\end{align}	
arises only when $\Delta\mathcal{H}$ is not null. Here,
\begin{align}	
	|dR| &= \sqrt{q^{ab} D_a R D_b R} = \dot{R}/\sqrt{\tilde{\alpha}}, \\
	\zeta^a &= \tilde{q}^{ab} \hat{r}^c \nabla_c l_b = \tilde{\omega}^a + \tilde{D}^a \ln |dR|, \label{eqn:zeta_vec}
\end{align}
$\sigma_{ab}$ is defined in Eq.~\eqref{eqn:shear}, and
$d^3V$ is the volume element determined by $q_{ab}$. It
is feasible but inconvenient to use the energy flux
$\mathcal{F}_g(\Delta\mathcal{H})$ in numerical studies, because the
expression depends on two simulation times, $t_1$ for $\mathcal{S}_1$
and $t_2$ for $\mathcal{S}_2$. A more practical choice is the
time derivative
\begin{align}
	\frac{d\mathcal{F}_g}{dt} = \frac{d}{dt} \mathcal{F}_g(\Delta\mathcal{H}) = \lim_{t_2\rightarrow t_1} \frac{\mathcal{F}_g(\Delta\mathcal{H})}{t_2 - t_1}.
\end{align}
We call $d\mathcal{F}_g/dt$ the \textit{energy flux rate} and may regard it as the energy flux across a common horizon. Its constituents $d\mathcal{F}_{g,\sigma}/dt$ and $d\mathcal{F}_{g,\zeta}/dt$ can be defined similarly.

The difference between the areal radii $R_1$ (of $\mathcal{S}_1$) and $R_2$ (of $\mathcal{S}_2$) is proportional to the energy flux \cite{gr-qc/0207080, gr-qc/0308033, gr-qc/0407042}:\footnote{In a non-vacuum spacetime, matter fields would have contribution to the right-hand side.}
\begin{align}
	R_2-R_1= 2\mathcal{F}_g = \frac{1}{8\pi} \int_{\Delta\mathcal{H}} |dR| \left( \sigma_{ab} \sigma^{ab} + 2\zeta_a\zeta^a \right) d^3V. \label{eqn:radiusbal}
\end{align}
This is the area balance law for areal radii. The differential version is more convenient in numerical studies:
\begin{align}
	\frac{dR}{dt}= 2\frac{d\mathcal{F}_g}{dt}. \label{eqn:radiusbal_diff}
\end{align}

There are balance laws for multipole moments as well. The difference in $I_{\ell m}$ and $L_{\ell m}$ between $\mathcal{S}_1$ and $\mathcal{S}_2$ can also be expressed as a flux across $\Delta\mathcal{H}$ \cite{1306.5697}:
\begin{align}
	I_{\ell m}&[\mathcal{S}_2] - I_{\ell m}[\mathcal{S}_1] \nonumber \\
	= &\int_{\Delta\mathcal{H}} |dR| \left(\frac{1}{4\dot{R}} Y^*_{\ell m} \mathcal{L}_X \tilde{\mathcal{R}}+\frac{1}{R} \zeta^{a} \partial_{a} Y^*_{\ell m}\right) d^3V \nonumber \\
	&+\int_{\Delta\mathcal{H}} \frac{|dR|}{2 R} \left(\sigma_{ab} \sigma^{ab} + 2\zeta_a\zeta^a \right) Y^*_{\ell m} d^3V, \label{eqn:Ilmbal} \\
	L_{\ell m}&[\mathcal{S}_2] - L_{\ell m}[\mathcal{S}_1] \nonumber \\
	= &\frac{1}{2} \int_{\Delta\mathcal{H}} \left[ \left(K^{\mathcal{H}}\right)^{a b}-K^{\mathcal{H}} q^{a b}\right] D_a\left(\tilde{\epsilon}_{b c} \tilde{D}^{c} Y^*_{\ell m}\right) d^3V \label{eqn:Llmbal}.
\end{align}
Here, $\left(K^{\mathcal{H}}\right)_{ab} = q_{a}^{\ c} q_{b}^{\ d} \nabla_{c} \hat{\tau}_{d}$ is the extrinsic curvature of $\mathcal{H}$ within the spacetime $\mathscr{M}$, and $K^{\mathcal{H}} = \left( K^{\mathcal{H}} \right)^a_{\ a}$ is its trace. The differential versions of these two balance laws are
\begin{align}
	\frac{dI_{\ell m}}{dt} = \frac{d}{dt} &\int_{\Delta\mathcal{H}} |dR| \left(\frac{1}{4\dot{R}} Y^*_{\ell m} \mathcal{L}_X \tilde{\mathcal{R}}+\frac{1}{R} \zeta^{a} \partial_{a} Y^*_{\ell m}\right) d^3V \nonumber \\
	+&\frac{d}{dt} \int_{\Delta\mathcal{H}} \frac{|dR|}{2 R} \left(\sigma_{ab} \sigma^{ab} + 2\zeta_a\zeta^a \right) Y^*_{\ell m} d^3V, \label{eqn:Ilmbal_diff}  \\
	\frac{dL_{\ell m}}{dt} = \frac{1}{2} &\frac{d}{dt} \int_{\Delta\mathcal{H}} \left[ \left( K^{\mathcal{H}}\right)^{a b} - K^{\mathcal{H}} q^{a b} \right] \nonumber \\ 
	&\hspace{2.2cm} \times D_a\left(\tilde{\epsilon}_{b c} \tilde{D}^{c} Y^*_{\ell m}\right) d^3V. \label{eqn:Llmbal_diff}
\end{align}

All these balance laws, Eqs.~\eqref{eqn:radiusbal}-\eqref{eqn:Llmbal_diff}, offer internal checks on numerical simulations, because both sides of these equations can be calculated independently. We will use them to check the correctness of our simulation in Appendix~\ref{sec:app_ballaw_error}.

\subsection{Quasinormal modes} \label{sec:qnm}

Perturbations of the Kerr spacetime can be described
by the Teukolsky equation \cite{Teukolsky1_skip,
Teukolsky2_skip}. It was first derived using the Kinnersley
tetrad \cite{Kinnersley1_skip} in Boyer-Lindquist
coordinates \{$t_\mathrm{BL}, r_\mathrm{BL}, \theta_\mathrm{BL},
\phi_\mathrm{BL}$\}. In this paper, we will only be concerned with the Teukolsky equation governing gravitational perturbations.
Let $\Psi_0^{(1)}$ and $\Psi_4^{(1)}$
denote the first-order perturbation of the Weyl scalars $\Psi_0$
and $\Psi_4$. Then, $\psi = \Psi_0^{(1)}$ has spin weight $s=2$ and
describes the ingoing gravitation wave, while $\psi = \rho^{-4}
\Psi_4^{(1)}$ is a spin-weight $s=-2$ quantity representing
the outgoing gravitation wave, where $\rho$ is one of the spin
coefficients of the Kerr metric.

The Teukolsky equation is separable. With appropriate boundary
conditions imposed at horizons and spatial infinity, it admits
 solutions
\begin{align}
	\psi_{\ell mn} = e^{-i \omega_{\ell mn} t_\mathrm{BL}} R(r_\mathrm{BL})\, _s\mathcal{Y}_{\ell m}(\theta_\mathrm{BL}, \phi_\mathrm{BL}, a\omega_{\ell mn}) \label{eqn:qnmgeneral}.
\end{align}
The indices $\ell, m$ represent angular modes, while $n$ represents
overtones. The indices take on integer values and satisfy $\ell\ge|s|$,
$|m|\le l$, and $n\ge0$. The quantity $\omega_{\ell mn}$ is
a complex number called the quasinormal mode
frequency,\footnote{There are two distinct families of QNMs:
the prograde modes, $\omega^+_{\ell m n}$, that corotate with
the BH, and the retrograde modes, $\omega^-_{\ell m n}$, that
counterrotate with the BH. They are related by $\omega^-_{\ell
(-m) n} = -(\omega^+_{\ell m n})^*$ \cite{Leaver1_skip, 0905.2975,
2107.05609}. In this paper, we will only consider $\omega^+_{\ell
m n}$ for $m \neq0 $, but we will use both $\omega^+_{\ell m n}$
and $\omega^-_{\ell m n}$ for $m = 0$. For the sake of readability,
we drop these superscripts and keep using the notation $\omega_{\ell
mn}$ throughout the paper. The meaning should be clear from the
context.} which necessarily has a negative imaginary component
\cite{Whiting1_skip, 1302.6902, 1910.02854, 2105.13329} because
the perturbed BH system is dissipative. Besides ($\ell, m,n$),
the frequency $\omega_{\ell m n}$ also depends on the spin weight
$s$, the mass $M_f$,\footnote{The final Kerr BH mass, $M_f$, is
smaller than the initial total ADM mass of the system, $M$. See
Sec.~\ref{sec:bbh_sim} for the numerical value of their ratio
in our simulation.} and the dimensionless spin $\chi_f$ of the
unperturbed Kerr BH. We calculate the values of $\omega_{\ell mn}$
using the \texttt{qnm} package \cite{1908.10377}. The functions
$_s\mathcal{Y}_{\ell m} (\theta_\mathrm{BL}, \phi_\mathrm{BL},
a\omega_{\ell mn})$ are the spin-weighted spheroidal harmonics,
where $a=\chi_f M_f$ is the dimensionful spin (i.e., spin angular
momentum per unit mass). They reduce to the spin-weighted spherical
harmonics $_sY_{\ell m} (\theta_\mathrm{BL}, \phi_\mathrm{BL})$
\cite{Goldberg1_skip} if $a=0$, which further reduce to the
usual spherical harmonics $Y_{\ell m} (\theta_\mathrm{BL},
\phi_\mathrm{BL})$ if $s=0$. The radial part $R(r_\mathrm{BL})$ is
not important in this paper. For further discussion on the Teukolsky
equation, see Refs.~\cite{Teukolsky1_skip, Teukolsky2_skip,
Teukolsky3_skip, Hartle2_skip}. Also, see Ref.~\cite{0905.2975}
for a review of QNMs and Ref.~\cite{Breuer1_skip} for details of
the spin-weighted spheroidal harmonics.

In a BBH simulation, one often expands a physical quantity on a
2-sphere $S^2$ into angular modes using spherical harmonics. If one
performs such an expansion on a time collection of 2-spheres, then each
angular mode is a function of simulation time $t$. To investigate
potential quasinormal behavior of a mode in the ringdown phase, one
then decomposes the mode into several damped sinusoids of $t$. For
example, strain $h$ is usually expanded into $h_{\ell m}$ using
the $s=-2$ spin-weighted spherical harmonics. Then, the ringdown
portion of $h_{22}$ can be modeled as a linear combination of $e^{-i
\omega_{22n} t}$ \cite{1903.08284, 1910.08708, 2010.08602}. Also,
see Ref.~\cite{2010.15186} for the QNM description of the shear spin
coefficient $\sigma$ on the horizon of a merged BH. Note that the
spherical harmonics used in simulations are constructed with respect
to some specifically chosen angular coordinates, and different
literature in general uses different sets of angular coordinates.

Several groups have studied the quasinormal behavior of mass
moments \cite{0907.0280,  1801.07048, 2006.03940, 2010.15186}. They
either consider head-on collisions of two BHs or use definitions of
multipole moments \textit {without referring to the connection among
MOTSs} (i.e., no Lie dragging along the vector $X^a$). In contrast,
we will investigate the quasinormal behavior of multipole moments
for an orbiting BBH system, and the definition of our multipole
moments does take into account the relation among MOTSs. We will
model mass and spin moments as linear combinations of QNMs, and
choose different models for different moments. We will describe
these models explicitly in Sec.~\ref{sec:results}, but no matter
what models we apply, we determine coefficients in these models by
unweighted least square linear fitting.

\section{Numerical implementation} \label{sec:numeric}

\subsection{Binary-black-hole simulation} \label{sec:bbh_sim}

We simulate the BBH system using the Spectral Einstein Code (SpEC) \cite{spec_skip}, which adopts the first order generalized harmonic formalism \cite{gr-qc/0512093}. SpEC constructs quasi-equilibrium initial data that is given by a Gaussian-weighted superposition of two single-BH analytic solutions \cite{0805.4192}. Spacetime quantities are evolved in the damped harmonic gauge after a smooth transition from the quasi-equilibrium initial gauge \cite{0909.3557}. SpEC uses excision boundaries that are placed slightly inside apparent horizons \cite{gr-qc/0407110, 1211.6079, 1412.1803}, and imposes constraint-preserving conditions on the outer boundary \cite{gr-qc/0512093, 0704.0782}. Apparent horizons are calculated using the fastflow method \cite{gr-qc/9707050}. A SpEC simulation starts with a spectral grid containing two excised regions (within two apparent horizons), and switches to a new grid that has only one excised region (within the common horizon) after merger. We consider the merger as the instant when the common horizon first appears. SpEC uses a dual-frame configuration \cite{gr-qc/0607056} whose domain arrangement is described in Ref.~\cite{1206.3015}. The adaptive mesh refinement algorithm, which SpEC uses to dynamically control grid resolutions and domain arrangement, is discussed in Refs.~\cite{1010.2777, 1405.3693}.

We evolve an equal-mass, non-spinning, noneccentric \cite{1012.1549}
BBH system. We use the same configuration as SXS:BBH:0389 in the
SXS catalog \cite{1904.04831} and record the simulation parameters
in Table~\ref{tbl:bbh_param}. We simulate the BBH system at two
resolutions. The target truncation errors of the adaptive mesh
refinement algorithm are $\sim 5 \times 10^{-8}$ for
the higher resolution and $\sim 2 \times 10^{-7}$
for the other resolution. Unless specified, the results in this
paper are generated from the higher resolution run. We only focus
on the post-merger portion of our BBH simulation. We set $t=0$ at
the merger (i.e., when the common horizon first appears). We assume
the merged BH settles down to the Kerr state at $t_f=500M$ (where $M$ is the initial total ADM mass of the BBH system), and we
shall see in Sec.~\ref{sec:i22} that this is a good assumption. The
final Kerr BH has dimensionless spin $\chi_f=0.68644$ (measured by
the method of approximate Killing vectors \cite{0805.4192}) and
mass $M_f=0.95162M$. Table~\ref{tbl:sample_omega} shows several
$\omega_{\ell mn}(\chi_f, M_f)$ that are used in this paper.

\begin{table}
	\caption{Parameters for the BBH simulation studied in this paper. The symbols $q$, $D_0$, $\Omega_0$, $\dot{a}_0$, and $e$ represent the mass ratio, initial coordinate separation, initial orbital frequency, initial rate of change of separation, and eccentricity. The symbols $\vec{\chi}_{A,B}$ stand for the dimensionless spin vectors of the two BHs. We choose the initial free data to be the Gaussian-weighted superposition of two BHs in the Kerr-Schild coordinates, and this is called \textit{superposed Kerr-Schild} in SpEC \cite{0805.4192}.}
	\label{tbl:bbh_param}
	\begin{ruledtabular}
		\begin{tabular}{@{\hspace{3em}} cc @{\hspace{3em}}}
			Parameter & Value \\
			\hline 
			Initial free data & superposed Kerr-Schild \\
			$q$ & 1 \\
			$D_0$ & 15.43$M$ \\
			$\Omega_0$ & 0.01525 \\
			$\dot{a}_0$ & $-$0.00003721  \\
			$\vec{\chi}_{A,B}$ & $(0,0,0)$ \\
			$e$ & $\sim$0.0009\\
			Number of orbits & 18.6\\
		\end{tabular}
	\end{ruledtabular}
\end{table}

\begin{table}
	\caption{The values of several spin-weight-2 QNM frequencies $\omega_{\ell mn}$ used in this paper. They are generated by the \texttt{qnm} package \cite{1908.10377}, based on the remnant parameter $M_f = 0.95162M$ and $\chi_f = 0.68644$. QNM frequencies are complex numbers. The real part, Re($\omega_{\ell mn}$), is the oscillation frequency, while the inverse imaginary part, $-1/$Im($\omega_{\ell mn}$), is the characteristic decay time. Note that we express the QNM frequencies in the unit of $M$, instead of $M_f$.}
	\label{tbl:sample_omega}
	\begin{ruledtabular}
		\begin{tabular}{@{\hspace{3em}} ccc @{\hspace{2em}}}
			$\ell,m,n$ & Re($\omega_{\ell mn}$) [$M^{-1}$] & $-1/$Im($\omega_{\ell mn}$) [$M$] \\
			\hline 
			2, 2, 0 & 0.5535 & 11.707 \\
			2, 2, 1 & 0.5410 & 3.8713 \\
			2, 2, 2 & 0.5180 & 2.2923 \\
			3, 2, 0 & 0.7920 & 11.235 \\
			4, 2, 0 & 1.0172 & 10.938 \\
			2, 0, 0 & 0.4132 & 11.236 \\
		\end{tabular}
	\end{ruledtabular}
\end{table}

We process the simulation following the procedure described
in Sec.~\ref{sec:moment_def}. We first calculate the invariant
spherical coordinates $(\theta, \phi)$ on the common horizon at
$t=t_f$, when the common horizon is axisymmetric. With $(\theta,
\phi)$, we immediately obtain a set of spherical harmonics
$Y_{\ell m}$ by Eq.~\eqref{eqn:ylm} at $t=t_f$. We then find
$V^a$ by $V^a\perp \mathcal{S}$ and Eq.~\eqref{eqn:V}, find
$\beta^a$ by Eqs.~\eqref{eqn:beta_g} and \eqref{eqn:gPDE2},
and construct the stitching vector $X^a$ on $\mathcal{H}$ by
Eq.~\eqref{eqn:X2}. Next, we Lie drag $Y_{\ell m}$ along $X^a$
[Eq.~\eqref{eqn:liedrag}] backward in time, from the final state
$t=t_f$ to the merger $t=0$. Finally, we calculate the mass and
spin moments by Eqs.~\eqref{eqn:DHIlm} and \eqref{eqn:DHLlm}.
Because of the symmetry of the BBH configuration, the mass moments $I_{\ell
m}$ are nonvanishing only for even $\ell$ and even $m$, while the
spin moments $L_{\ell m}$ are nonvanishing only for odd $\ell$
and even $m$. To fix the rotational degree of freedom mentioned in
Sec.~\ref{sec:moment_def}, we multiply $I_{\ell m}$ and $L_{\ell
m}$ by an $m$-dependent phase factor $e^{im\eta}$, where $\eta$
is some real constant, such that $I_{22}$ is real at $t=0$. Under this convention, the even-$m$ modes are
unambiguous, but the odd-$m$ modes are still determined up to
a sign. We do not choose a further convention to fix this sign,
because all odd-$m$ modes are trivial in this paper.

Besides the coordinates $\{t, \theta, \phi\}$ used above, we sometimes need the notion of simulation coordinates $\{t, \grave{x}, \grave{y}, \grave{z}\}$ in this paper. These are the horizon-penetrating Cartesian coordinates used directly to simulate the BBH system in SpEC, and they are called the inertial coordinates in Ref.~\cite{1211.6079}. We also construct the simulation spherical coordinates $\{t, \grave{r}, \grave{\theta}, \grave{\phi}\}$ such that 
\begin{align}
	\grave{x} &= \grave{r} \sin\grave{\theta} \cos\grave{\phi}, \\
	\grave{y} &= \grave{r} \sin\grave{\theta} \sin\grave{\phi}, \\
	\grave{z} &= \grave{r} \cos\grave{\theta}.
\end{align}
On a dynamical horizon, which is a 3D object, we only need $\{t, \grave{\theta}, \grave{\phi}\}$. Note that in general, $\grave{\theta} \neq \theta$ and $\grave{\phi} \neq \phi$.

\subsection{Rotation procedure on multipole moments} \label{sec:rot_proc}

To compare multipole moments with QNMs in this simulation, we
need to apply one more procedure on these multipole moments. In
Sec.~\ref{sec:2+1}, by Lie dragging a spherical harmonic basis
as in Eq.~\eqref{eqn:liedrag}, we construct an invariant basis
of $Y_{\ell m}$'s and use it to define multipole moments. While this
construction leads to an invariantly defined set of multipole
moments, this basis of $Y_{\ell m}$'s is not well adapted for the
QNM analysis. In particular, as the dynamical horizon $\mathcal{H}$
approaches the Kerr horizon, the Lie dragged $Y_{\ell m}$'s are
rotating with respect to the Kerr-Schild coordinates. This rotation can be understood from the following chain of arguments.
	\begin{enumerate}
		\item In the limit at equilibrium, the right hand side of Eq.~\eqref{eqn:gPDE2} vanishes, so $X^a$ approaches $V^a$.
		\item Because $V^a$ is tangent to the horizon and perpendicular to the foliation, it must be a null normal of the Kerr horizon. This implies
		\begin{equation}
			X^a = f(t^a + \Omega_{H} \phi^a),
		\end{equation}
		where $t^a$ and $\phi^a$ are the timelike and rotational Killing vector fields of the Kerr spacetime, $\Omega_{H}$ is the horizon angular velocity \cite{Wald1_skip}, and $f$ is some function.
		\item This function $f$ is actually a constant, since $X^a$
                  preserves the foliation and the foliation is known to become
                  stationary at late times. Moreover, the simulation coordinates
                  in SpEC are remarkably close to the Kerr-Schild coordinates at
                  late times,\footnote{Ref.~\cite{1808.07490} found
                      that an isolated BH in damped harmonic gauge has lapse,
                      shift, and extrinsic curvature nearly identical to that of
                      Kerr-Schild coordinates, only the spatial metric is
                      different.} which fixes the normalization $f\approx
                  1$. Thus, we have
		\begin{equation}
			X^a \partial_a \approx \partial_t + \Omega_{H} \partial_{\grave{\phi}}, \label{eqn:l_comp}
		\end{equation}
		where we write the Killing vector fields explicitly in the simulation coordinates $t, \grave{\phi}$. [Note that the normalization is irrelevant to the Lie dragging procedure, Eq.~\eqref{eqn:liedrag}.]
	\end{enumerate}

We now see that the azimuthal coordinate $\phi$, being Lie dragged along $X^a$, is rotating with frequency $\Omega_{H}$, relative to the Kerr-Schild azimuthal coordinate. In Kerr perturbation theory, one uses a Kerr-Schild-like coordinate systems to obtain QNM frequencies. If we use an azimuthal coordinate that is Lie dragged along $X^a$, we expect different frequencies in the temporal behaviors of perturbed quantities. We can, however, simply undo this rotation by the transformation $\phi \to \phi - \Omega t$, which yields the transformation $Y_{\ell m}\to Y_{\ell m}e^{-im\Omega t}$. Crucially, this transformation changes the temporal behaviors of horizon multipole moments, and makes them more suitable for the QNM analysis. However, we note that the transformed $\phi$ is not covariantly defined, because the transformation depends on the simulation time.

We will apply this procedure on multipole moments in Sec.~\ref{sec:results}, specifically in Eqs.~\eqref{eqn:ilm_rot_corr} and \eqref{eqn:Llm_rot_corr}. Note that we use a symbol $\Omega$ here instead of $\Omega_{H}$, because we will choose a frequency value slightly different from $\Omega_{H}$. See Sec.~\ref{sec:i22} for the detail of this choice of $\Omega$.

\section{Results} \label{sec:results}

In this section, we analyze in detail both the mass and spin moments extracted
from the BBH simulation described in Sec.~\ref{sec:numeric}. In particular, we investigate the
dominant mass moment ($I_{22}$) in Sec.~\ref{sec:i22}, the dominant spin moment
($L_{32}$) in Sec.~\ref{sec:l32}, and the $I_{20}$ multipole moment in Sec.~\ref{sec:i20}. We summarize the behaviors of other multipole moments up to $\ell=6$ in Sec.~\ref{sec:other_moments}. For those readers interested in the correctness of our simulation, we numerically confirm the balance laws and demonstrate the error convergence in Appendix~\ref{sec:app_ballaw_error}.

\subsection{($\mathbf{2,2}$) mass moment} \label{sec:i22}

\begin{figure}[t]
	\centering 
	\includegraphics[width=\linewidth]{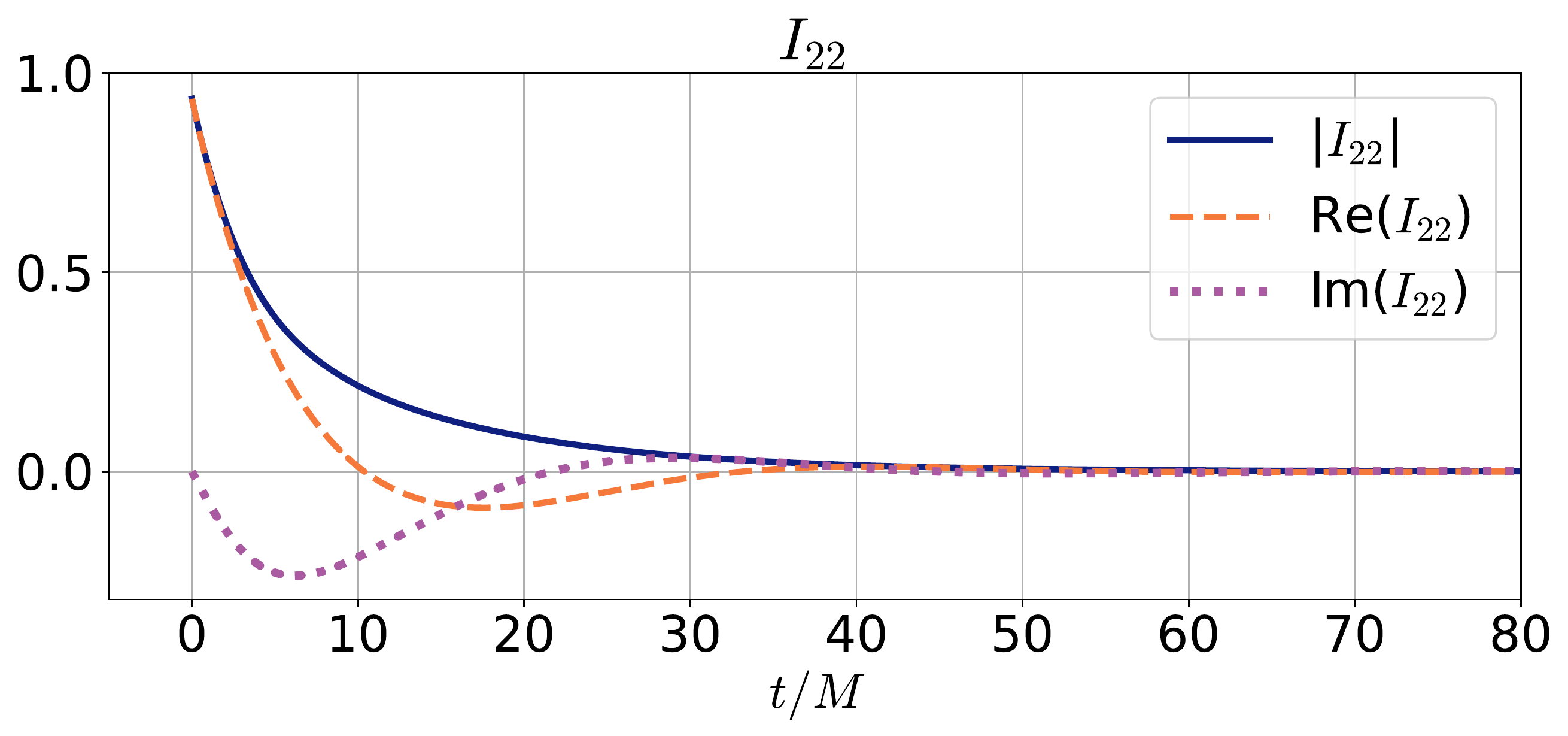}
	\includegraphics[width=\linewidth]{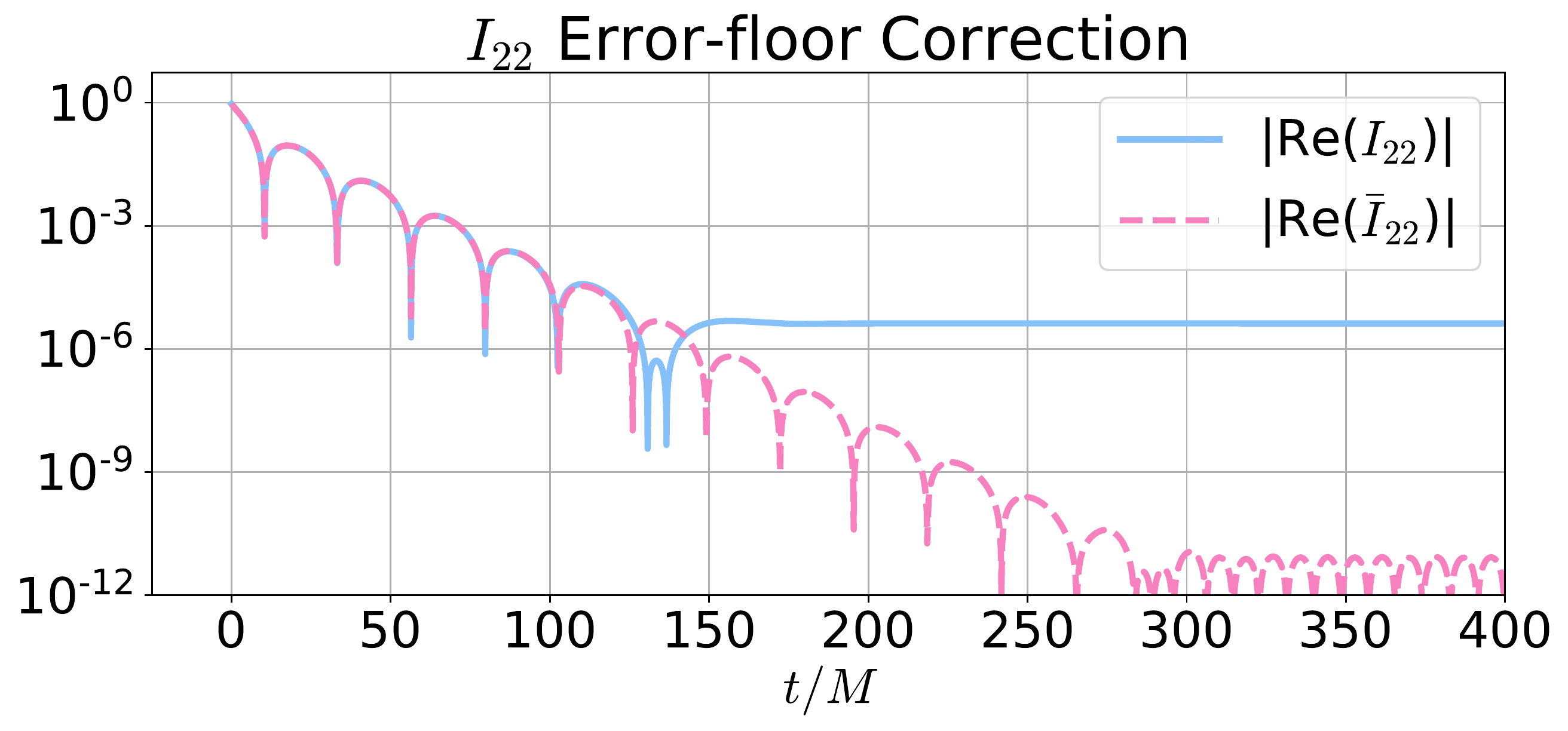}
	\caption{The mass moment $I_{22}$ and its floor correction. The top panel shows $|I_{22}|$ in blue/solid, Re($I_{22}$) in orange/dashed, and Im($I_{22}$) in purple/dotted. The bottom panel shows $|\mathrm{Re}(I_{22})|$ in cyan/solid. This curve directly demonstrates the damped oscillation pattern of $I_{22}$. It also reveals a numerical floor at the level $4\times10^{-6}$ after $t\sim150M$. Subtracting this floor from $I_{22}$, we obtain the floor-corrected mass moment $\bar{I}_{22}$, which is shown in pink/dashed in the bottom panel. The pattern of damped oscillation extends to $t\sim280M$.}
	\label{fig:i22_mag}
\end{figure}

The $(2,2)$ mass moment $I_{22}$ is the dominant mode among the
$I_{\ell m}$ with nonzero $m$. Figure~\ref{fig:i22_mag} shows the $(2,2)$
mass moment as a complex function of $t$. In the top panel, the
magnitude (absolute value), the real part, and the imaginary part of
$I_{22}$ are plotted in blue (solid), orange (dashed), and purple
(dotted). We use a linear scale to demonstrate that both real and
imaginary parts alternate between positive and negative values. The
linear scale also provides a better reading on the magnitude of these
curves before $t<30M$. Note that the imaginary part of $I_{22}$
is 0 at $t=0$, since we choose the convention that $I_{22}$ is
real at $t=0$ (see Sec.~\ref{sec:numeric}). In the bottom panel,
we show $|\mathrm{Re}(I_{22})|$, i.e., the absolute value of the
real part of $I_{22}$, in cyan (solid). We use a logarithmic scale
in this panel to show the manifest pattern of damped oscillations of
$I_{22}$. This curve decays exponentially until reaching a floor
at the level $4\times10^{-6}$ after $t\sim150M$. Because $I_{22}$
(and other $I_{\ell m}$ with nonzero $m$) should approach 0 because
of the axisymmetry of the remnant BH, the floor provides a measure of numerical error for
$I_{22}$. We can remove this numerical floor by subtracting it from
$I_{22}$. Specifically, we define
\begin{align}
	\bar{I}_{22} = I_{22} - \mathrm{mean}[I_{22}(t\ge400M)], \label{eqn:i22_floor_corr}
\end{align}
where $\mathrm{mean}[I_{22}(t\ge400M)]$ refers to the average
value\footnote{Even more specifically, $I_{22}(t)$ is a series
of discrete data points generated from the simulation. They are
equally spaced by $0.1M$ in $400M\le t \le 500M$. The quantity
$\mathrm{mean}[I_{22}(t\ge400M)]$ is the unweighted mean of these
data points, which is of the order of $10^{-6}$ in our simulation.}
of $I_{22}$ over the range $400M\le t \le 500M$. The bottom panel
displays $|\mathrm{Re}(\bar{I}_{22})|$ in a pink dashed style. We
observe that $|\mathrm{Re}(\bar{I}_{22})|$ also possesses a pattern
of damped oscillation, but now the pattern extends to $t\sim280M$. As
$\bar{I}_{22}$ has a longer-lasting nontrivial behavior, we will use
$\bar{I}_{22}$ instead of $I_{22}$ from now on. However, we keep in
mind that the $t>150M$ portion of $\bar{I}_{22}$ is within numerical
uncertainty, so we will only focus on $t\le150M$ from now on.

\begin{figure}[t]
	\centering 
	\includegraphics[width=\linewidth]{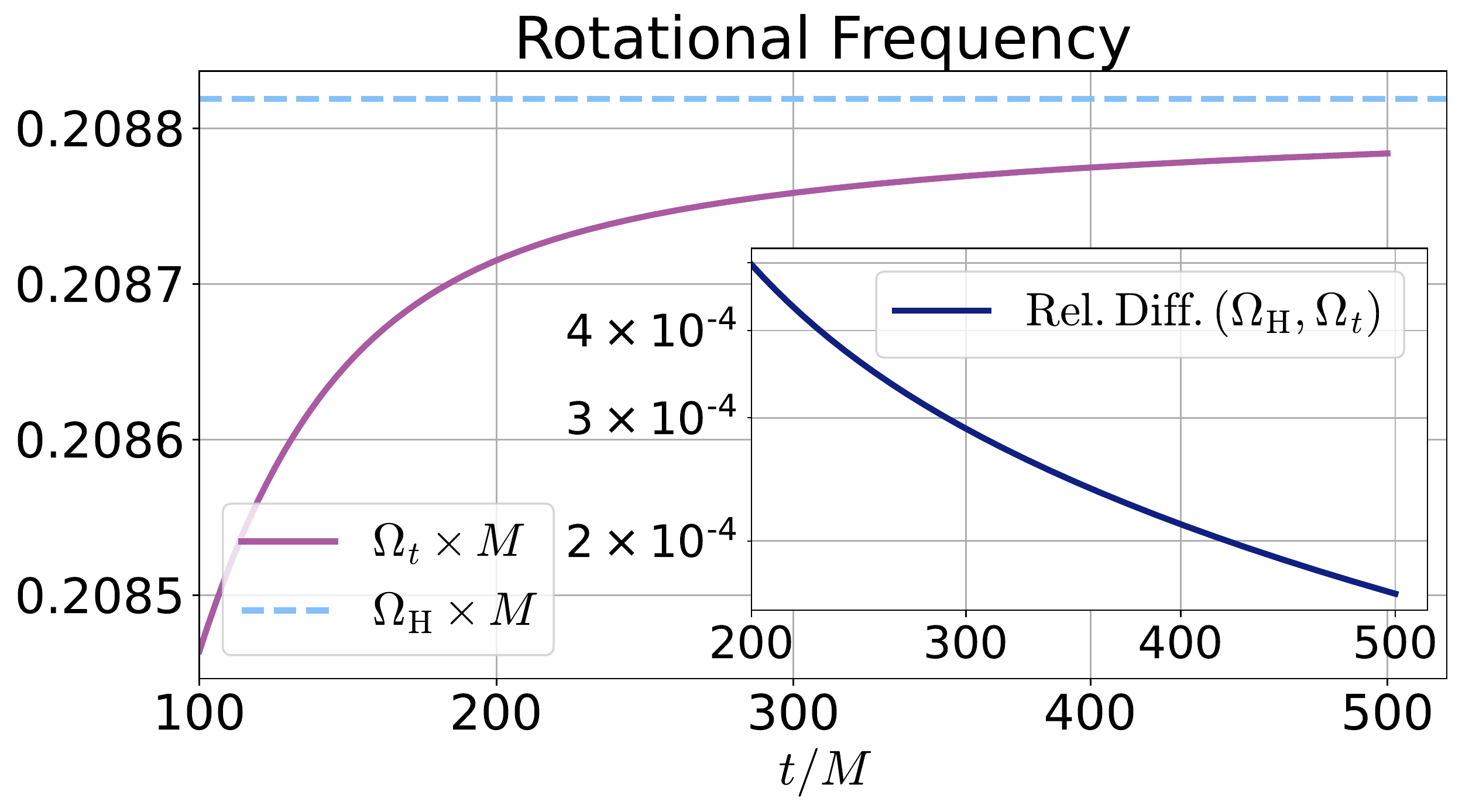}
	\includegraphics[width=\linewidth]{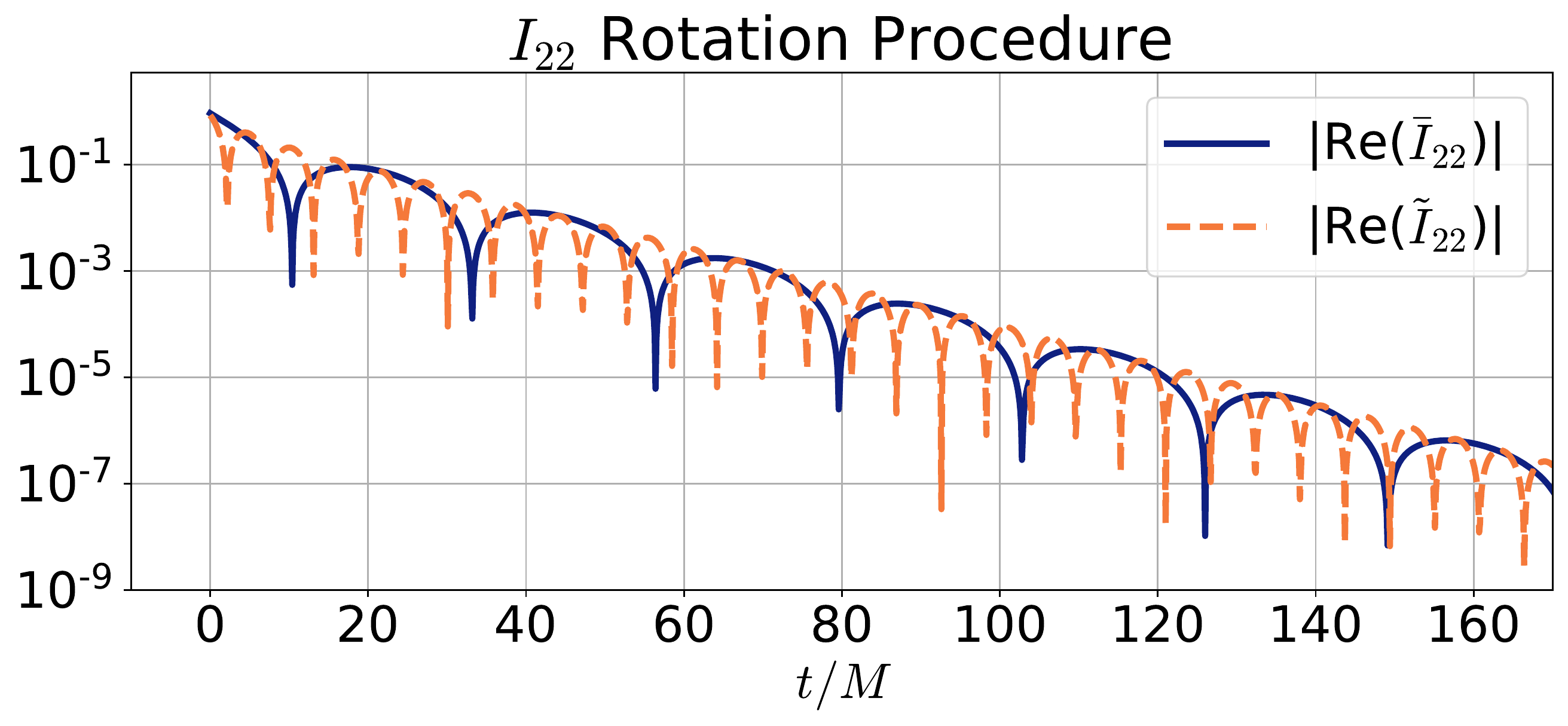}
	\caption{The rotational frequency of $Y_{\ell m}$ and the rotated mass moment. The top panel shows $\Omega_t$ (purple/solid), the rotational frequency of $Y_{\ell m}$, as a function of time. The curve does not settle down to a constant even at a very late time. This panel also shows $\Omega_H$ (cyan/dashed), the horizon angular velocity, as a reference. The relative difference between $\Omega_t$ and $\Omega_H$ is given in the inset. We show the comparison between $\bar{I}_{22}$ (blue/solid) and its rotated version $\tilde{I}_{22}$ (orange/dashed) in the bottom panel. Applying the rotation does not alter the decay rate, but it increases the frequency significantly.}
	\label{fig:i22_rot}
\end{figure}

To further analyze the behavior of this mass moment,
we will implement the rotation procedure outlined in
Sec.~\ref{sec:rot_proc}. We first check the validity of
Eq.~\eqref{eqn:l_comp} in the simulation at late times
by comparing $\Omega_{H}$ with $\Omega_t$. Here, $\Omega_t$
is defined as the average value of $X^{\grave{\phi}}$ (the $\grave{\phi}$-component of $X^a$) over the common horizon
$\mathcal{S}$ at time $t$, i.e.,
\begin{align}
	\Omega_t = \underset{\mathcal{S}}{\text{mean}} \left(X^{\grave{\phi}}\right).
\end{align}
Note that in the simulation, the
maximum deviation of $X^{\grave{\phi}}$ from $\Omega_t$ on every
$\mathcal{S}$ is within $10^{-5}$ for $t\ge300M$, as expected. What
is unexpected is shown in the top panel of Fig.~\ref{fig:i22_rot}:
Although we expect $\Omega_t$ to approach the horizon angular
velocity \cite{Wald1_skip},
\begin{align}
	\Omega_H = \frac{\chi_f}{2M_f\left(1+\sqrt{1-\chi_f^2}\right)} = 0.208819M^{-1},
\end{align}
it does not completely settle down even at $t=t_f=500M$. Nevertheless, as $\Omega_t$ varies gradually near $t=500M$, we set the rotational frequency of the transformation $\phi \to \phi - \Omega t$ in this paper to be
\begin{align}
	\Omega = \Omega_{t=500M} = 0.208784M^{-1}.
\end{align}
All results in the following sections are based on this choice. We also show the relative difference\footnote{In this paper, the relative difference/error between any two numbers, $f$ and $g$, is defined as $2|f-g|/|f+g|$. The relative difference between $\Omega_H$ and $\Omega$ is $1.7\times 10^{-4}$. This is the same as the difference between the surface gravity for $V^a$ and the Kerr surface gravity, introduced in Appendix~\ref{sec:app_surf_grav}.} between $\Omega_t$ and $\Omega_H$ in the inset.

We rotate the mass moments by defining
\begin{align}
	\tilde{I}_{\ell m}(t) = \bar{I}_{\ell m}(t) e^{-im\Omega t}. \label{eqn:ilm_rot_corr}
\end{align}
The bottom panel of
Fig.~\ref{fig:i22_rot} compares the rotated mass moment
$|\mathrm{Re}(\tilde{I}_{22})|$ (orange/dashed) with the nonrotated
one $|\mathrm{Re}(\bar{I}_{22})|$ (blue/solid). The rotation does
not change the decay rate of the mass moment but greatly increases
its oscillation frequency: $\tilde{I}_{22}$ oscillates almost four
times as quickly as $\bar{I}_{22}$. Thus, the use of $\bar{I}_{22}$ or $\tilde{I}_{22}$ may
lead to very different conclusions. In this paper,
we choose to investigate the behavior of $\tilde{I}_{22}$, namely
the rotated, error-floor-corrected $(2,2)$ mass moment. As we will
see, the behavior of this
mass moment resembles that of a gravitational waveform.

\begin{figure}[t]
	\centering
	\includegraphics[width=\linewidth]{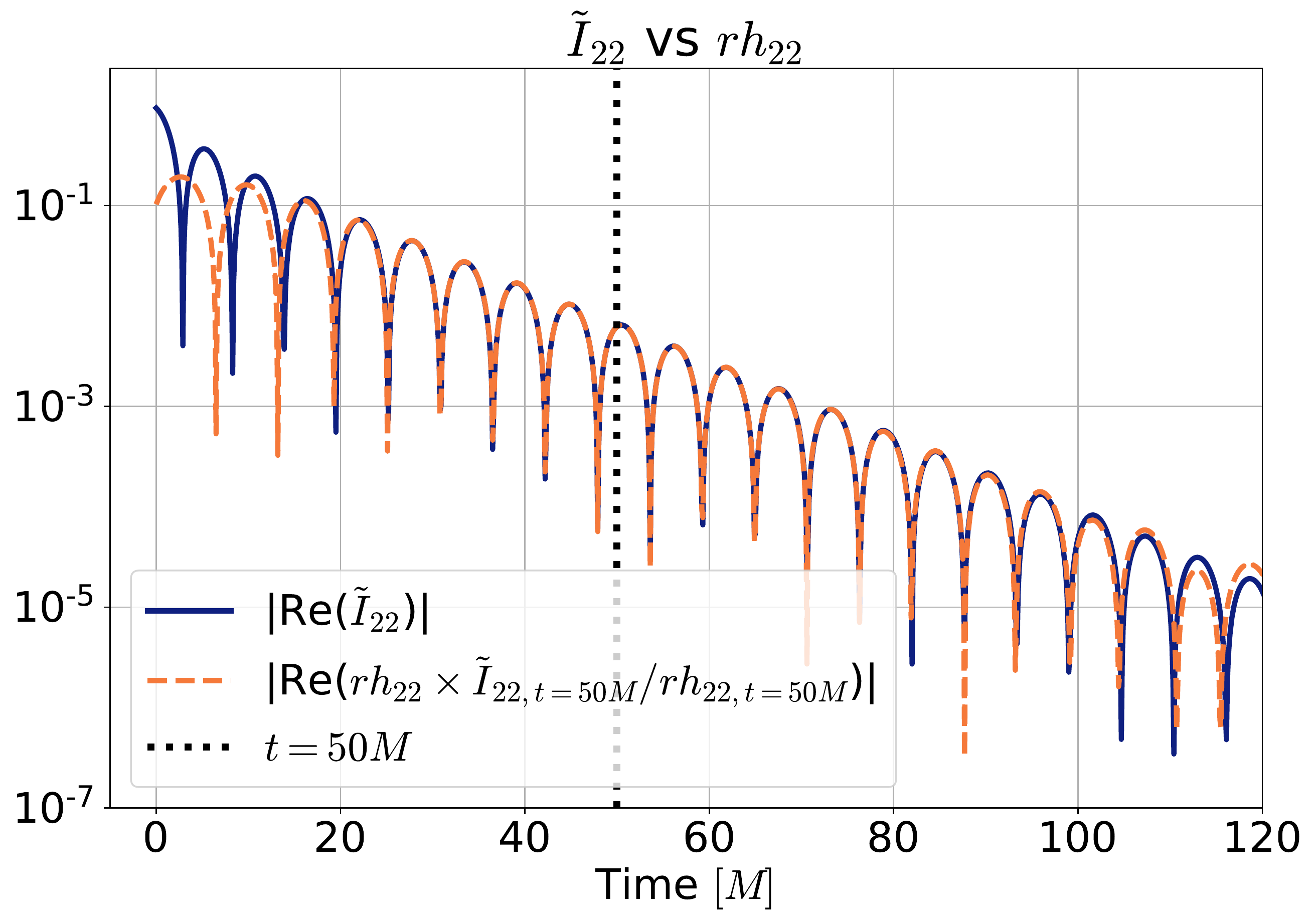}
	\caption{The comparison between the mass moment $\tilde{I}_{22}$ (blue/solid) and the waveform $rh_{22}$ (orange/dashed). The mass moment is plotted as a function of simulation time $t$, while the waveform is of retarded time $t_\mathrm{ret}$. The waveform is time shifted and multiplied by a constant factor, as described in the main text. The black dotted vertical line marks $t=50M$, at which the values of two curves are matched.	We see strong correlation between these two quantities in $20M\le t\le 120M$. }
	\label{fig:i22_h22}
\end{figure}

Our first step in the analysis of $\tilde{I}_{22}$ is to compare it
with the waveform strain $h$.\footnote{Comparison between horizon data and asymptotic data in SpEC BBH simulations is not new. Reference~\cite{2104.07052} is such an example that compares masses, spins, and recoil velocities of remnant BHs.} We extract $h$ on the surfaces of
multiple concentric spherical shells of finite Euclidean radii $r$,
and extrapolate $rh$ to $\mathscr{I}^+$ as a function of retarded
time $t_\mathrm{ret}$ \cite{0905.3177, 1302.2919, 1409.4431,
1509.00862, 2010.15200}. Then, $rh_{22}$ is the $(\ell=2,m=2)$
coefficient in the $s=-2$ spin-weighted spherical harmonic expansion
of $rh$. Note that $rh_{22}$ is both time shifted and phase shifted
in this paper: We set $t_\mathrm{ret}=0$ when $|rh_{22}|$ (not
necessarily $|rh|$) reaches its maximum. We also multiply $rh_{22}$ by
a constant complex factor such that $rh_{22}(t_\mathrm{ret}=50M)$
matches $\tilde{I}_{22}(t=50M)$\footnote{Matching at any time between $25M$ and $95M$ yields a very similar
result.}. We show both $\tilde{I}_{22}$ (blue/solid) and $rh_{22}$
(orange/dashed) in Fig.~\ref{fig:i22_h22}. The graph displays the
absolute values of their real parts, so that we can compare the decay
and oscillation between the two curves simultaneously. The horizontal
axes represent the simulation time $t$ for $\tilde{I}_{22}$
and the retarded time $t_\mathrm{ret}$ for $rh_{22}$. We see
from the graph that $\tilde{I}_{22}$ and $rh_{22}$ are strongly
correlated. Specifically, in the range $20M\le t\le 120M$, they
share the same decay constant and oscillation frequency. For $t>120M$
(not shown), the comparison becomes meaningless, because the strain
reaches its level of numerical error. For $t<20M$, $\tilde{I}_{22}$
and $rh_{22}$ are less correlated, possibly because the meaning
of time (or the behavior of the lapse) in the strong field regime is
substantially different from at infinity. 

Figure~\ref{fig:i22_h22} strongly suggests that the mass moment $\tilde{I}_{22}$,
like $h_{22}$, is described by the QNM of spin-weight $s=-2$ or
$s=2$. We include the possibility $s=2$ here, because the frequency
of an $s=2$ QNM is the same as that of $s=-2$. Knowing
that $\tilde{I}_{22}$ has spin weight 0, one might be curious about
why $\tilde{I}_{22}$ is described by $s=\pm2$ QNMs. The expression
of $\tilde{R}$'s first order perturbation, Eq.~(2.21) in Hartle's
\cite{Hartle1_skip}, provides a potential explanation. 

In the following
sections, we investigate the quasinormal pattern of $\tilde{I}_{22}$
quantitatively, by linearly fitting $\tilde{I}_{22}$ to multiple
QNMs of spin weight $s=2$ (or equivalently $s=-2$).

\subsubsection{Mode mixing} \label{sec:i22_mix}

We start with a model with only fundamental modes,
\begin{align}
	\tilde{I}_{22} = \sum_{\ell=2}^{L} C_{\ell 20} e^{-i \omega_{\ell 20} (t-t_0)}, \label{eqn:i22_fitmodel_l}
\end{align}
with a fitting time range $t_0 \le t \le 120M$. We choose
$120M$ as the end fitting time, when the mass moment is
still slightly above the numerical error of $\tilde{I}_{22}$
(see Fig.~\ref{fig:i22_mag}). The parameters $C_{\ell20}$ are to
be determined by a linear fit. (All the symbols $C_{\ell mn}$ in
this paper should be understood as fitting parameters.) We consider
several $L\ge2$ and allow $t_0$ to vary. We measure the error of fit
by the mismatch between $\tilde{I}_{22}$ and its fit. The mismatch
between two complex-valued functions $f(t)$ and $g(t)$ is defined as
\begin{align}
	\mathcal{M}(f,g) = 1 - \frac{\mathrm{Re}(\langle f|g\rangle)}{\sqrt{\langle f|f\rangle \langle g|g\rangle}}, \label{eqn:mismatch}
\end{align}
where
\begin{align}
	\langle f|g\rangle = \int f(t) g^*(t) dt,
\end{align}
with integration domain over the fitting time range.

\begin{figure}[t]
	\centering 
	\includegraphics[width=\linewidth]{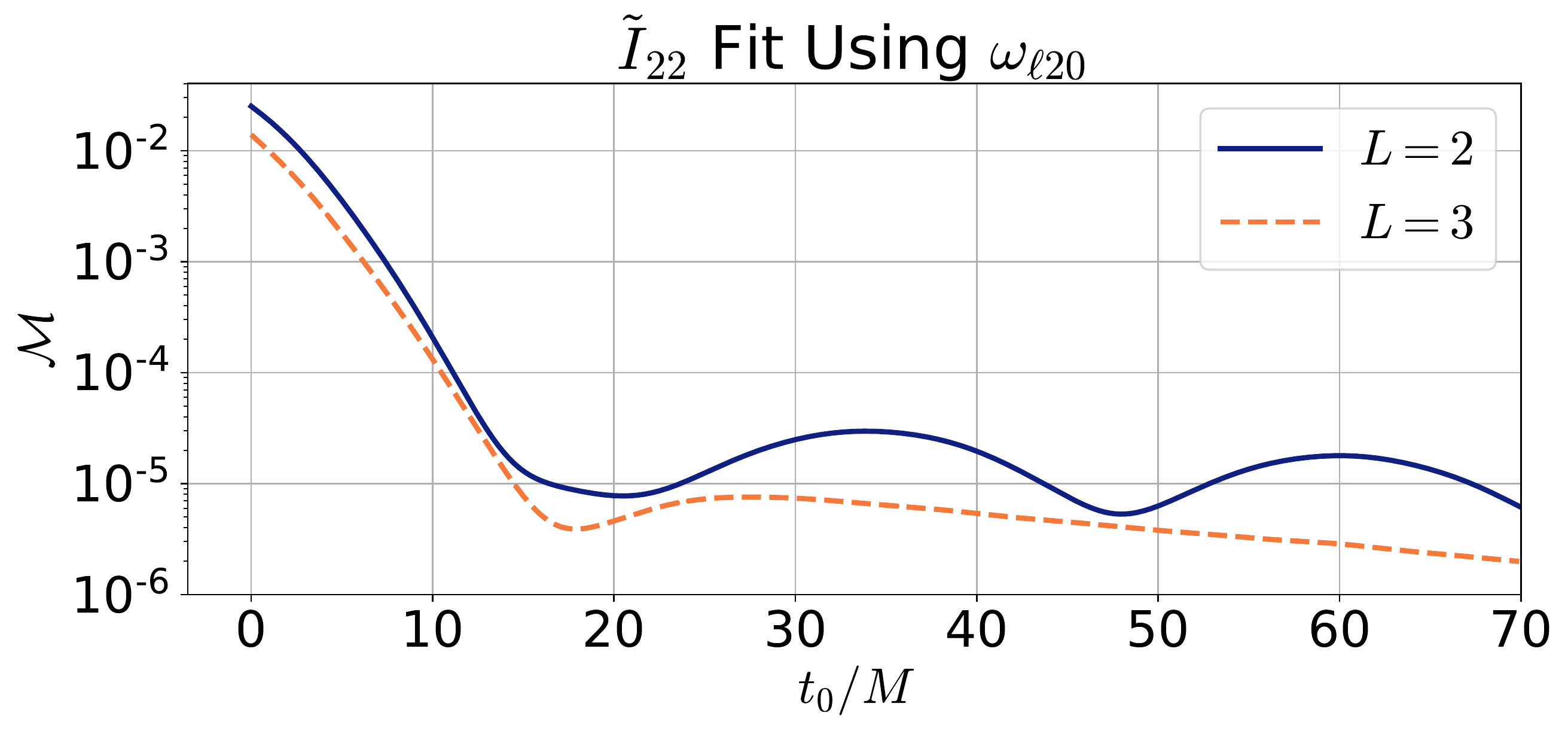}
	\caption{The mismatch between $\tilde{I}_{22}$ and its fit using $\omega_{\ell 20}$ QNMs [Eq.~\eqref{eqn:i22_fitmodel_l}], plotted as a function of the initial fitting time $t_0$. Both the $L=2$ (blue/solid) and $L=3$ (orange/dashed) curves decay sharply before $t_0=18M$, because overtones are not included in the model. The $L=2$ curve, which only uses the $\omega_{220}$ QNM, contains a persistent oscillatory pattern after $t_0=18M$. This is a beat pattern formed by the $\omega_{220}$ and $\omega_{320}$ QNMs, and is removed in the $L=3$ curve.}
	\label{fig:i22_qnm_varyt0_l20}
\end{figure}

We first consider the simplest choice $L=2$ in this model, which means we fit $\tilde{I}_{22}$ using only the fundamental tone of $(2,2)$ QNMs. The mismatch $\mathcal{M}$ as a function of the initial fitting time $t_0$ is shown in blue (solid) in Fig.~\ref{fig:i22_qnm_varyt0_l20}. The curve decays from $10^{-2}$ to $10^{-5}$ before $t_0=18M$. This decay is expected, because the current model does not include overtones, which are strongly excited near the merger. However, it is surprising to see a wavy pattern in the curve after $t_0=18M$, since the QNM fit of $rh_{22}$ does not have such a feature \cite{1903.08284, 1910.08708}. This oscillatory pattern extends well beyond $t_0=70M$, which is not shown.

This oscillatory pattern suggests that the $L=2$ model does not
capture an essential feature of $\tilde{I}_{22}$. We can rule out
the following two possibilities for this missing feature. First,
this feature is not related to the oscillation of $\bar{I}_{22}$,
i.e., the nonrotated mass moment. This is because the period of
the oscillatory pattern in the $L=2$ mismatch curve ($\sim$$26M$)
differs from the period of $\bar{I}_{22}$. Second, the missing
feature is not related to the $\omega_{22n}$ overtones either,
because the oscillatory pattern cannot be eliminated by including
them in the $L=2$ model (not shown). Accordingly, we consider one
more possibility: There is another fundamental tone, other than
$\omega_{220}$, that contributes to $\tilde{I}_{22}$. Indeed,
$\omega_{220}$ and $\omega_{320}$ share a similar decay
rate, and they can generate a beat period of $26.3M$ (see
Table~\ref{tbl:sample_omega}), which is close to the period of
the oscillatory pattern ($\sim$$26M$). So we now examine the model
Eq.~\eqref{eqn:i22_fitmodel_l} with $L=3$. The orange dashed curve
in Fig.~\ref{fig:i22_qnm_varyt0_l20} represents the mismatch using
this model. It contains no oscillatory pattern at late times, confirming the nonnegligible contribution of the $(3,2)$
fundamental tone to $\tilde{I}_{22}$. The curve decreases steadily
after the local maximum at $t=27.4M$, so we may treat $t=27.4M$ as
the instant when overtones are negligible, and only two fundamental
tones dominate. We have also investigated the $L=4$ and
$L=5$ cases,
but they hardly improve the fit (not shown).

We now connect the presence of the $(3,2)$ QNM
in the description of $\tilde{I}_{22}$ to the concept
of mode mixing. In  BH perturbation theory, the natural
angular basis for strain $h$ (whose second time derivative is
$\Psi_4$) is the spin-weighted \textit{spheroidal} harmonics
(Sec.~\ref{sec:qnm}). However, 
the natural angular basis for $h$ at future null infinity $\mathcal{I}^+$ is
the basis of the
spin-weighted \textit{spherical} harmonics \cite{1904.04831}. 
This is the basis used, for example, in LIGO waveform
analysis. The
use of spherical harmonics intertwines spheroidal modes of the
same $m$ but different $\ell$ \cite{1408.1860}. For example,
the spherical mode $h_{22}$ (i.e., the expansion coefficient
corresponding to $_{-2}Y_{22}$) can be decomposed into not only
the $\omega_{22n}$ modes, but also the $\omega_{32n}$ modes,
etc. This phenomenon is called \textit{mode mixing}. In our BBH
configuration (equal-mass, non-spinning), modes other than $\omega_{22n}$
may be ignored in $h_{22}$'s decomposition. This is because the
$\omega_{22n}$ modes are strongly dominant \cite{gr-qc/0703053},
and the mixing of spheroidal and spherical harmonics is tiny
\cite{1408.1860}. However, this argument does not apply
to mass moments $I_{\ell m}$. The natural angular basis of the
perturbed $\tilde{\mathcal{R}}$ in Eq.~\eqref{eqn:DHIlm} is
neither spheroidal nor spherical harmonics, but a complicated
function of angles $(\theta, \phi)$ instead.\footnote{The angular dependence of the perturbed
$\tilde{\mathcal{R}}$ is a surface derivative of spheroidal
harmonics in certain coordinates. See Ref.~\cite{Hartle1_skip} for expressions of the
perturbed $\tilde{\mathcal{R}}$.} The mixing of this complicated
angular function and spherical harmonics, if nonnegligible, would
lead to the presence of $(3,2)$ QNMs in $\tilde{I}_{22}$. In this
paper, we refer to this phenomenon as \textit{mode mixing} as well,
but in a somewhat broader sense.

\begin{figure}[t]
	\centering 
	\includegraphics[width=\linewidth]{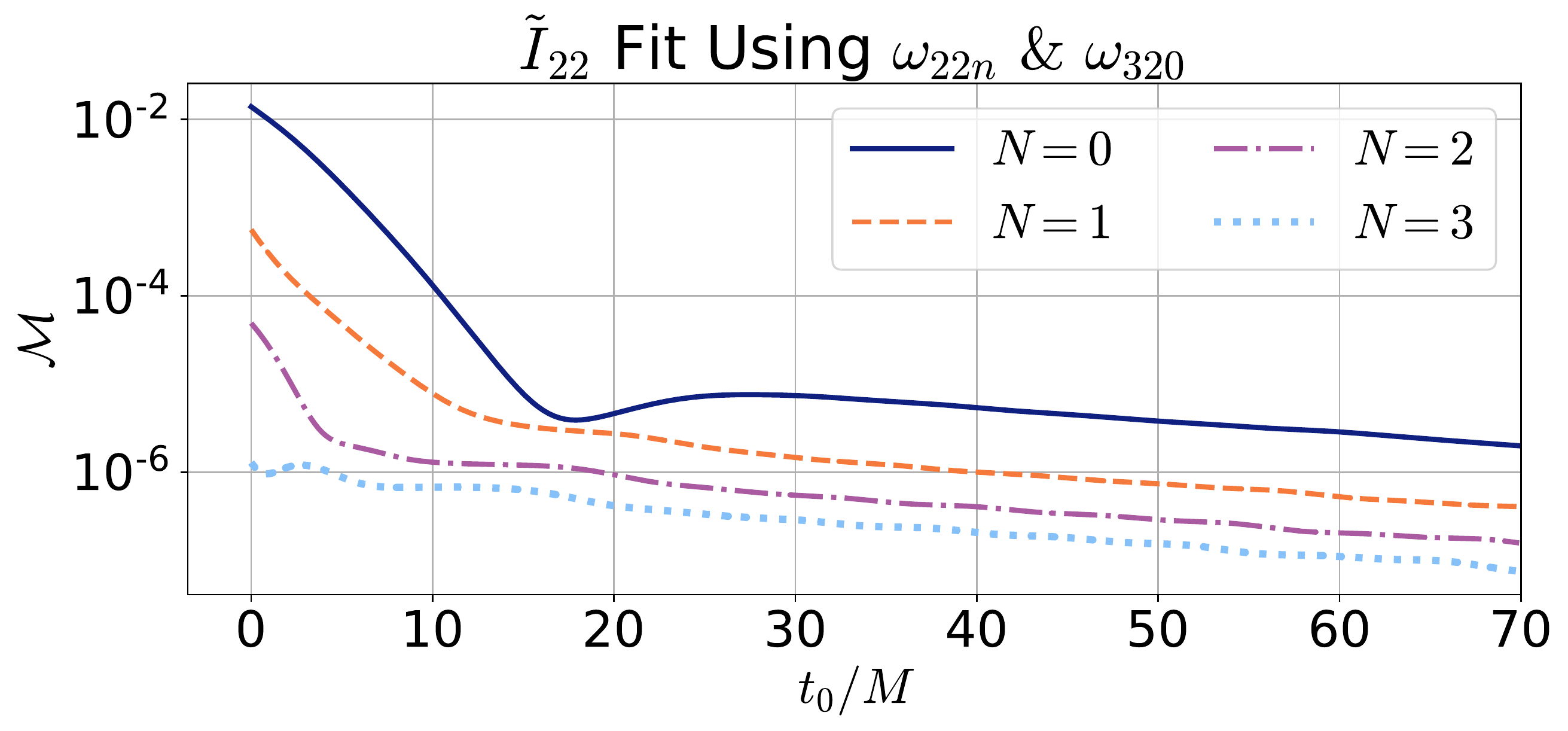}
	\caption{The mismatch between $\tilde{I}_{22}$ and its fit using $\omega_{22n}$ and $\omega_{320}$ QNMs [Eq.~\eqref{eqn:i22_fitmodel_n}]. The $N=0$ curve is, by construction, the same as the $L=3$ curve in Fig.~\ref{fig:i22_qnm_varyt0_l20}. Adding higher overtones renders a better fit for all $t_0$, and specifically, brings down the portion of large mismatch before $t\sim10M$. This figure demonstrates the important contribution of overtones to the mass moment.}
	\label{fig:i22_qnm_varyt0_320_22n}
\end{figure}

Now that we know $\tilde{I}_{22}$ can be well approximated by the
fundamental tones of $(2,2)$ and $(3,2)$ QNMs after $t=27.4M$, we
shall analyze the effect of overtones on $\tilde{I}_{22}$. Inspired
by the use of overtones in the QNM fit of waveforms and horizon moments
in Refs.~\cite{1903.08284, 1910.08708, 2010.15186}, we consider
the following model,
\begin{align}
	\tilde{I}_{22} = C_{320} e^{-i \omega_{320} (t-t_0)} + \sum_{n=0}^{N} C_{2 2n} e^{-i \omega_{22n} (t-t_0)}, \label{eqn:i22_fitmodel_n}
\end{align}
with the same fitting time range $t_0\le
t\le120M$. Figure~\ref{fig:i22_qnm_varyt0_320_22n} shows the
mismatch of this model as a function of $t_0$ for multiple $N$
($0\le N\le 3$). By construction, the $N=0$ curve is the same
as the $L=3$ curve in Fig.~\ref{fig:i22_qnm_varyt0_l20}. As
more overtones are included, the mismatch curve becomes flatter
and lower, and the initial damping part shrinks and
ends earlier. For $N=3$, we no longer see the initial damping
part. This means that the overtones $\omega_{22n}$ (at least for
$1\le n\le 3$) do contribute to $\tilde{I}_{22}$, and the fitting
model Eq.~\eqref{eqn:i22_fitmodel_n} indeed captures them. Note
that compared to the $N=0$ model, those $N\ge1$ models improve the
accuracy even after the overtones are supposed to damp away. This
might be caused by overfitting to numerical noise. We also checked
several $N\ge4$ models, but they do not display much improvement
(not shown) compared to the $N=3$ model.

\subsubsection{Fit using fundamental tones} \label{sec:i22_n0}

In this section, we will have a closer look at the late-time QNM description of $\tilde{I}_{22}$. We continue using the model Eq.~\eqref{eqn:i22_fitmodel_l} with $L=3$, which reads
\begin{align}
	\tilde{I}_{22} = C_{220} e^{-i \omega_{220} (t-t_0)} + C_{320} e^{-i \omega_{320} (t-t_0)}. \label{eqn:i22_n0}
\end{align}
Instead of varying $t_0$ as in the previous section, we now fix the value of $t_0$. In particular, we choose $t_0 = 50M$, at which all overtones have decayed sufficiently.\footnote{At $t_0 = 50M$, the mismatch of this model (Fig.~\ref{fig:i22_qnm_varyt0_l20}) has decreased below $4\times10^{-6}$, which is the numerical error of $\tilde{I}_{22}$ estimated by the numerical floor in Fig.~\ref{fig:i22_mag}.} 

\begin{figure*}[t]
	\centering 
	\includegraphics[width=0.49\linewidth]{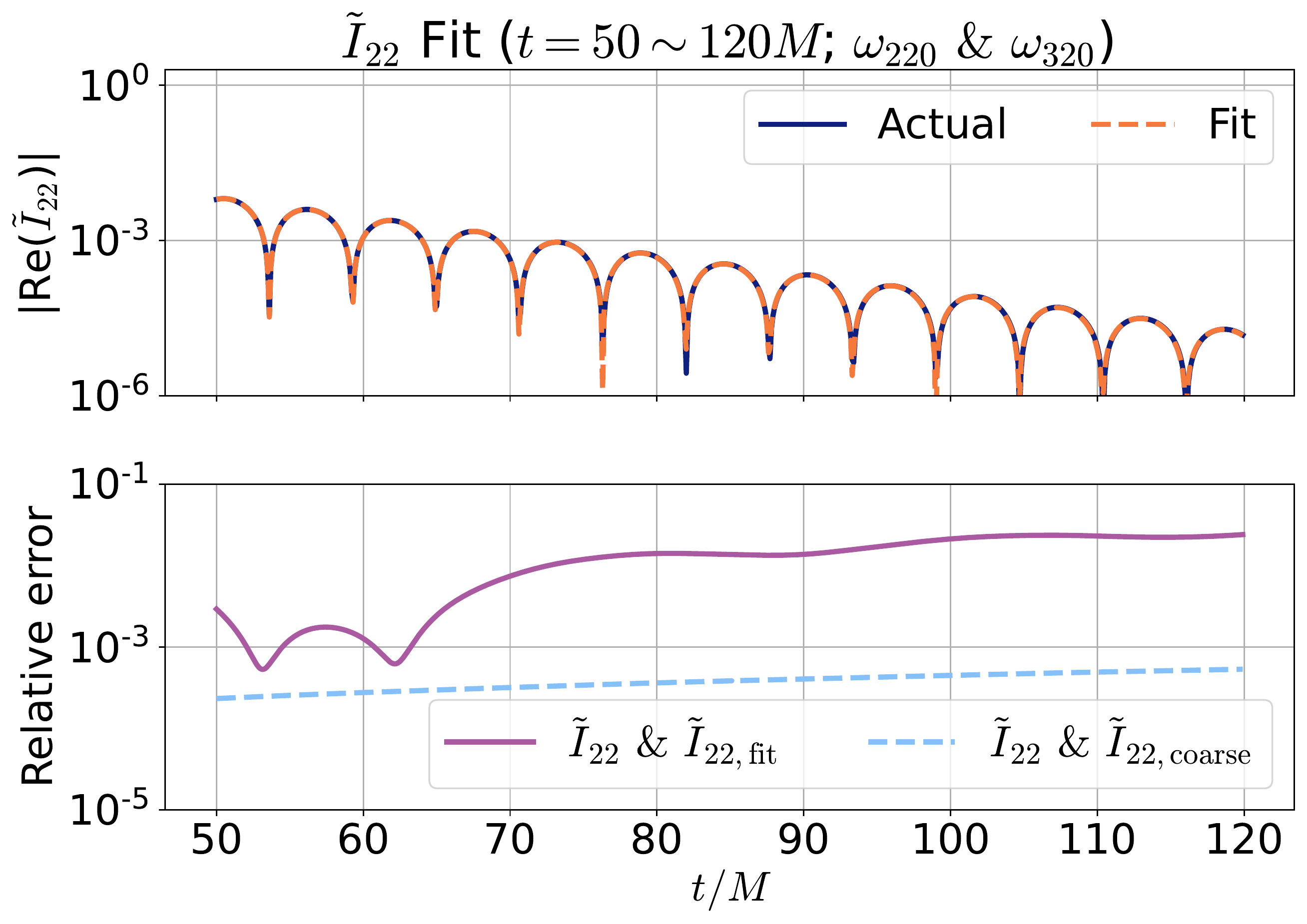}
	\includegraphics[width=0.49\linewidth]{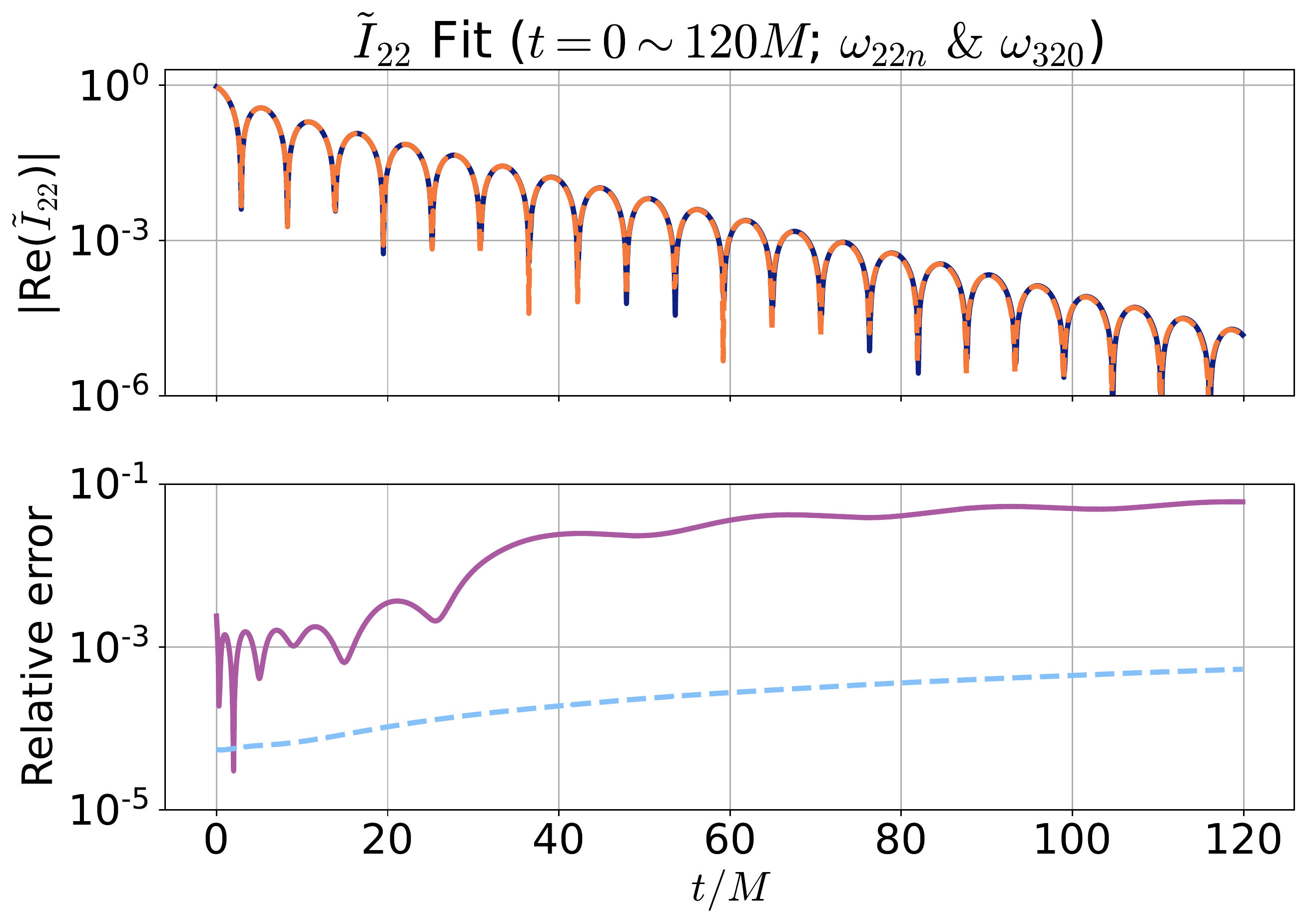}
	\caption{The comparison between $\tilde{I}_{22}$ and its fit. The left two panels are based on the fit using $\omega_{220}$ and $\omega_{320}$ [Eq.~\eqref{eqn:i22_n0}], in the time range $50M\le t\le120M$. The right two panels are based on the fit using \{$\omega_{220}$, $\omega_{221}$, $\omega_{222}$, $\omega_{223}$, $\omega_{320}$\} [Eq.~\eqref{eqn:i22_fitmodel_n} with $N=3$], in the time range $0\le t\le120M$. The top two panels show the absolute real parts of $\tilde{I}_{22}$ (blue/solid) and its fit (orange/dashed). In either top panel, the two curves overlap very well. The bottom two panels show the relative difference between $\tilde{I}_{22}$ and the fit in purple/solid, and the difference in $\tilde{I}_{22}$ between two resolutions in cyan/dashed. The quantity $\tilde{I}_{22, \mathrm{coarse}}$ refers to the $(2,2)$ mass moment extracted from the low-resolution simulation.}
	\label{fig:i22_qnm_n}
\end{figure*}

The top left panel of Fig.~\ref{fig:i22_qnm_n} shows the fit using this model with the fitting time range $50M\le t\le120M$. The blue solid curve represents the actual mass moment $\tilde{I}_{22}$, while the orange dashed curve represents the fit. They are both plotted in the magnitude of their real parts. We see that the two curves overlap very well, so the model Eq.~\eqref{eqn:i22_n0} indeed provides a good description of $\tilde{I}_{22}$. The relative difference between $\tilde{I}_{22}$ and its fit (including their imaginary parts) is plotted in purple (solid) in the bottom panel of the same figure. For reference, the cyan dashed curve in this panel is the relative difference in $\tilde{I}_{22}$ between the two resolutions used in our simulation (Sec.~\ref{sec:bbh_sim}), which provides another estimate of the numerical error of $\tilde{I}_{22}$. Note that both curves in the bottom panel have an increasing trend, as $\tilde{I}_{22}$ gets closer to the level of numerical uncertainty. After $t\ge80M$, the relative error of the QNM fit is larger than the numerical error of $\tilde{I}_{22}$ by about two orders of magnitude. This means the model is good but not perfect, and there is room for improvement in the future.

\begin{figure*}[t]
\centering 
\includegraphics[width=0.485\linewidth]{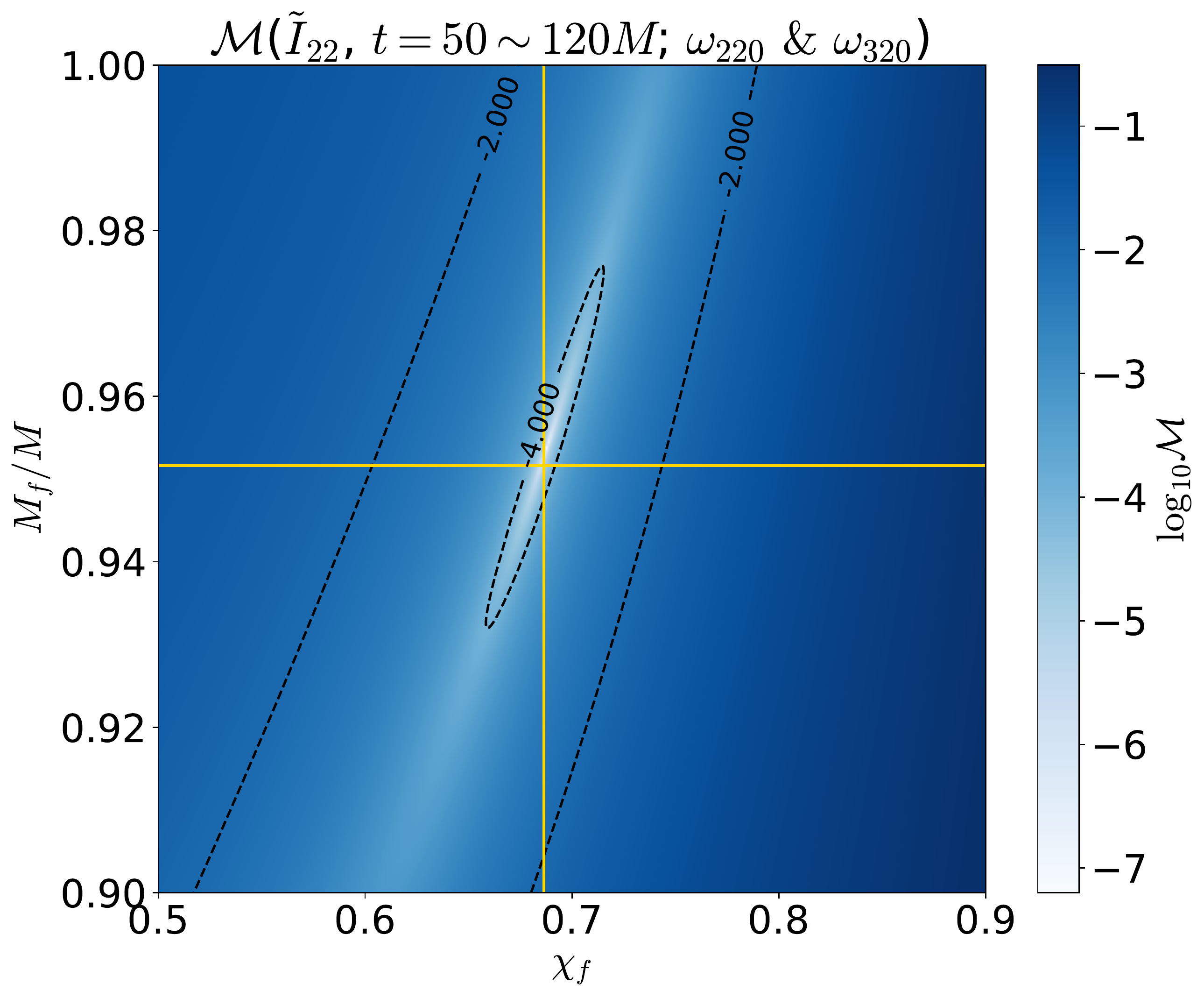}
\includegraphics[width=0.5\linewidth]{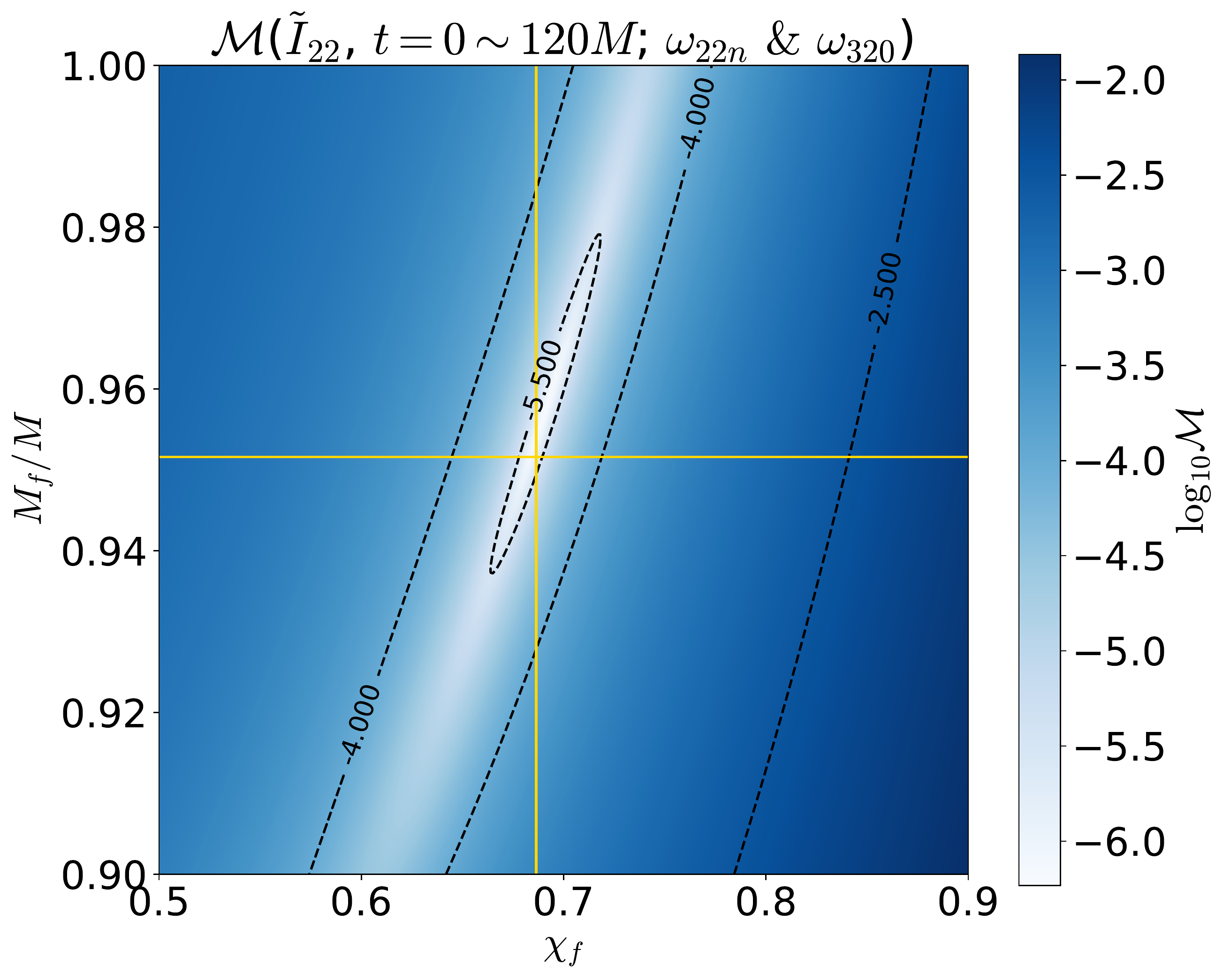}
\caption{Heat maps of the mismatch $\log_{10}\mathcal{M}$ over the ($M_f, \chi_f$) parameter space. The left panel is based on the model
Eq.~\eqref{eqn:i22_n0}, while the right one on the model
Eq.~\eqref{eqn:i22_fitmodel_n} with $N=3$: the lighter the shading,
the smaller the mismatch. In each panel,
we use two golden lines to represent the true values of
$M_f$ and $\chi_f$. The dashed curves are the contour lines of constant mismatch. The deep minimum of the mismatch is
located close to the golden cross, which means that the
QNM model can be used to recover the true values of the
remnant parameters.}
\label{fig:i22_qnm_n_heatmap}
\end{figure*}

Once we accept that the model Eq.~\eqref{eqn:i22_n0} can describe
the mass moment at late times, we may use it to estimate
the final mass and spin of the remnant. The QNM frequencies
$\omega_{220}$ and $\omega_{320}$ used to generate the left panels
of Fig.~\ref{fig:i22_qnm_n} are calculated based on $M_f$ and
$\chi_f$ that are measured by SpEC (Sec.~\ref{sec:bbh_sim}). In
the following discussion, we regard the SpEC values of $M_f$
and $\chi_f$ as their \textit{true} values. Now, we allow $M_f$
and $\chi_f$ to deviate from the true values, and repeat the
QNM fit over the ($M_f, \chi_f$) parameter space (similar
to the procedure in Ref.~\cite{1903.08284}). For each ($M_f,
\chi_f$) combination, we measure the error of the fit by the
mismatch, Eq.~\eqref{eqn:mismatch}. The result is visualized
as a heat map of $\log_{10}\mathcal{M}$ in the left
panel of Fig.~\ref{fig:i22_qnm_n_heatmap}: the lighter
the shading, the smaller the mismatch. We also show the true values of $M_f$ and $\chi_f$ in
golden (solid) lines for reference. We see from the plot that
not only does the mismatch have a deep minimum over the ($M_f, \chi_f$)
parameter space, but also the minimum approximately recovers the
true values. In particular, the best estimates of the mass and
spin (i.e., their values at the minimum) are $M_f'=0.95390M$ and
$\chi_f'=0.68825$. We can assess the goodness of these estimates
by the error,
\begin{align}
	\epsilon_f = \sqrt{(M_f' - M_f)^2/M^2 + (\chi_f' - \chi_f)^2},
\end{align}
as proposed in Ref.~\cite{1903.08284}. The error of these estimates is
$\epsilon_f = 2.9\times10^{-3}$, compared to a difference between the two resolutions, $3\times10^{-6}$. Note that the minimum mismatch
does not necessarily make ($M_f'$, $\chi_f'$) a better pair of candidates for
the final mass and spin, because as we will see, different QNM models produce
different ($M_f', \chi_f'$) combinations, and there is no consistent choice
among these models to determine mass and spin yet.

\subsubsection{Fit using overtones} \label{sec:i22_n7}

We extend the analysis in the previous section to the early-time portion of
$\tilde{I}_{22}$, by including overtones up to $n=3$. In particular, we
investigate the model Eq.~\eqref{eqn:i22_fitmodel_n} with $N=3$, and fix the
fitting time range as $0\le t\le 120M$. The right panels of
Fig.~\ref{fig:i22_qnm_n} shows the comparison between the actual
$\tilde{I}_{22}$ and its QNM fit using this $N=3$ model. We see from the top
panel that the QNM description of the $(2,2)$ mass moment is valid even near the
merger. The relative error of this fit is $10^{-3}$ -- $10^{-2}$, which is about
two orders of magnitude greater than the numerical error measured by the
difference in $\tilde{I}_{22}$ between two resolutions, as shown in the bottom
panel. Again, this means the model could be improved in the future.

This model also provides an estimate of the final mass and spin of the
remnant. The right panel of Fig.~\ref{fig:i22_qnm_n_heatmap} shows the mismatch
heat map over the ($M_f, \chi_f$) parameter space, together with a golden cross
representing the true $M_f$ and $\chi_f$. Once more, we see a deep minimum near
the golden cross. The mass ($M_f'=0.95699M$) and spin ($\chi_f'=0.69066$) at the
minimum reproduce the true values, with error $\epsilon_f =
6.8\times10^{-3}$. This result also rules out overfitting
partially, because
almost any ($M_f, \chi_f$) combination yields a worse fit than the true
values. We cannot completely rule out overfitting since the five complex
  frequencies represent 10 real degrees of freedom, and we only vary two (final
  mass and spin).

\subsection{($\mathbf{3,2}$) spin moment} \label{sec:l32}

\begin{figure}[t]
	\centering 
	\includegraphics[width=\linewidth]{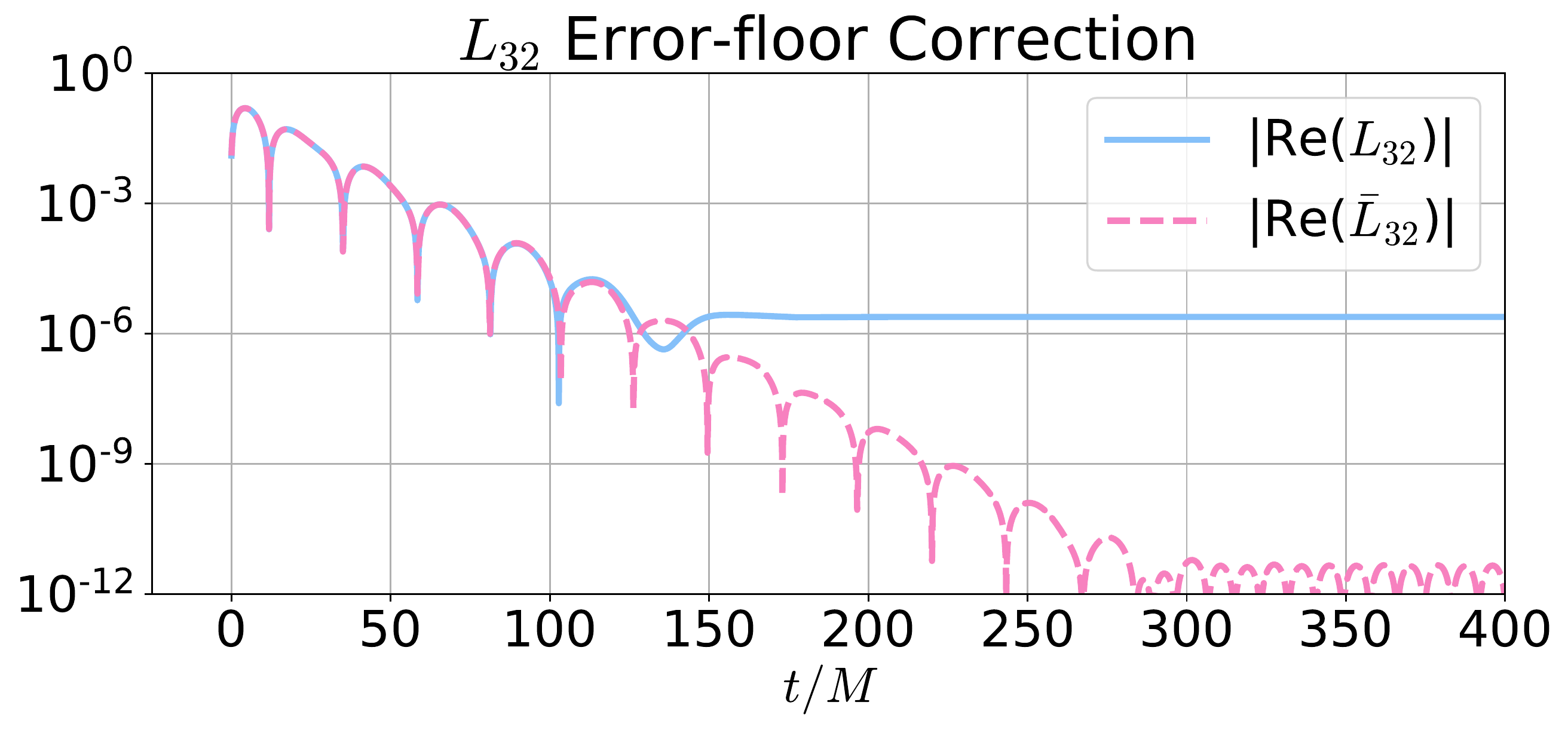}
	\caption{The spin moment $L_{32}$ and its floor correction. The original $(3,2)$ spin moment (cyan/solid) reaches a numerical floor at the level $5\times10^{-6}$ after $t\sim150M$. We define the floor-corrected spin moment $\bar{L}_{32}$ (pink/dashed) by subtracting the floor from $L_{32}$. The damped oscillatory pattern of $\bar{L}_{32}$ extends to $t\sim280M$. We also observe that the first several cycles are stretched wider near the local maxima.}
	\label{fig:l32_mag}
\end{figure}

The $(3,2)$ spin moment $L_{32}$ is the dominant mode among $L_{\ell m}$ with nonzero $m$. Figure~\ref{fig:l32_mag} shows the value of $|\mathrm{Re}(L_{32})|$, i.e., the magnitude of the real part of $L_{32}$, in cyan (solid). Similar to the $|\mathrm{Re}(I_{22})|$ curve in Fig.~\ref{fig:i22_mag}, this curve has a pattern of damped oscillation before $t=150M$, and then stays unchanged on a $5\times10^{-6}$ numerical error floor after $t=150M$. We subtract this floor from $L_{32}$ and define the floor-corrected spin moment
\begin{align}
	\bar{L}_{32} = L_{32} - \mathrm{mean}[L_{32}(t\ge400M)]. \label{eqn:l32_floor_corr}
\end{align}
The pink dashed curve in Fig.~\ref{fig:l32_mag} represents the value of $|\mathrm{Re}(\bar{L}_{32})|$. After the error floor correction, the damped oscillation extends to $t=280M$. Nevertheless, we will only focus on the portion $t\le150M$ of $\bar{L}_{32}$ henceforth. In Fig.~\ref{fig:l32_mag}, we also observe that the early-time portion of both curves does not follow a normal damped-oscillatory pattern: the first 3 -- 4 cycles are stretched wider at the local maxima, especially near $t\sim25M$ and $t\sim50M$. This is caused by mode mixing, as we shall see in the following subsection. This feature is not visible in Fig.~\ref{fig:i22_mag}, where the mixing of modes is relatively small.

\subsubsection{Mode mixing}

\begin{figure}[t]
\centering 
\includegraphics[width=\linewidth]{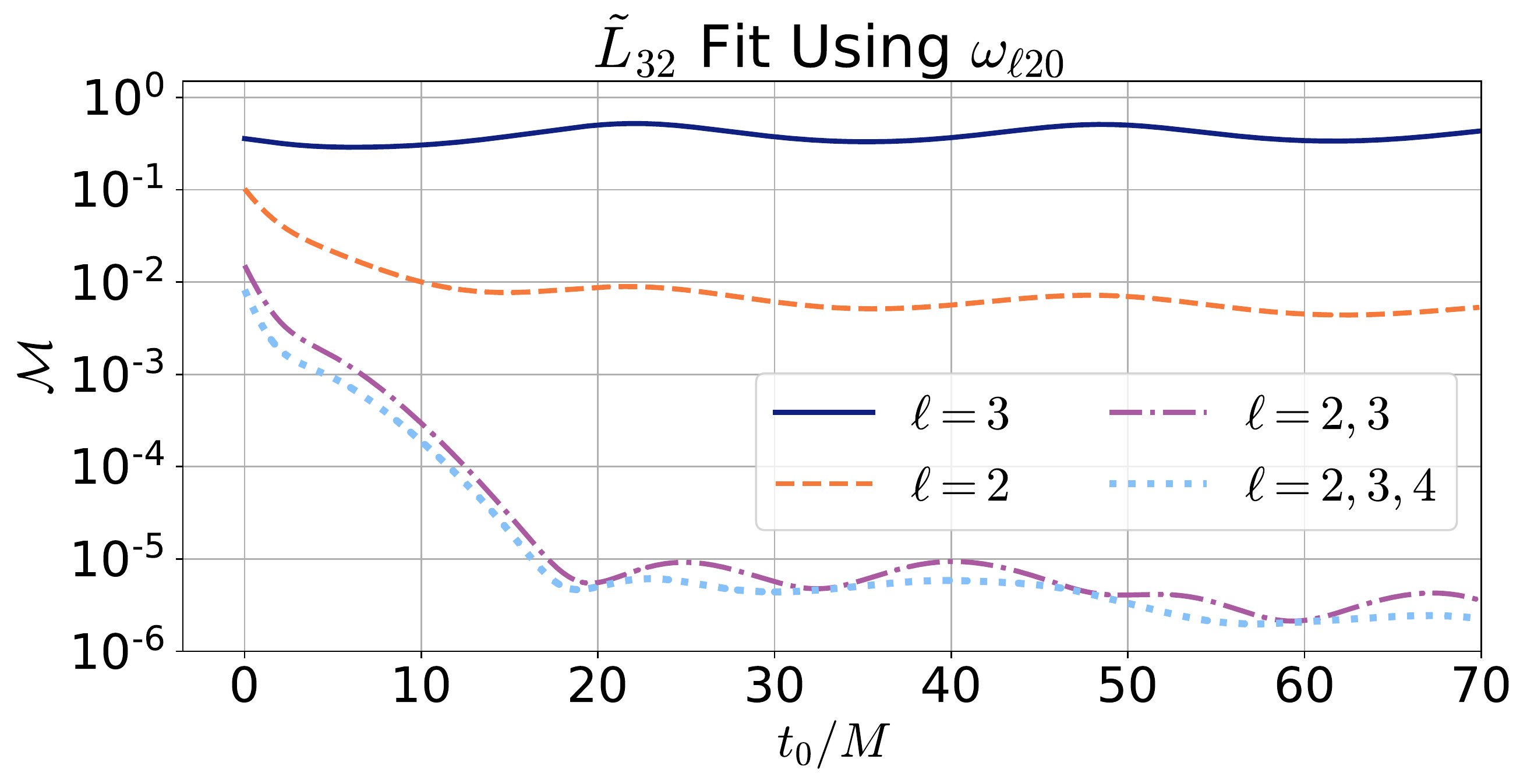}
\caption{The mismatch between $\tilde{L}_{32}$ and its fit
using the model Eq.~\eqref{eqn:l32_fitmodel_L}. The intuitive
choice $Q=\{3\}$ (blue/solid) actually produces
the QNM fit with the largest
mismatch. The best single-$\ell$ model uses $\ell=2$
(orange/dashed), producing a mismatch $\sim$$10^{-2}$. The
best two-$\ell$ model uses $\ell=2,3$ (purple/dash-dot),
which decreases the mismatch by a factor of $\sim$1000
compared to the orange dashed curve after $t_0=20M$. The
$\ell=2,3$ curve exhibits a wavy pattern, which can be
reduced by using the $\ell=2,3,4$ model (cyan/dotted).}
\label{fig:l32_qnm_varyt0_l20}
\end{figure}

Following the rotation procedure in Sec.~\ref{sec:rot_proc}, we define the rotated spin moments,
\begin{align}
	\tilde{L}_{\ell m}(t) = \bar{L}_{\ell m}(t) e^{-im\Omega t}, \label{eqn:Llm_rot_corr}
\end{align}
and investigate the mode mixing in $\tilde{L}_{32}$. We perform a QNM fit of $\tilde{L}_{32}$ using the following model:
\begin{align}
	\tilde{L}_{32} = \sum_{\ell\in Q} C_{\ell 20} e^{-i \omega_{\ell 20} (t-t_0)} \label{eqn:l32_fitmodel_L}.
\end{align}
We choose the fitting time range to be $t_0\le t\le 120M$, with $t_0$ varying, and assess the goodness of fit by mismatch [Eq.~\eqref{eqn:mismatch}]. The set $Q$ consists of integers to be specified. Since we are investigating the $(\ell=3, m=2)$ spin moment, the most intuitive choice of $Q$ is the singleton \{3\}, i.e., only considering the $(3,2)$ QNM. However, this choice completely fails the QNM fit with mismatch always above 0.1, as indicated by the blue solid curve in Fig.~\ref{fig:l32_qnm_varyt0_l20}. The best single-$\ell$ model is actually of $\ell=2$ (the orange dashed curve in the same graph), whose mismatch is smaller than the $\ell=3$ curve (blue/dashed) by a factor of 10 after $t_0=10M$. Thus, the $(2,2)$ QNM is the actual dominant mode in $\tilde{L}_{32}$. This is not unreasonable, because the perturbation of $\tilde{D}_a (\tilde{\epsilon}^{ab} \omega_b)$ [see $L_{\ell m}$'s definition, Eq.~\eqref{eqn:DHLlm}] is not guaranteed to satisfy the Teukolsky equation.

From Fig.~\ref{fig:l32_qnm_varyt0_l20}, we see that even the
best single-$\ell$ model has poor performance with mismatch
$\sim$$10^{-2}$. Thus, we move on to models
using two different $\ell$'s. In particular, we consider
all possible pairs of $\ell$ among $\{2,3,4,5\}$. The pair
$\ell=2,3$ yields the best QNM fit, as shown in purple/dash-dot in
Fig.~\ref{fig:l32_qnm_varyt0_l20}, while all other pairs produce
much worse mismatch (not shown).\footnote{The pairs $\ell=2,4$
and $\ell=2,5$ have mismatch close to the orange dashed curve in
Fig.~\ref{fig:l32_qnm_varyt0_l20}, while the remaining pairs close
to the blue dashed curve.} The mismatch of the $\ell=2,3$ curve is
much smaller than the $\ell=2$ curve (orange/dashed), by a factor of
$\sim$1000 after $t=20M$. This means that the $(2,2)$ and $(3,2)$
QNMs are the first two dominant modes in $\tilde{L}_{32}$. It also
demonstrates that a two-$\ell$ model can outperform any single-$\ell$
model when mode mixing is significant.

The purple dash-dot curve in Fig.~\ref{fig:l32_qnm_varyt0_l20}
has a wavy pattern after $t=20M$, similar to the $L=2$ curve
in Fig.~\ref{fig:i22_qnm_varyt0_l20}, which suggests a further
mode mixing. This oscillatory feature is indeed reduced by using
the $\ell=2,3,4$ model, as shown by the cyan dotted curve in
Fig.~\ref{fig:l32_qnm_varyt0_l20}. We continued expanding the model
to include more $\ell$, but we found the improvement negligible
(not shown). Hence, our $(3,2)$ spin moment is best described by
a linear combination of the $(2,2)$, $(3,2)$ and $(4,2)$ QNMs at
late times ($t\ge20M$).

\begin{figure}[t]
	\centering 
	\includegraphics[width=\linewidth]{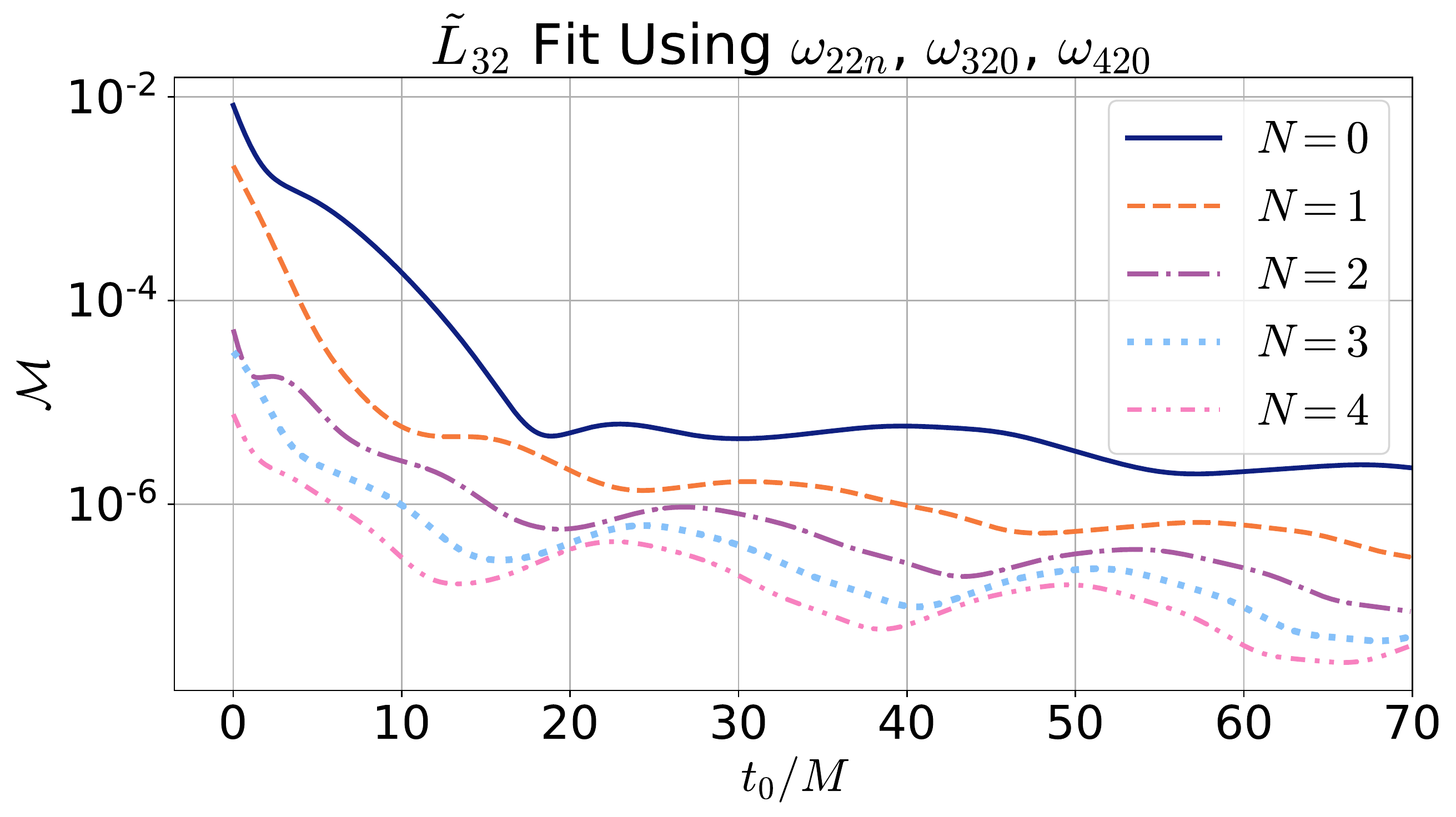}
	\caption{The mismatch between $\tilde{L}_{32}$ and its fit using the model Eq.~\eqref{eqn:l32_fitmodel_n}. By construction, the $N=0$ curve is the same as the cyan dotted curve in Fig.~\ref{fig:l32_qnm_varyt0_l20}. Including higher overtones brings down the mismatch, but also reveals a new oscillatory pattern. Unless this pattern is resolved, the effect of overtones on $\tilde{L}_{32}$ remains unclear.}
	\label{fig:l32_qnm_varyt0_320_420_22n}
\end{figure}

For $t\le20M$, the mismatch of the $\ell=2,3,4$ model (cyan/dotted) decays sharply from $10^{-2}$ to $10^{-5}$. To probe the effect of overtones on the early-time behavior of $\tilde{L}_{32}$, we consider the following fitting model,
\begin{align}
	\tilde{L}_{32} &= C_{320} e^{-i \omega_{320} (t-t_0)} + C_{420} e^{-i \omega_{420} (t-t_0)} \nonumber \\ 
	&+ \sum_{n=0}^{N} C_{2 2n} e^{-i \omega_{22n} (t-t_0)}, \label{eqn:l32_fitmodel_n}
\end{align}
with the fitting range $t_0\le t\le 120M$. We plot the mismatch as a function of
$t_0$ in Fig.~\ref{fig:l32_qnm_varyt0_320_420_22n} for five different $N$. By
construction, the $N=0$ curve (blue/solid) is identical to the cyan dotted curve
in Fig.~\ref{fig:l32_qnm_varyt0_l20}. As more overtones are included, the
mismatch decreases, and the initial decay pattern fades. However, it is yet
unclear whether the decay completely disappears, because a newly emerging wavy
pattern overshadows this decay. The wavy pattern is manifest in all four $N>0$
curves and persists for even higher $N$ (not shown). This suggests more
potential mixing from other QNMs, which we do not pursue further in this paper.\footnote{We have tried including an $\omega_{520}$ term in the fitting model Eq.~\eqref{eqn:l32_fitmodel_n}. This only improves the mismatch little and generates a figure similar to Fig.~\ref{fig:l32_qnm_varyt0_320_420_22n}.}

\subsection{($\mathbf{2,0}$) mass moment} \label{sec:i20}

There are two major differences between multipole moments of $m=0$ and those of $m\ne0$. First, an $m=0$ multipole moment is real-valued, while an $m\ne0$ mode is complex-valued. Second, as the remnant BH settles down, a nontrivial $m=0$ mode tends to a nonzero constant, while a nontrivial $m\ne0$ mode always tends to 0. Because of these distinctions, it is instructive to discuss $m=0$ multipole moments separately. We apply the techniques used in the previous two sections (Secs.~\ref{sec:i22} and \ref{sec:l32}) on $I_{20}$, but with slight modification. 

Mass and spin moments of a Kerr BH can be calculated
theoretically given its mass and spin \cite{1801.07048}. Let
$I_{20, \mathrm{theory}}$ be the theoretical value of the $(2,0)$
mass moment of a Kerr BH. We find that the
relative difference between $I_{20}$ and $I_{20, \mathrm{theory}}$
always lies below $4\times10^{-6}$ after $t=150M$, so our $I_{20}$
indeed approaches the expected value. To investigate the possible
QNM description of $I_{20}$, we subtract its
asymptotic value and define
\begin{align}
	\bar{I}_{20} = I_{20} - \mathrm{mean}[I_{20}(t\ge400M)].
\end{align}
This is similar to Eq.~\eqref{eqn:i22_floor_corr}, except that the nonzero value
of $I_{20}$ at a late time is related to the horizon geometry instead of numerical
errors. Note that for $m=0$, there is no need to rotate $\bar{I}_{20}$, and we
can directly set $\tilde{I}_{20} = \bar{I}_{20}$ [see
Eq.~\eqref{eqn:ilm_rot_corr}].

\begin{figure}[t]
	\centering 
	\includegraphics[width=\linewidth]{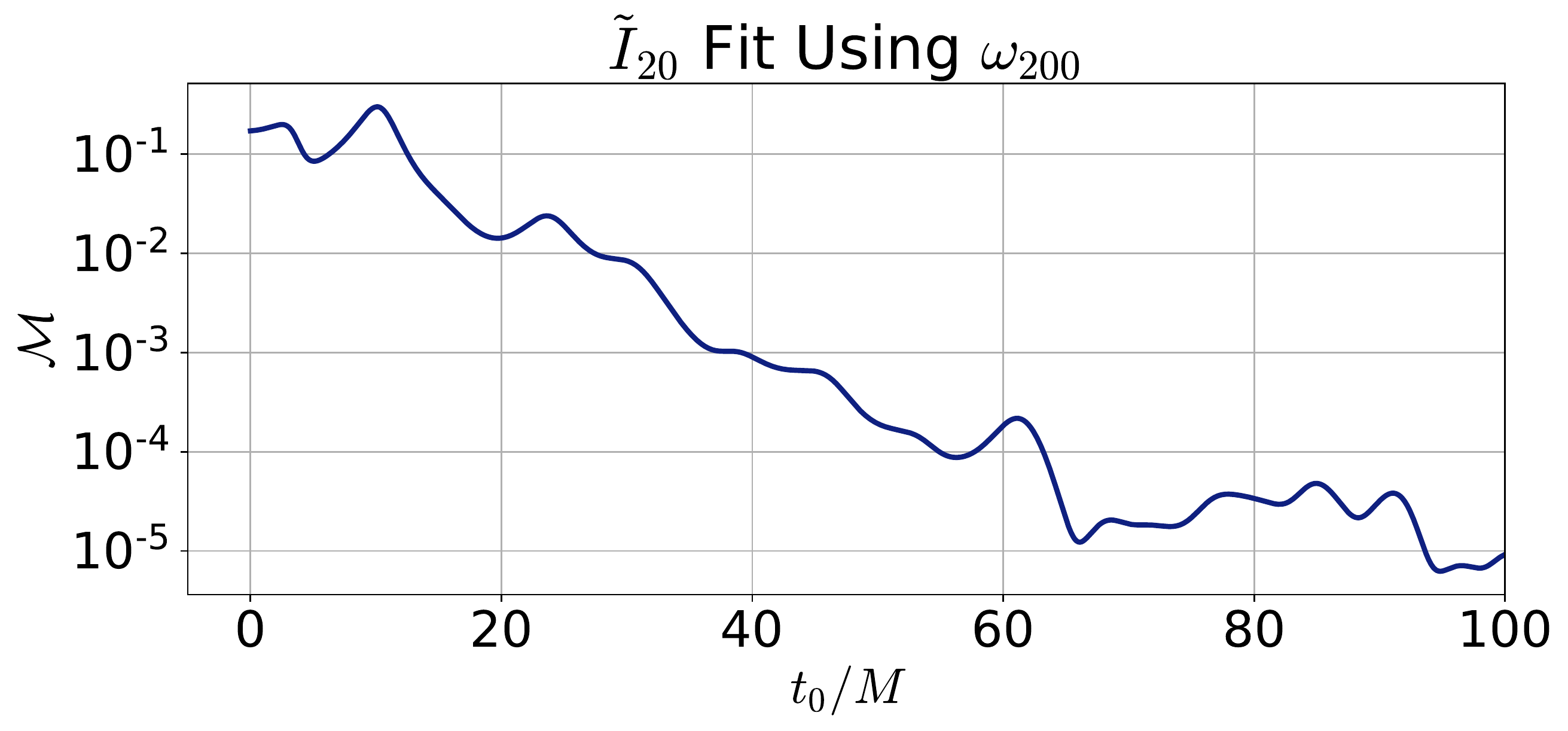}
	\caption{The mismatch between $\tilde{I}_{20}$ and its fit using the $\omega_{200}$ QNM [Eq.~\eqref{eqn:i20_fitmodel}]. The mismatch decays to the $10^{-5}$ level very slowly, unlike the $\tilde{I}_{22}$ case. There are irregular bumps along the curve, which is in stark contrast to the smooth curves in Figs.~\ref{fig:i22_qnm_n} and \ref{fig:l32_qnm_varyt0_l20}. The origin of these bumps is unknown.} 
	\label{fig:i20_vary_t0}
\end{figure}

We expect $\tilde{I}_{20}$ to be described by the fundamental tone of the $(2,0)$ QNM at late times. Because $\omega_{200}$ is a complex number while $\tilde{I}_{20}$ is real-valued, we use the following fitting model for $\tilde{I}_{20}$,\footnote{This model can be regarded as a linear combination of the prograde mode with the frequency $\omega^+_{200}$ and the retrograde mode with $\omega^-_{200}$.}
\begin{align}
	\tilde{I}_{20} = e^{-\lambda_1 (t-t_0)} [A_1\cos\lambda_2(t-t_0) + A_2 \sin\lambda_2(t-t_0)], \label{eqn:i20_fitmodel}
\end{align}
where $\lambda_1$ and $\lambda_2$ are the real and imaginary
parts of $-\omega_{200}$. The real parameters $A_1$ and $A_2$
are to be determined by a linear fit. The fitting range is
$t_0\le t\le 120M$ as usual. We first vary $t_0$ and analyze
the mismatch Eq.~\eqref{eqn:mismatch} as a function of $t_0$
in Fig.~\ref{fig:i20_vary_t0}. This curve ultimately reaches the
level of $10^{-5}$, but very gradually. This is different from
the mismatch curve of  $\tilde{I}_{22}$ fit by the $\omega_{220}$
mode (the blue solid curve in Fig.~\ref{fig:i22_qnm_varyt0_l20}),
which damps sharply to the $10^{-5}$ level before $t_0=20M$. Such
a distinction is unexpected, because the decay rates of
$\omega_{200}$ and $\omega_{220}$ differ by only a few percent
(see Table~\ref{tbl:sample_omega}). This suggests that the
model Eq.~\eqref{eqn:i20_fitmodel} may not be appropriate for
$\tilde{I}_{20}$ before $t_0=70M$ (at which $\tilde{I}_{20}$ drops to
near $10^{-5}$).

\begin{figure}[t]
	\centering 
	\includegraphics[width=\linewidth]{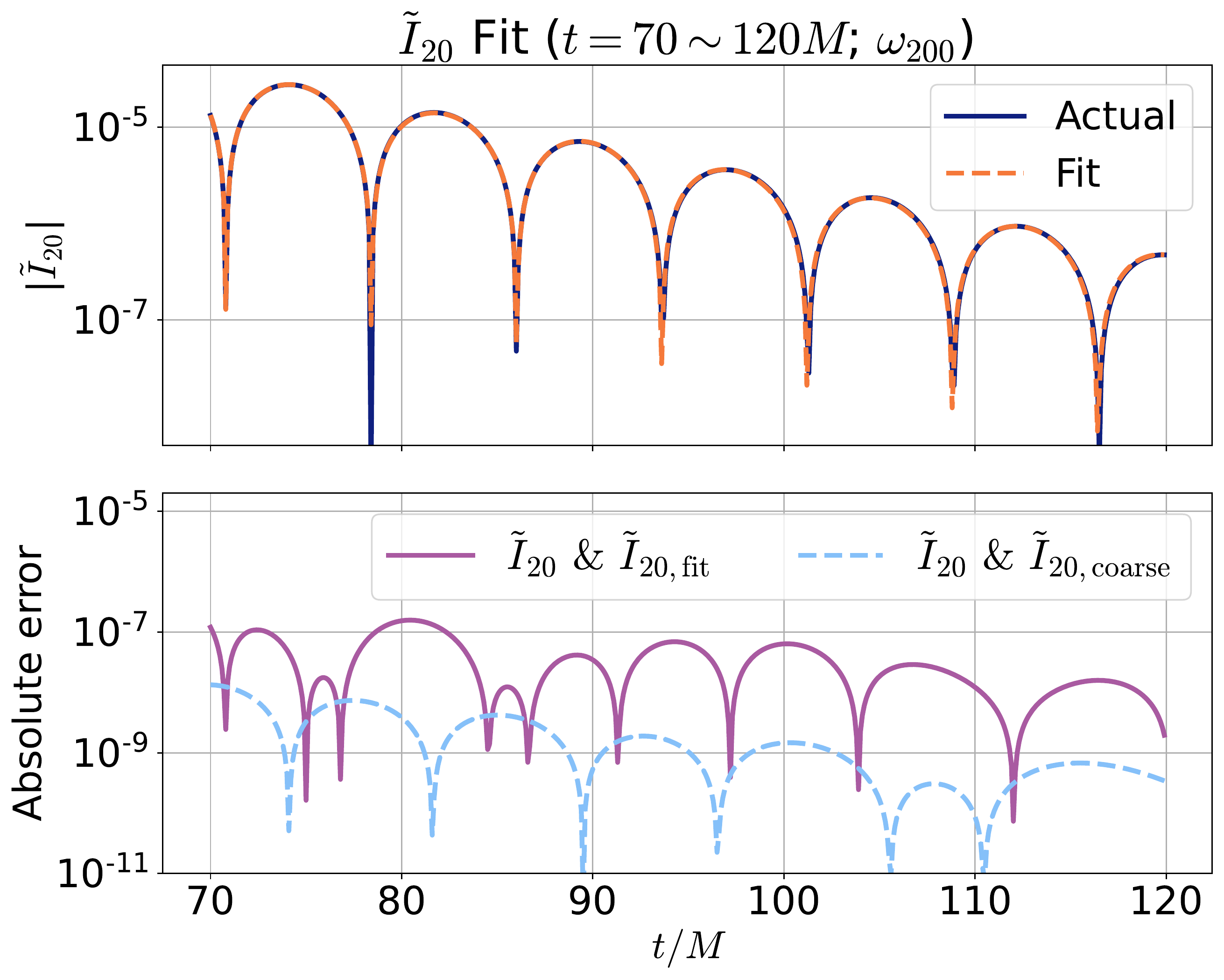}
	\caption{The comparison between $\tilde{I}_{20}$ and its fit based on the model Eq.~\eqref{eqn:i20_fitmodel}. The top panel shows the absolute values of $\tilde{I}_{20}$ (blue/solid) and the fit (orange/dashed), and these two curves overlap well. The bottom panel shows the absolute difference between $\tilde{I}_{20}$ and the fit in purple/solid, and the difference in $\tilde{I}_{20}$ between two resolution in cyan/dashed. }
	\label{fig:i20_n0}
\end{figure}

Next, we examine the performance of the model after $t=70M$, by fitting
$\tilde{I}_{20}$ with the $\omega_{200}$ mode in the time range $70M\le t\le
120M$. The top panel of Fig.~\ref{fig:i20_n0} displays both $\tilde{I}_{20}$ and
its fit, which overlap to within about 1\% relative error. The absolute difference between these two curves
is shown in purple (solid) in the bottom panel. Here, we use the absolute difference
instead of relative difference to measure error, because $\tilde{I}_{20}$
crosses zero periodically. The amplitude of the purple solid curve stays near
the level $10^{-7}$, which means the relative error is at the level $10^{-2}$ --
$10^{-1}$, after we take into account the magnitude of $\tilde{I}_{20}$. The
bottom panel also shows the absolute difference in $\tilde{I}_{20}$ between two
resolutions for reference (cyan/dashed). The figure indicates that
$\tilde{I}_{20}$ can be reasonably described by the $\omega_{200}$ mode at
sufficiently late times.

\begin{figure}[t]
	\centering 
	\includegraphics[width=\linewidth]{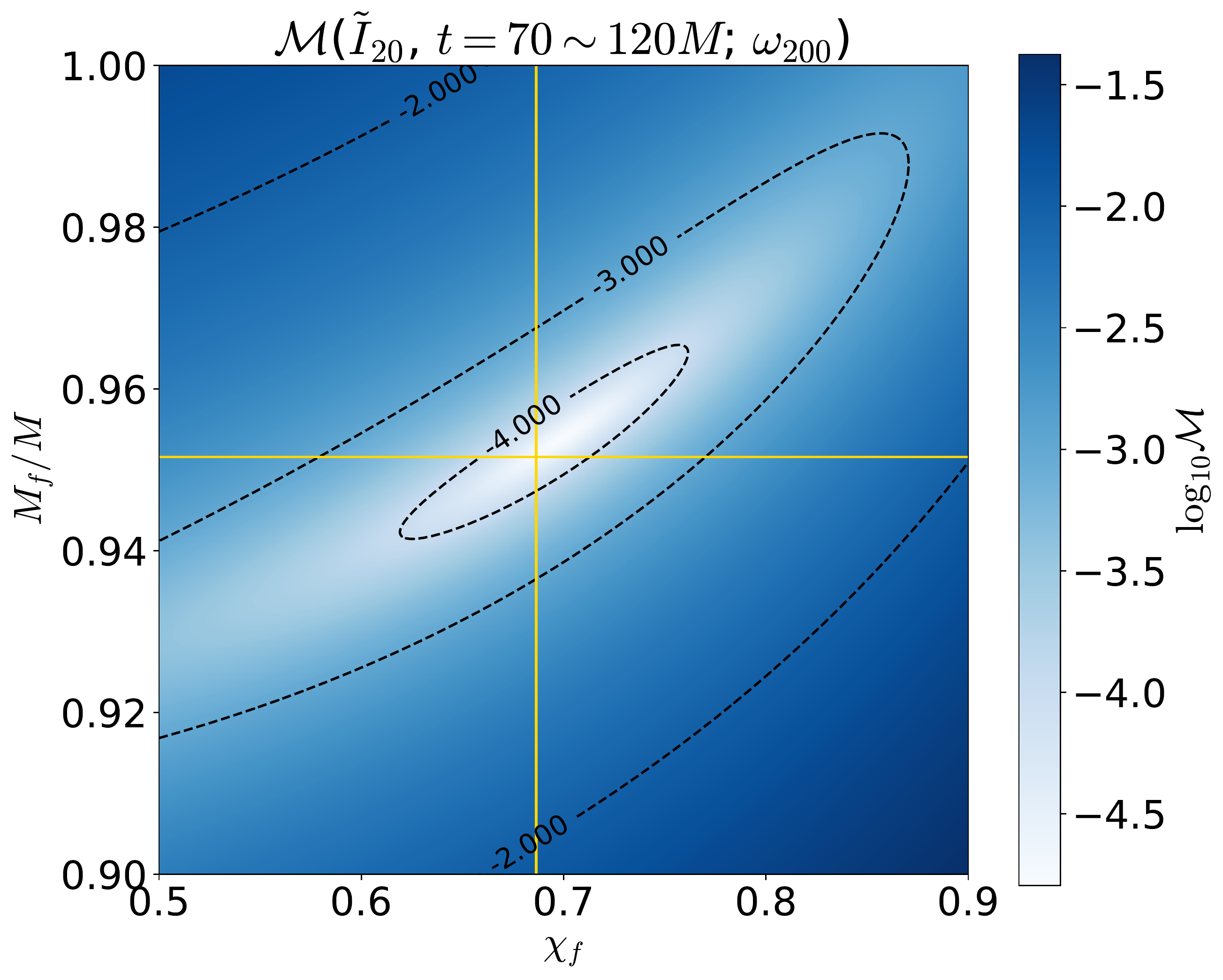}
	\caption{Heat map of the mismatch
            $\log_{10}\mathcal{M}$ over the ($M_f, \chi_f$) parameter
          space. This is generated based on the fit of $\tilde{I}_{20}$ using
          the model Eq.~\eqref{eqn:i20_fitmodel}. The fitting time range is
          $70M<t<120M$. The color representation is similar to
          Fig.~\ref{fig:i22_qnm_n_heatmap}, and we again use two golden lines to
          represent the true values. The dashed curves are the contour lines of constant mismatch. Although the minimum mismatch is
          located near the golden cross, the minimum is shallow, as discussed in
          Sec.~\ref{sec:i20}.}
	\label{fig:i20_qnm_n0_heatmap}
\end{figure}

Knowing that the model Eq.~\eqref{eqn:i20_fitmodel} can describe
the late-time behavior of $\tilde{I}_{20}$, we would like to
estimate the final mass and spin by minimizing the mismatch of
the fit. The outcome is not so satisfactory compared to the previous
cases. Figure~\ref{fig:i20_qnm_n0_heatmap} shows the mismatch
of the QNM fit (with the fitting range $70M\le t\le120M$), as both
the final mass and spin vary. Again, the golden lines represent the
true mass and spin, and a lighter-shaded region has lower mismatch. The
local minimum is achieved at $M_f' = 0.95374M$ and $\chi_f' =
0.69868$, which yields an error $\epsilon_f = 1.2\times10^{-2}$,
about 4 times the error $\epsilon_f$ in Sec.~\ref{sec:i22_n0}. This
means that, with regard to the performance of mass or spin estimate,
fitting $\tilde{I}_{20}$ is inferior to fitting $\tilde{I}_{22}$. To
understand why the $\omega_{200}$ model for $\tilde{I}_{20}$ is less
faithful, we should realize that this model is not very sensitive
to the remnant parameters. This can be seen from
Fig.~\ref{fig:i20_qnm_n0_heatmap}, where
the local minimum of the mismatch is shallow. Specifically, the minimum
mismatch is $1.61\times10^{-5}$, which is very close to the mismatch
at the true mass and spin, $1.87\times10^{-5}$. There is actually
a fundamental reason for the weakness of this model: the variation
in the values of $\omega_{200}$ versus spin is much smaller than the
one of $\omega_{220}$. In particular, as the spin ranges from 0.5 to
0.9, Re($\omega_{220}$) increases by 45\%, while Re($\omega_{200}$)
by only 7\%. In summary, the $\omega_{200}$ model is a reasonable
but spin-insensitive model for $\tilde{I}_{20}$ at late times.

\subsection{Other multipole moments} \label{sec:other_moments}

Here, we briefly summarize the results for those multipole moments that have not been  discussed previously. We will focus on the nontrivial $\tilde{I}_{\ell m}$ and $\tilde{L}_{\ell m}$ up to $\ell=6$. Note that these multipole moments are all floor-corrected and rotated. 

We start with the multipole moments with $\ell=m$, specifically,
$\tilde{I}_{44}$ and $\tilde{I}_{66}$.\footnote{The moment $I_{00}$
has a constant value.} Fitting $\tilde{I}_{44}$ or $\tilde{I}_{66}$
with a single-$\ell$ QNM model results in a beat pattern at late
times, so there is mode mixing in both
cases. The best\footnote{The \textit{best} model includes all
$\ell$ that can appreciably improve the QNM fit, and excludes those
$\ell$ that produce negligible improvement.} multi-$\ell$ model
(with $m$ fixed) for the late-time behavior of $\tilde{I}_{44}$
consists of the $\omega_{440}$ and $\omega_{540}$ modes, while
the best model for $\tilde{I}_{66}$ consists of $\omega_{660}$ and
$\omega_{760}$. We have not found any good model that describes the
early-time behavior of $\tilde{I}_{44}$ and $\tilde{I}_{66}$. For
example, simply including $\omega_{44n}$ (or $\omega_{66n}$)
overtones in a QNM model does not eliminate the initial decay of
$\tilde{I}_{44}$ (or $\tilde{I}_{66}$).

Next, we consider the nontrivial multipole moments with $0 < m
<\ell$: $\tilde{I}_{42}$, $\tilde{I}_{62}$, $\tilde{I}_{64}$,
$\tilde{L}_{52}$, and $\tilde{L}_{54}$. Their behaviors are very
similar to that of $\tilde{L}_{32}$. Mode mixing is significant
for these multipole moments, and the best multi-$\ell$ models for
them are comprised of three or four fundamental tones of different
$\ell$. For example, $\tilde{I}_{42}$, $\tilde{I}_{62}$, and
$\tilde{L}_{52}$ are all best described by the \{$\omega_{220}$,
$\omega_{320}$, $\omega_{420}$, $\omega_{520}$\} model at late
times. For early-time behavior, adding overtones does greatly
reduce the initial decay pattern, but this comes with the emergence
of additional oscillatory patterns whose origin is unclear at
this time.

Finally, we study the multipole moments with $m=0$:
$\tilde{I}_{40}$, $\tilde{I}_{60}$, $\tilde{L}_{30}$,
and $\tilde{L}_{50}$.\footnote{The moment $L_{10}$ is
proportional to the angular momentum of the merged BH.} They
all approach their respective theoretical values with error
below $1.2\times10^{-5}$. The best multi-$\ell$ model [by
extending Eq.~\eqref{eqn:i20_fitmodel}] for $\tilde{L}_{30}$
uses \{$\omega_{200}$, $\omega_{300}$\}, while the best model
for $\tilde{I}_{40}$, $\tilde{L}_{50}$, and $\tilde{I}_{60}$ uses
\{$\omega_{200}$, $\omega_{300}$, $\omega_{400}$\}. A common feature
shared by these models is their failure to describe the multipole
moments before $t\sim 60$ -- $80M$. At sufficiently late times, these
models do produce a good description of the respective multipole
moments. However, we should keep in mind that the $m=0$ QNMs used
in these models are not as sensitive to the
remnant spin as the $m\ne0$ QNMs,
so the models might not be very precise.

\section{Conclusion} \label{sec:conclu}

In this paper, we numerically construct the multipole moments on
the common horizon of an equal-mass BBH system on a sequence of
time slices. The construction
process captures the connection among the common horizons on
different time slices, which ensures that this set of multipole
moments is spatially gauge independent. We apply
a geometrically motivated rotation to the multipole moments,
which turns out to simplify
the analysis. We compare the
multipole moments of the horizons with those of the gravitational
waveform, and see a strong correlation between the $(\ell=2,m=2)$
mass multipole moment and the strain $(2,2)$-mode. Specifically,
they share the same oscillation frequency and decay constant at
late times. This suggests the possible QNM description of
horizon multipole moments, which we pursue next.

We consider all nontrivial multipole moments up to $\ell=6$,
and model each multipole moment as a linear combination of
spin-weight-2 QNMs. At sufficiently late times, these multipole
moments are well described by the fundamental tones of QNMs: not
only do the true values overlap with the predicted values fit to the
QNM models, but also the mismatch between them is small. However,
the multipole moments do not match one-to-one with the fundamental
tones, and we actually see a manifest mode-mixing phenomenon in all
the multipole moments. For example, our best QNM model for the late-time
behavior of the $(2,2)$ mass moment consists of the $\omega_{220}$
and $\omega_{320}$ QNMs, where the $\omega_{320}$ mode has a tiny
but nonnegligible contribution. A more counter-intuitive example is
the $(3,2)$ spin moment, in which the $\omega_{220}$ mode dominates
over the $\omega_{320}$ mode, instead of vice versa. We find that in
general, the $(\ell,m)$ multipole moment at late times is described
by a QNM model consisting of the $(\ell',m)$ fundamental tones for
the first several possible $\ell'$.

We also explore the possibility of QNM modeling for the early-time
behavior of multipole moments by including overtones. We find 
that the inclusion of $\omega_{22n}$ overtones up to $n=3$ is
sufficient to provide an accurate representation of the $(2,2)$
mass moment immediately after the merger. This extends the power
of BH perturbation theory back to the time of coalescence. However,
this picture does not apply to other multipole moments: a QNM model
with overtones does reduce the mismatch significantly, but at the
same time, it also unveils further mixing of modes. As a consequence,
a more careful modeling with overtones is needed in the future to
describe the early-time behavior of multipole moments other than
the $(2,2)$ mass moment.

Taking into account the effect of mode mixing, we find that the
QNM models using fundamental tones at late times provide a fairly
faithful estimate of the remnant mass and spin, especially for those
multipole moments of nonzero $m$. Furthermore, in the case of the
$(2,2)$ mass moment, the QNM model with overtones also recovers the
true mass and spin at the minimum mismatch. We also note that for
the $m=0$ multipole moments, the performance of these estimates is
not as good as in the $m\ne0$ cases. This is interpreted as
resulting from the weaker dependence of the $m=0$ mode frequencies
on the spin.

In summary, this paper provides promising evidence for the QNM
description of horizon multipole moments of a remnant BH in the
ringdown phase of an equal-mass non-spinning BBH system. These
multipole moments are spatially gauge independent, as we take into
account the relation among apparent horizons in the construction
step. Such gauge independence, along with the accuracy of
the SpEC code, allows these multipole moments to be
described with QNMs much more accurately than those horizon multipole moments constructed in
previous literature (e.g., \cite{1801.07048, 2109.01193}).

As future
work, one can consider more generic BBH systems whose progenitors
have different masses or nonzero spins, and then construct horizon
multipole moments as outlined in this paper. One may also define
a similar set of horizon multipole moments for the progenitor BHs,
and investigate their possible imprint on the common horizon
multipole moments. Note that Ref.~\cite{2109.01193} discusses the
multipole moments of the progenitors, but the construction there does
not yet capture the connection among the apparent horizons. Regarding
the QNM models, one can continue improving them to mitigate the
effect of mode mixing. Such improvement should reveal a clearer
pattern in the early-time portion of horizon multipole moments.

\section*{Acknowledgments}

We thank Abhay Ashtekar, Bangalore Sathyaprakash, Ssohrab Borhanian, Leo Stein, and Robert Owen for useful discussions. Computations for this work were performed with the Wheeler cluster at Caltech and the Bridges system (and XSEDE) at the Pittsburgh Supercomputing Center (PSC). This work was supported in part by the Sherman Fairchild Foundation and by NSF Grants PHY-2011961, PHY-2011968, and OAC-1931266 at Caltech, as well as NSF Grants PHY-1912081, OAC-1931280, and PHY-2209655 at Cornell. This work was also supported by NSF grant PHY-1806356, the Eberly Chair funds of Penn State University, and the Mebus Fellowship to N.K. P.K. acknowledges support of the Department of Atomic Energy, Government of India, under project no. RTI4001, and of the Ashok and Gita Vaish Early Career Faculty Fellowship at the International Centre for Theoretical Sciences.

\appendix
\section{Balance laws and error convergence} \label{sec:app_ballaw_error}

\begin{figure}[t]
	\centering
	\includegraphics[width=\linewidth]{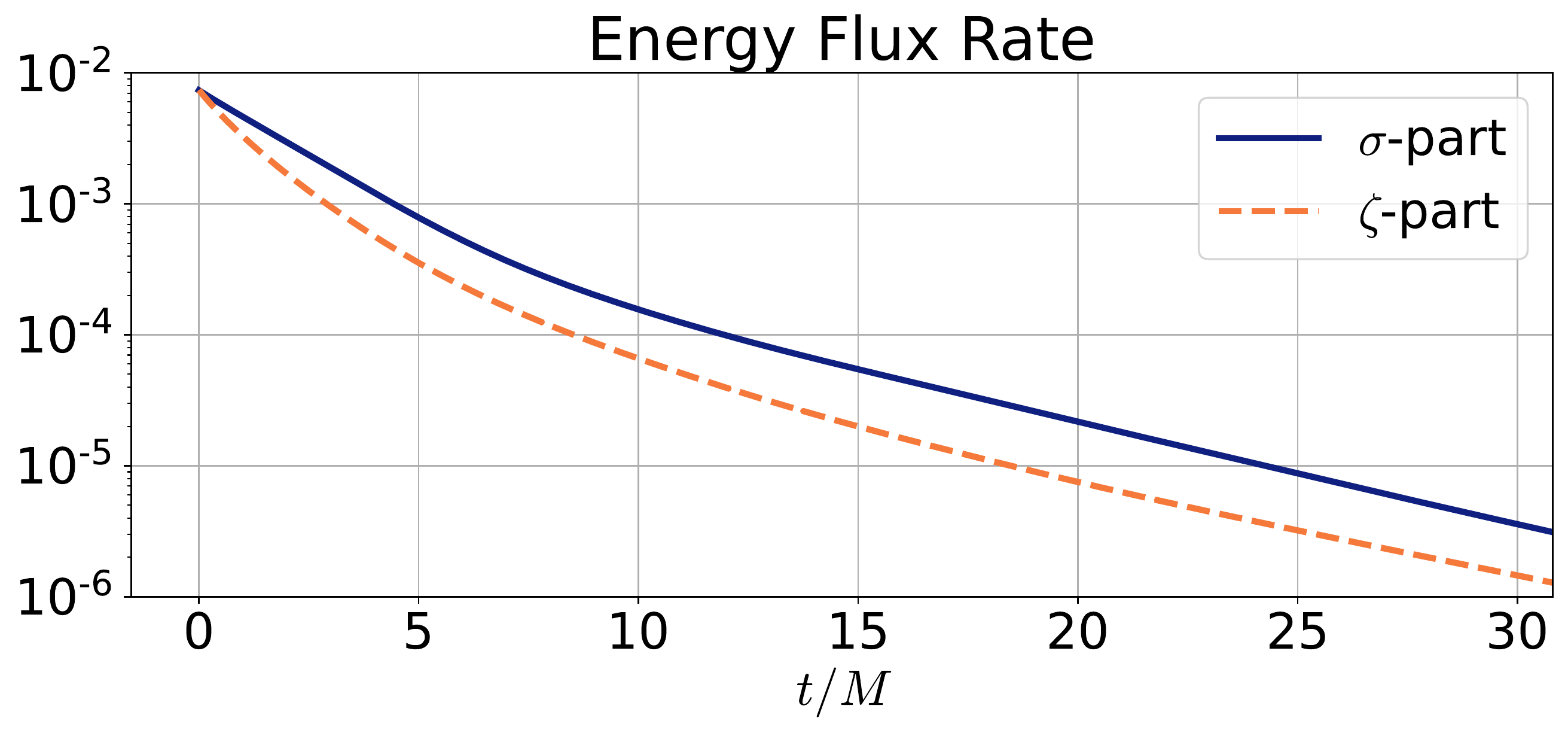}
	\caption{The energy flux rate $d\mathcal{F}_g/dt$. The rate consists of two parts, and we show the $\sigma$-part ($d\mathcal{F}_{g,\sigma}/dt$) in blue/solid and the $\zeta$-part ($d\mathcal{F}_{g,\zeta}/dt$) in orange/dashed. We only consider the time range $0\le t\le 30.8M$. Except at the merger, the $\sigma$-part is always greater than the $\zeta$-part, but the difference is not substantial: the $\sigma$-part is at most 2 -- 3 times as much as the $\zeta$-part. }
	\label{fig:flux_rate}
\end{figure}

As mentioned in Sec.~\ref{sec:ballaw}, the balance laws,
Eqs.~\eqref{eqn:radiusbal_diff}, \eqref{eqn:Ilmbal_diff}, and
\eqref{eqn:Llmbal_diff}, provide internal consistency checks for BH
simulations. In this section, we use them to test the correctness of
the BBH simulation in Sec.~\ref{sec:bbh_sim}. We start by showing the
energy flux rate $d\mathcal{F}_g/dt$ in Fig.~\ref{fig:flux_rate},
as it is relevant to the area balance law. The graph displays the
$\sigma$-part ($d\mathcal{F}_{g,\sigma}/dt$) in blue (solid) and
the $\zeta$-part ($d\mathcal{F}_{g,\zeta}/dt$) in orange (dashed),
as a function of simulation time $t$. We only show the time
range $t\leq 30.8M$, since the calculation of the $\zeta$-part
is numerically unstable at late times because of the divergence of
the components of
$\hat{r}^a$. Both curves decay exponentially, with
higher decay rates near the merger. We see that the $\sigma$-part
always dominates the $\zeta$-part, except at the merger. They differ
by a factor of 2 -- 3 after $t=5M$, which is not significant.

Next, we investigate the numerical violations of these three
balance laws as functions of simulation time ($t\leq 30.8M$). The
violations are measured by the relative difference between the left-
and right-hand sides of their respective equations. We find that
the area balance law [Eq.~\eqref{eqn:radiusbal_diff}] always holds
within $10^{-4}$, and for most of the time within $10^{-5}$. The
mass moment balance law [Eq.~\eqref{eqn:Ilmbal_diff}] always
holds within $3\times 10^{-6}$ for all nontrivial mass moments
with $1\le\ell\le8$,\footnote{We did not check the balance law for
$I_{00}$, even though it is nontrivial. This is because $I_{00}$
is equal to the constant $\sqrt{\pi}$ (which we checked), and both
sides of the differential balance law should vanish.} and the spin
moment balance law [Eq.~\eqref{eqn:Llmbal_diff}] always holds to within
$10^{-5}$ for all nontrivial spin moments up to $\ell=8$.

To demonstrate the convergence of relative errors in the balance
laws, we perform simulations of the same BBH system as described in
Sec.~\ref{sec:bbh_sim}, but at four additional resolutions. Including
the two resolutions used in the main text, we have six resolutions
in total. These resolutions are labeled ``Lev-$i$", where
$i=1,2,\cdots,6$. For a fixed $i$, the target truncation error
of the adaptive mesh refinement algorithm is $\sim 2\times 4^{-i}
\times 10^{-4}$. Note that Lev-6 corresponds to the higher resolution
in the main text, while Lev-5 corresponds to the lower one.

\begin{figure}[t]
	\centering 
	\includegraphics[width=\linewidth]{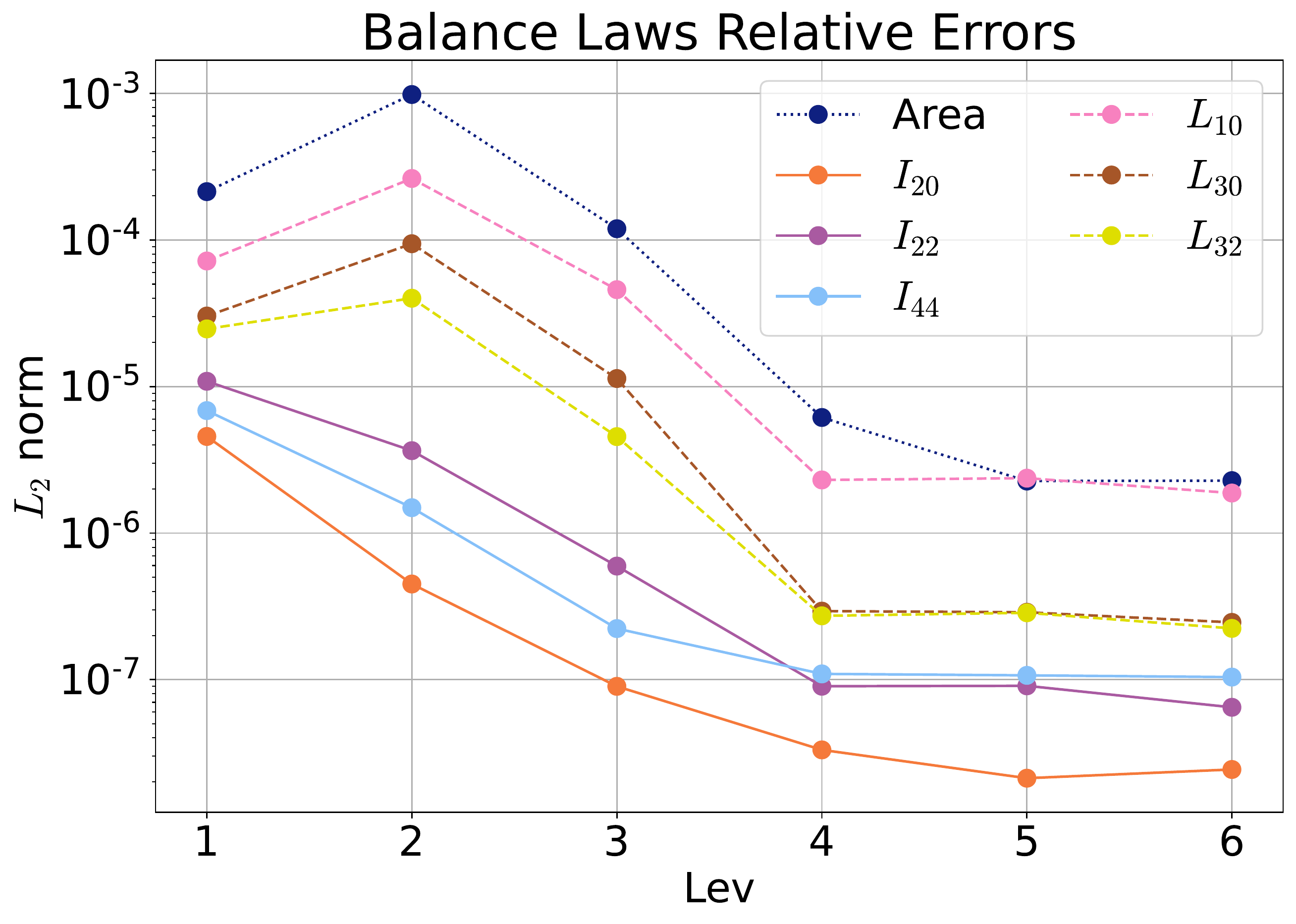}
	\caption{The convergence of relative errors in the balance laws. The horizontal axis represents the resolution labeled by ``Lev", and the vertical axis represents the $L_2$ norm of the relative errors in these balance laws. The blue dotted line stands for the area balance law. The solid lines are for the mass moment balance law, while the dashed lines for the spin moment balance law.}
	\label{fig:bal_laws_levs}
\end{figure}

Figure~\ref{fig:bal_laws_levs} shows the $L_2$ norm\footnote{Specifically, the relative error in a balance law is a time series in $0\leq t\leq 30.8M$. The $L_2$ norm here refers to the Euclidean $L_2$ norm of this time series, then divided by the square root of the length of the series.} of the relative errors in the balance laws. The blue dotted line represents the area balance law, while the solid lines stand for the mass moment balance law, and the dashed lines for the spin moment balance law. We only show three mass moments and three spin moments here, but we checked that these curves are representative\ of the behaviors of other nontrivial horizon moments. We can see from the graph that the errors converge as the resolution increases from Lev-2 to Lev-5, and they reach floors around Lev-5. Therefore, we conclude that the balance laws for the area, mass moments, and spin moments are accurate and satisfied in our simulation.

\section{Surface gravity} \label{sec:app_surf_grav}

In this section, we briefly investigate the surface gravity on a dynamical horizon \cite{gr-qc/0308033, gr-qc/0610032, 1306.5697},
\begin{align}
	\kappa_V = -n_b V^a\nabla_a V^b.
\end{align}
Here, $n^a$ is the ingoing null normal to the common horizon ($t = \text{constant}$ slice) on $\mathcal{H}$, satisfying $V^a n_a = -1$. As the dynamical horizon approaches an isolated horizon, $V^a$ becomes null and this surface gravity coincides with the one on an isolated horizon. Because $\kappa_V$ is a function on a dynamical horizon, it is more convenient to consider the average value of $\kappa_V$ over each common horizon $\mathcal{S}$, which we denote as $\kappa_{V,t}$. 

\begin{figure}[t]
	\centering 
	\includegraphics[width=\linewidth]{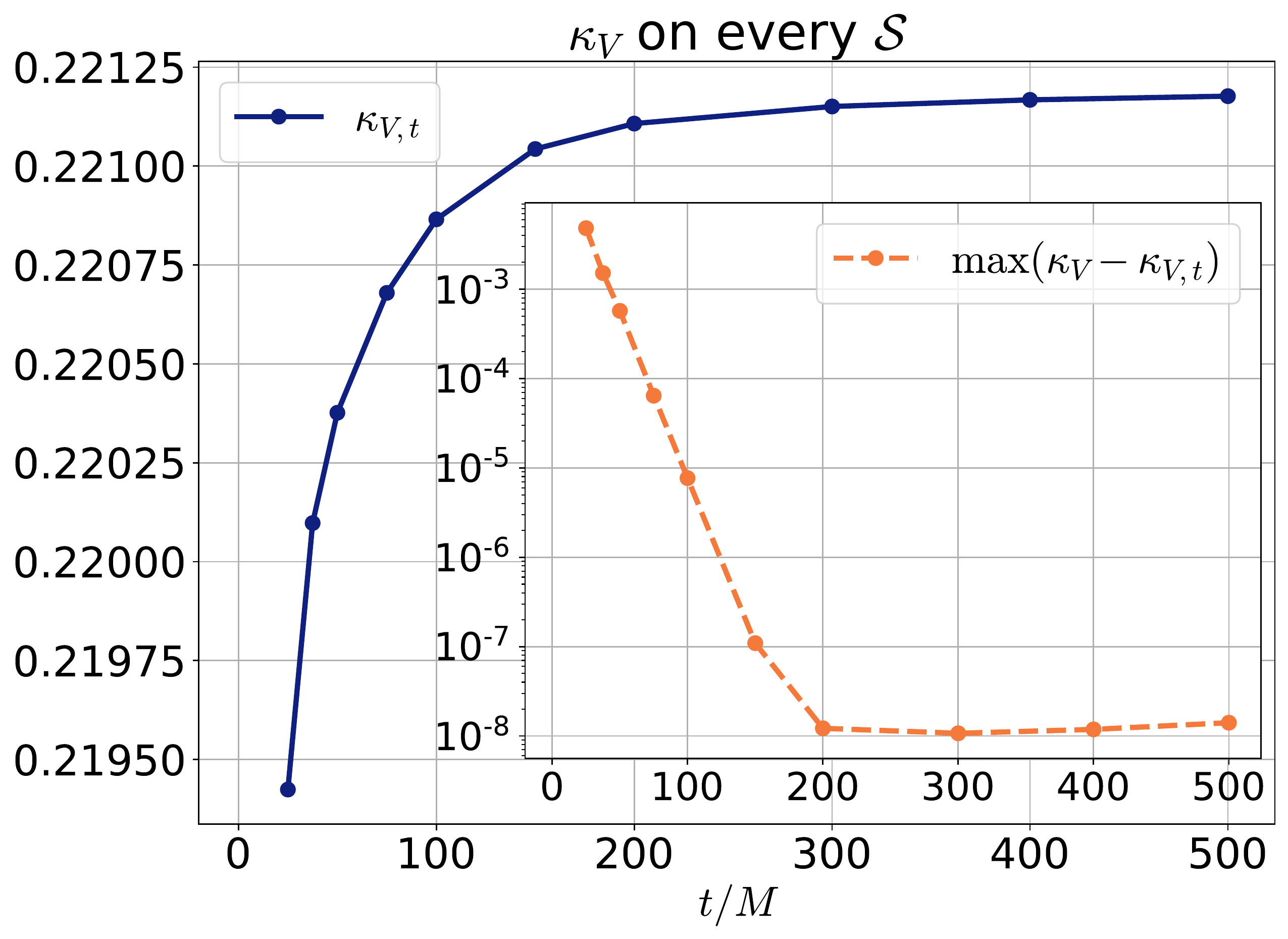}
	\caption{The temporal behavior of the surface gravity $\kappa_V$ on the dynamical horizon. The average value of $\kappa_V$, denoted by $\kappa_{V,t}$, is shown in blue/solid. The maximum deviation of $\kappa_V$ from $\kappa_{V,t}$ on every common horizon $\mathcal{S}$ is shown in orange/dashed in the inset. We see that $\kappa_V$ becomes a constant at $t=500M$.}
	\label{fig:kappa}
\end{figure}

In Fig.~\ref{fig:kappa}, we show $\kappa_V$ as a function of the simulation time $t$, starting from $t=25M$. The blue solid curve represents $\kappa_{V,t}$, and the orange dashed curve represents $\max(\kappa_V - \kappa_{V,t})$, i.e., the maximum deviation of $\kappa_V$ from its average value on every $\mathcal{S}$. We see from the blue curve that $\kappa_{V,t}$ is settling down, and we check that the absolute difference between $\kappa_{V,t=400M}$ and $\kappa_{V,t=500M}$ is $\sim 10^{-5}$. The orange curve tells us that $\kappa_V$ is a constant on every common horizon after $t=200M$, with error $\sim 10^{-8}$. From this, we conclude that $\kappa_V$ already reaches a constant on the dynamical horizon at $t=500M$, with error $\sim 10^{-5}$.

The final value of $\kappa_V$ in our simulation is
\begin{align}
	\kappa_{V,t=500M} = 0.221177 M^{-1},
\end{align}
which is very close to the Kerr surface gravity \cite{Wald1_skip, 1412.5432},
\begin{align}
	\kappa_{\mathrm{Kerr}} = \frac{1}{4M_f} - M_f \Omega_{H}^2 = 0.221214 M^{-1}.
\end{align}
Note that this expression for $\kappa_{\mathrm{Kerr}}$ is calculated
using the canonical null Killing vector of the Kerr solution on the horizon. The relative difference between $\kappa_{V,t=500M}$
and $\kappa_{\mathrm{Kerr}}$ is
$1.7\times 10^{-4}$. This confirms the approximation $f\approx 1$ in Sec.~\ref{sec:rot_proc}, and is related to the slight deviation of $\Omega_{t=500M}$ from $\Omega_{H}$ seen in Sec.~\ref{sec:i22}.

\newpage

\bibliography{library,library2}

\begin{thebibliography}{80}%
\makeatletter
\providecommand \@ifxundefined [1]{%
 \@ifx{#1\undefined}
}%
\providecommand \@ifnum [1]{%
 \ifnum #1\expandafter \@firstoftwo
 \else \expandafter \@secondoftwo
 \fi
}%
\providecommand \@ifx [1]{%
 \ifx #1\expandafter \@firstoftwo
 \else \expandafter \@secondoftwo
 \fi
}%
\providecommand \natexlab [1]{#1}%
\providecommand \enquote  [1]{``#1''}%
\providecommand \bibnamefont  [1]{#1}%
\providecommand \bibfnamefont [1]{#1}%
\providecommand \citenamefont [1]{#1}%
\providecommand \href@noop [0]{\@secondoftwo}%
\providecommand \href [0]{\begingroup \@sanitize@url \@href}%
\providecommand \@href[1]{\@@startlink{#1}\@@href}%
\providecommand \@@href[1]{\endgroup#1\@@endlink}%
\providecommand \@sanitize@url [0]{\catcode `\\12\catcode `\$12\catcode
  `\&12\catcode `\#12\catcode `\^12\catcode `\_12\catcode `\%12\relax}%
\providecommand \@@startlink[1]{}%
\providecommand \@@endlink[0]{}%
\providecommand \url  [0]{\begingroup\@sanitize@url \@url }%
\providecommand \@url [1]{\endgroup\@href {#1}{\urlprefix }}%
\providecommand \urlprefix  [0]{URL }%
\providecommand \Eprint [0]{\href }%
\providecommand \doibase [0]{https://doi.org/}%
\providecommand \selectlanguage [0]{\@gobble}%
\providecommand \bibinfo  [0]{\@secondoftwo}%
\providecommand \bibfield  [0]{\@secondoftwo}%
\providecommand \translation [1]{[#1]}%
\providecommand \BibitemOpen [0]{}%
\providecommand \bibitemStop [0]{}%
\providecommand \bibitemNoStop [0]{.\EOS\space}%
\providecommand \EOS [0]{\spacefactor3000\relax}%
\providecommand \BibitemShut  [1]{\csname bibitem#1\endcsname}%
\let\auto@bib@innerbib\@empty
\bibitem [{\citenamefont {Israel}(1967)}]{Israel1_skip}%
  \BibitemOpen
  \bibfield  {author} {\bibinfo {author} {\bibfnamefont {W.}~\bibnamefont
  {Israel}},\ }\bibfield  {title} {\bibinfo {title} {Event horizons in static
  vacuum space-times},\ }\href {https://doi.org/10.1103/PhysRev.164.1776}
  {\bibfield  {journal} {\bibinfo  {journal} {Phys. Rev.}\ }\textbf {\bibinfo
  {volume} {164}},\ \bibinfo {pages} {1776} (\bibinfo {year}
  {1967})}\BibitemShut {NoStop}%
\bibitem [{\citenamefont {Carter}(1971)}]{Carter1_skip}%
  \BibitemOpen
  \bibfield  {author} {\bibinfo {author} {\bibfnamefont {B.}~\bibnamefont
  {Carter}},\ }\bibfield  {title} {\bibinfo {title} {Axisymmetric black hole
  has only two degrees of freedom},\ }\href
  {https://doi.org/10.1103/PhysRevLett.26.331} {\bibfield  {journal} {\bibinfo
  {journal} {Phys. Rev. Lett.}\ }\textbf {\bibinfo {volume} {26}},\ \bibinfo
  {pages} {331} (\bibinfo {year} {1971})}\BibitemShut {NoStop}%
\bibitem [{\citenamefont {Scheel}\ \emph {et~al.}(2009)\citenamefont {Scheel},
  \citenamefont {Boyle}, \citenamefont {Chu}, \citenamefont {Kidder},
  \citenamefont {Matthews},\ and\ \citenamefont {Pfeiffer}}]{0810.1767}%
  \BibitemOpen
  \bibfield  {author} {\bibinfo {author} {\bibfnamefont {M.~A.}\ \bibnamefont
  {Scheel}}, \bibinfo {author} {\bibfnamefont {M.}~\bibnamefont {Boyle}},
  \bibinfo {author} {\bibfnamefont {T.}~\bibnamefont {Chu}}, \bibinfo {author}
  {\bibfnamefont {L.~E.}\ \bibnamefont {Kidder}}, \bibinfo {author}
  {\bibfnamefont {K.~D.}\ \bibnamefont {Matthews}},\ and\ \bibinfo {author}
  {\bibfnamefont {H.~P.}\ \bibnamefont {Pfeiffer}},\ }\bibfield  {title}
  {\bibinfo {title} {{High-accuracy waveforms for binary black hole inspiral,
  merger, and ringdown}},\ }\href {https://doi.org/10.1103/PhysRevD.79.024003}
  {\bibfield  {journal} {\bibinfo  {journal} {Phys. Rev. D}\ }\textbf {\bibinfo
  {volume} {79}},\ \bibinfo {pages} {024003} (\bibinfo {year} {2009})},\
  \Eprint {https://arxiv.org/abs/0810.1767} {arXiv:0810.1767 [gr-qc]}
  \BibitemShut {NoStop}%
\bibitem [{\citenamefont {Campanelli}\ \emph {et~al.}(2009)\citenamefont
  {Campanelli}, \citenamefont {Lousto},\ and\ \citenamefont
  {Zlochower}}]{0811.3006}%
  \BibitemOpen
  \bibfield  {author} {\bibinfo {author} {\bibfnamefont {M.}~\bibnamefont
  {Campanelli}}, \bibinfo {author} {\bibfnamefont {C.~O.}\ \bibnamefont
  {Lousto}},\ and\ \bibinfo {author} {\bibfnamefont {Y.}~\bibnamefont
  {Zlochower}},\ }\bibfield  {title} {\bibinfo {title} {{Algebraic
  Classification of Numerical Spacetimes and Black-Hole-Binary Remnants}},\
  }\href {https://doi.org/10.1103/PhysRevD.79.084012} {\bibfield  {journal}
  {\bibinfo  {journal} {Phys. Rev. D}\ }\textbf {\bibinfo {volume} {79}},\
  \bibinfo {pages} {084012} (\bibinfo {year} {2009})},\ \Eprint
  {https://arxiv.org/abs/0811.3006} {arXiv:0811.3006 [gr-qc]} \BibitemShut
  {NoStop}%
\bibitem [{\citenamefont {Owen}(2010)}]{1004.3768}%
  \BibitemOpen
  \bibfield  {author} {\bibinfo {author} {\bibfnamefont {R.}~\bibnamefont
  {Owen}},\ }\bibfield  {title} {\bibinfo {title} {{Degeneracy measures for the
  algebraic classification of numerical spacetimes}},\ }\href
  {https://doi.org/10.1103/PhysRevD.81.124042} {\bibfield  {journal} {\bibinfo
  {journal} {Phys. Rev. D}\ }\textbf {\bibinfo {volume} {81}},\ \bibinfo
  {pages} {124042} (\bibinfo {year} {2010})},\ \Eprint
  {https://arxiv.org/abs/1004.3768} {arXiv:1004.3768 [gr-qc]} \BibitemShut
  {NoStop}%
\bibitem [{\citenamefont {Bhagwat}\ \emph {et~al.}(2018)\citenamefont
  {Bhagwat}, \citenamefont {Okounkova}, \citenamefont {Ballmer}, \citenamefont
  {Brown}, \citenamefont {Giesler}, \citenamefont {Scheel},\ and\ \citenamefont
  {Teukolsky}}]{1711.00926}%
  \BibitemOpen
  \bibfield  {author} {\bibinfo {author} {\bibfnamefont {S.}~\bibnamefont
  {Bhagwat}}, \bibinfo {author} {\bibfnamefont {M.}~\bibnamefont {Okounkova}},
  \bibinfo {author} {\bibfnamefont {S.~W.}\ \bibnamefont {Ballmer}}, \bibinfo
  {author} {\bibfnamefont {D.~A.}\ \bibnamefont {Brown}}, \bibinfo {author}
  {\bibfnamefont {M.}~\bibnamefont {Giesler}}, \bibinfo {author} {\bibfnamefont
  {M.~A.}\ \bibnamefont {Scheel}},\ and\ \bibinfo {author} {\bibfnamefont
  {S.~A.}\ \bibnamefont {Teukolsky}},\ }\bibfield  {title} {\bibinfo {title}
  {{On choosing the start time of binary black hole ringdowns}},\ }\href
  {https://doi.org/10.1103/PhysRevD.97.104065} {\bibfield  {journal} {\bibinfo
  {journal} {Phys. Rev. D}\ }\textbf {\bibinfo {volume} {97}},\ \bibinfo
  {pages} {104065} (\bibinfo {year} {2018})},\ \Eprint
  {https://arxiv.org/abs/1711.00926} {arXiv:1711.00926 [gr-qc]} \BibitemShut
  {NoStop}%
\bibitem [{\citenamefont {Jaramillo}\ \emph {et~al.}(2012)\citenamefont
  {Jaramillo}, \citenamefont {Panosso~Macedo}, \citenamefont {Moesta},\ and\
  \citenamefont {Rezzolla}}]{1108.0060}%
  \BibitemOpen
  \bibfield  {author} {\bibinfo {author} {\bibfnamefont {J.~L.}\ \bibnamefont
  {Jaramillo}}, \bibinfo {author} {\bibfnamefont {R.}~\bibnamefont
  {Panosso~Macedo}}, \bibinfo {author} {\bibfnamefont {P.}~\bibnamefont
  {Moesta}},\ and\ \bibinfo {author} {\bibfnamefont {L.}~\bibnamefont
  {Rezzolla}},\ }\bibfield  {title} {\bibinfo {title} {{Black-hole horizons as
  probes of black-hole dynamics I: post-merger recoil in head-on collisions}},\
  }\href {https://doi.org/10.1103/PhysRevD.85.084030} {\bibfield  {journal}
  {\bibinfo  {journal} {Phys. Rev. D}\ }\textbf {\bibinfo {volume} {85}},\
  \bibinfo {pages} {084030} (\bibinfo {year} {2012})},\ \Eprint
  {https://arxiv.org/abs/1108.0060} {arXiv:1108.0060 [gr-qc]} \BibitemShut
  {NoStop}%
\bibitem [{\citenamefont {Gupta}\ \emph {et~al.}(2018)\citenamefont {Gupta},
  \citenamefont {Krishnan}, \citenamefont {Nielsen},\ and\ \citenamefont
  {Schnetter}}]{1801.07048}%
  \BibitemOpen
  \bibfield  {author} {\bibinfo {author} {\bibfnamefont {A.}~\bibnamefont
  {Gupta}}, \bibinfo {author} {\bibfnamefont {B.}~\bibnamefont {Krishnan}},
  \bibinfo {author} {\bibfnamefont {A.}~\bibnamefont {Nielsen}},\ and\ \bibinfo
  {author} {\bibfnamefont {E.}~\bibnamefont {Schnetter}},\ }\bibfield  {title}
  {\bibinfo {title} {{Dynamics of marginally trapped surfaces in a binary black
  hole merger: Growth and approach to equilibrium}},\ }\href
  {https://doi.org/10.1103/PhysRevD.97.084028} {\bibfield  {journal} {\bibinfo
  {journal} {Phys. Rev. D}\ }\textbf {\bibinfo {volume} {97}},\ \bibinfo
  {pages} {084028} (\bibinfo {year} {2018})},\ \Eprint
  {https://arxiv.org/abs/1801.07048} {arXiv:1801.07048 [gr-qc]} \BibitemShut
  {NoStop}%
\bibitem [{\citenamefont {Prasad}(2021)}]{2109.01193}%
  \BibitemOpen
  \bibfield  {author} {\bibinfo {author} {\bibfnamefont {V.}~\bibnamefont
  {Prasad}},\ }\bibfield  {title} {\bibinfo {title} {{Generalized source
  multipole moments of dynamical horizons in binary black hole mergers}},\
  }\href@noop {} {\  (\bibinfo {year} {2021})},\ \Eprint
  {https://arxiv.org/abs/2109.01193} {arXiv:2109.01193 [gr-qc]} \BibitemShut
  {NoStop}%
\bibitem [{\citenamefont {Ashtekar}\ \emph {et~al.}(2004)\citenamefont
  {Ashtekar}, \citenamefont {Engle}, \citenamefont {Pawlowski},\ and\
  \citenamefont {Van Den~Broeck}}]{gr-qc/0401114}%
  \BibitemOpen
  \bibfield  {author} {\bibinfo {author} {\bibfnamefont {A.}~\bibnamefont
  {Ashtekar}}, \bibinfo {author} {\bibfnamefont {J.}~\bibnamefont {Engle}},
  \bibinfo {author} {\bibfnamefont {T.}~\bibnamefont {Pawlowski}},\ and\
  \bibinfo {author} {\bibfnamefont {C.}~\bibnamefont {Van Den~Broeck}},\
  }\bibfield  {title} {\bibinfo {title} {{Multipole moments of isolated
  horizons}},\ }\href {https://doi.org/10.1088/0264-9381/21/11/003} {\bibfield
  {journal} {\bibinfo  {journal} {Class. Quant. Grav.}\ }\textbf {\bibinfo
  {volume} {21}},\ \bibinfo {pages} {2549} (\bibinfo {year} {2004})},\ \Eprint
  {https://arxiv.org/abs/gr-qc/0401114} {arXiv:gr-qc/0401114} \BibitemShut
  {NoStop}%
\bibitem [{\citenamefont {Schnetter}\ \emph {et~al.}(2006)\citenamefont
  {Schnetter}, \citenamefont {Krishnan},\ and\ \citenamefont
  {Beyer}}]{gr-qc/0604015}%
  \BibitemOpen
  \bibfield  {author} {\bibinfo {author} {\bibfnamefont {E.}~\bibnamefont
  {Schnetter}}, \bibinfo {author} {\bibfnamefont {B.}~\bibnamefont
  {Krishnan}},\ and\ \bibinfo {author} {\bibfnamefont {F.}~\bibnamefont
  {Beyer}},\ }\bibfield  {title} {\bibinfo {title} {{Introduction to dynamical
  horizons in numerical relativity}},\ }\href
  {https://doi.org/10.1103/PhysRevD.74.024028} {\bibfield  {journal} {\bibinfo
  {journal} {Phys. Rev. D}\ }\textbf {\bibinfo {volume} {74}},\ \bibinfo
  {pages} {024028} (\bibinfo {year} {2006})},\ \Eprint
  {https://arxiv.org/abs/gr-qc/0604015} {arXiv:gr-qc/0604015} \BibitemShut
  {NoStop}%
\bibitem [{\citenamefont {Ashtekar}\ \emph {et~al.}(2013)\citenamefont
  {Ashtekar}, \citenamefont {Campiglia},\ and\ \citenamefont
  {Shah}}]{1306.5697}%
  \BibitemOpen
  \bibfield  {author} {\bibinfo {author} {\bibfnamefont {A.}~\bibnamefont
  {Ashtekar}}, \bibinfo {author} {\bibfnamefont {M.}~\bibnamefont
  {Campiglia}},\ and\ \bibinfo {author} {\bibfnamefont {S.}~\bibnamefont
  {Shah}},\ }\bibfield  {title} {\bibinfo {title} {{Dynamical Black Holes:
  Approach to the Final State}},\ }\href
  {https://doi.org/10.1103/PhysRevD.88.064045} {\bibfield  {journal} {\bibinfo
  {journal} {Phys. Rev. D}\ }\textbf {\bibinfo {volume} {88}},\ \bibinfo
  {pages} {064045} (\bibinfo {year} {2013})},\ \Eprint
  {https://arxiv.org/abs/1306.5697} {arXiv:1306.5697 [gr-qc]} \BibitemShut
  {NoStop}%
\bibitem [{\citenamefont {Teukolsky}(1972)}]{Teukolsky1_skip}%
  \BibitemOpen
  \bibfield  {author} {\bibinfo {author} {\bibfnamefont {S.~A.}\ \bibnamefont
  {Teukolsky}},\ }\bibfield  {title} {\bibinfo {title} {{Rotating black holes -
  separable wave equations for gravitational and electromagnetic
  perturbations}},\ }\href {https://doi.org/10.1103/PhysRevLett.29.1114}
  {\bibfield  {journal} {\bibinfo  {journal} {Phys. Rev. Lett.}\ }\textbf
  {\bibinfo {volume} {29}},\ \bibinfo {pages} {1114} (\bibinfo {year}
  {1972})}\BibitemShut {NoStop}%
\bibitem [{\citenamefont {{Teukolsky}}(1973)}]{Teukolsky2_skip}%
  \BibitemOpen
  \bibfield  {author} {\bibinfo {author} {\bibfnamefont {S.~A.}\ \bibnamefont
  {{Teukolsky}}},\ }\bibfield  {title} {\bibinfo {title} {{Perturbations of a
  Rotating Black Hole. I. Fundamental Equations for Gravitational,
  Electromagnetic, and Neutrino-Field Perturbations}},\ }\href
  {https://doi.org/10.1086/152444} {\bibfield  {journal} {\bibinfo  {journal}
  {\apj}\ }\textbf {\bibinfo {volume} {185}},\ \bibinfo {pages} {635} (\bibinfo
  {year} {1973})}\BibitemShut {NoStop}%
\bibitem [{\citenamefont {{Press}}\ and\ \citenamefont
  {{Teukolsky}}(1973)}]{Press1_skip}%
  \BibitemOpen
  \bibfield  {author} {\bibinfo {author} {\bibfnamefont {W.~H.}\ \bibnamefont
  {{Press}}}\ and\ \bibinfo {author} {\bibfnamefont {S.~A.}\ \bibnamefont
  {{Teukolsky}}},\ }\bibfield  {title} {\bibinfo {title} {{Perturbations of a
  Rotating Black Hole. II. Dynamical Stability of the Kerr Metric}},\ }\href
  {https://doi.org/10.1086/152445} {\bibfield  {journal} {\bibinfo  {journal}
  {\apj}\ }\textbf {\bibinfo {volume} {185}},\ \bibinfo {pages} {649} (\bibinfo
  {year} {1973})}\BibitemShut {NoStop}%
\bibitem [{\citenamefont {Teukolsky}\ and\ \citenamefont
  {Press}(1974)}]{Teukolsky3_skip}%
  \BibitemOpen
  \bibfield  {author} {\bibinfo {author} {\bibfnamefont {S.~A.}\ \bibnamefont
  {Teukolsky}}\ and\ \bibinfo {author} {\bibfnamefont {W.~H.}\ \bibnamefont
  {Press}},\ }\bibfield  {title} {\bibinfo {title} {{Perturbations of a
  rotating black hole. III - Interaction of the hole with gravitational and
  electromagnet ic radiation}},\ }\href {https://doi.org/10.1086/153180}
  {\bibfield  {journal} {\bibinfo  {journal} {Astrophys. J.}\ }\textbf
  {\bibinfo {volume} {193}},\ \bibinfo {pages} {443} (\bibinfo {year}
  {1974})}\BibitemShut {NoStop}%
\bibitem [{\citenamefont {Buonanno}\ \emph {et~al.}(2007)\citenamefont
  {Buonanno}, \citenamefont {Cook},\ and\ \citenamefont
  {Pretorius}}]{gr-qc/0610122}%
  \BibitemOpen
  \bibfield  {author} {\bibinfo {author} {\bibfnamefont {A.}~\bibnamefont
  {Buonanno}}, \bibinfo {author} {\bibfnamefont {G.~B.}\ \bibnamefont {Cook}},\
  and\ \bibinfo {author} {\bibfnamefont {F.}~\bibnamefont {Pretorius}},\
  }\bibfield  {title} {\bibinfo {title} {{Inspiral, merger and ring-down of
  equal-mass black-hole binaries}},\ }\href
  {https://doi.org/10.1103/PhysRevD.75.124018} {\bibfield  {journal} {\bibinfo
  {journal} {Phys. Rev. D}\ }\textbf {\bibinfo {volume} {75}},\ \bibinfo
  {pages} {124018} (\bibinfo {year} {2007})},\ \Eprint
  {https://arxiv.org/abs/gr-qc/0610122} {arXiv:gr-qc/0610122} \BibitemShut
  {NoStop}%
\bibitem [{\citenamefont {Berti}\ \emph {et~al.}(2007)\citenamefont {Berti},
  \citenamefont {Cardoso}, \citenamefont {Gonzalez}, \citenamefont {Sperhake},
  \citenamefont {Hannam}, \citenamefont {Husa},\ and\ \citenamefont
  {Bruegmann}}]{gr-qc/0703053}%
  \BibitemOpen
  \bibfield  {author} {\bibinfo {author} {\bibfnamefont {E.}~\bibnamefont
  {Berti}}, \bibinfo {author} {\bibfnamefont {V.}~\bibnamefont {Cardoso}},
  \bibinfo {author} {\bibfnamefont {J.~A.}\ \bibnamefont {Gonzalez}}, \bibinfo
  {author} {\bibfnamefont {U.}~\bibnamefont {Sperhake}}, \bibinfo {author}
  {\bibfnamefont {M.}~\bibnamefont {Hannam}}, \bibinfo {author} {\bibfnamefont
  {S.}~\bibnamefont {Husa}},\ and\ \bibinfo {author} {\bibfnamefont
  {B.}~\bibnamefont {Bruegmann}},\ }\bibfield  {title} {\bibinfo {title}
  {{Inspiral, merger and ringdown of unequal mass black hole binaries: A
  Multipolar analysis}},\ }\href {https://doi.org/10.1103/PhysRevD.76.064034}
  {\bibfield  {journal} {\bibinfo  {journal} {Phys. Rev. D}\ }\textbf {\bibinfo
  {volume} {76}},\ \bibinfo {pages} {064034} (\bibinfo {year} {2007})},\
  \Eprint {https://arxiv.org/abs/gr-qc/0703053} {arXiv:gr-qc/0703053}
  \BibitemShut {NoStop}%
\bibitem [{\citenamefont {Giesler}\ \emph {et~al.}(2019)\citenamefont
  {Giesler}, \citenamefont {Isi}, \citenamefont {Scheel},\ and\ \citenamefont
  {Teukolsky}}]{1903.08284}%
  \BibitemOpen
  \bibfield  {author} {\bibinfo {author} {\bibfnamefont {M.}~\bibnamefont
  {Giesler}}, \bibinfo {author} {\bibfnamefont {M.}~\bibnamefont {Isi}},
  \bibinfo {author} {\bibfnamefont {M.~A.}\ \bibnamefont {Scheel}},\ and\
  \bibinfo {author} {\bibfnamefont {S.}~\bibnamefont {Teukolsky}},\ }\bibfield
  {title} {\bibinfo {title} {{Black Hole Ringdown: The Importance of
  Overtones}},\ }\href {https://doi.org/10.1103/PhysRevX.9.041060} {\bibfield
  {journal} {\bibinfo  {journal} {Phys. Rev. X}\ }\textbf {\bibinfo {volume}
  {9}},\ \bibinfo {pages} {041060} (\bibinfo {year} {2019})},\ \Eprint
  {https://arxiv.org/abs/1903.08284} {arXiv:1903.08284 [gr-qc]} \BibitemShut
  {NoStop}%
\bibitem [{\citenamefont {Owen}(2009)}]{0907.0280}%
  \BibitemOpen
  \bibfield  {author} {\bibinfo {author} {\bibfnamefont {R.}~\bibnamefont
  {Owen}},\ }\bibfield  {title} {\bibinfo {title} {{The Final Remnant of Binary
  Black Hole Mergers: Multipolar Analysis}},\ }\href
  {https://doi.org/10.1103/PhysRevD.80.084012} {\bibfield  {journal} {\bibinfo
  {journal} {Phys. Rev. D}\ }\textbf {\bibinfo {volume} {80}},\ \bibinfo
  {pages} {084012} (\bibinfo {year} {2009})},\ \Eprint
  {https://arxiv.org/abs/0907.0280} {arXiv:0907.0280 [gr-qc]} \BibitemShut
  {NoStop}%
\bibitem [{\citenamefont {Pook-Kolb}\ \emph
  {et~al.}(2020{\natexlab{a}})\citenamefont {Pook-Kolb}, \citenamefont
  {Birnholtz}, \citenamefont {Jaramillo}, \citenamefont {Krishnan},\ and\
  \citenamefont {Schnetter}}]{2006.03940}%
  \BibitemOpen
  \bibfield  {author} {\bibinfo {author} {\bibfnamefont {D.}~\bibnamefont
  {Pook-Kolb}}, \bibinfo {author} {\bibfnamefont {O.}~\bibnamefont
  {Birnholtz}}, \bibinfo {author} {\bibfnamefont {J.~L.}\ \bibnamefont
  {Jaramillo}}, \bibinfo {author} {\bibfnamefont {B.}~\bibnamefont
  {Krishnan}},\ and\ \bibinfo {author} {\bibfnamefont {E.}~\bibnamefont
  {Schnetter}},\ }\bibfield  {title} {\bibinfo {title} {{Horizons in a binary
  black hole merger II: Fluxes, multipole moments and stability}},\ }\href@noop
  {} {\  (\bibinfo {year} {2020}{\natexlab{a}})},\ \Eprint
  {https://arxiv.org/abs/2006.03940} {arXiv:2006.03940 [gr-qc]} \BibitemShut
  {NoStop}%
\bibitem [{\citenamefont {Mourier}\ \emph {et~al.}(2021)\citenamefont
  {Mourier}, \citenamefont {Jim\'enez~Forteza}, \citenamefont {Pook-Kolb},
  \citenamefont {Krishnan},\ and\ \citenamefont {Schnetter}}]{2010.15186}%
  \BibitemOpen
  \bibfield  {author} {\bibinfo {author} {\bibfnamefont {P.}~\bibnamefont
  {Mourier}}, \bibinfo {author} {\bibfnamefont {X.}~\bibnamefont
  {Jim\'enez~Forteza}}, \bibinfo {author} {\bibfnamefont {D.}~\bibnamefont
  {Pook-Kolb}}, \bibinfo {author} {\bibfnamefont {B.}~\bibnamefont
  {Krishnan}},\ and\ \bibinfo {author} {\bibfnamefont {E.}~\bibnamefont
  {Schnetter}},\ }\bibfield  {title} {\bibinfo {title} {{Quasinormal modes and
  their overtones at the common horizon in a binary black hole merger}},\
  }\href {https://doi.org/10.1103/PhysRevD.103.044054} {\bibfield  {journal}
  {\bibinfo  {journal} {Phys. Rev. D}\ }\textbf {\bibinfo {volume} {103}},\
  \bibinfo {pages} {044054} (\bibinfo {year} {2021})},\ \Eprint
  {https://arxiv.org/abs/2010.15186} {arXiv:2010.15186 [gr-qc]} \BibitemShut
  {NoStop}%
\bibitem [{\citenamefont {Ashtekar}\ and\ \citenamefont
  {Galloway}(2005)}]{gr-qc/0503109}%
  \BibitemOpen
  \bibfield  {author} {\bibinfo {author} {\bibfnamefont {A.}~\bibnamefont
  {Ashtekar}}\ and\ \bibinfo {author} {\bibfnamefont {G.~J.}\ \bibnamefont
  {Galloway}},\ }\bibfield  {title} {\bibinfo {title} {{Some uniqueness results
  for dynamical horizons}},\ }\href
  {https://doi.org/10.4310/ATMP.2005.v9.n1.a1} {\bibfield  {journal} {\bibinfo
  {journal} {Adv. Theor. Math. Phys.}\ }\textbf {\bibinfo {volume} {9}},\
  \bibinfo {pages} {1} (\bibinfo {year} {2005})},\ \Eprint
  {https://arxiv.org/abs/gr-qc/0503109} {arXiv:gr-qc/0503109} \BibitemShut
  {NoStop}%
\bibitem [{\citenamefont {Baumgarte}\ \emph {et~al.}(1996)\citenamefont
  {Baumgarte}, \citenamefont {Cook}, \citenamefont {Scheel}, \citenamefont
  {Shapiro},\ and\ \citenamefont {Teukolsky}}]{gr-qc/9606010}%
  \BibitemOpen
  \bibfield  {author} {\bibinfo {author} {\bibfnamefont {T.~W.}\ \bibnamefont
  {Baumgarte}}, \bibinfo {author} {\bibfnamefont {G.~B.}\ \bibnamefont {Cook}},
  \bibinfo {author} {\bibfnamefont {M.~A.}\ \bibnamefont {Scheel}}, \bibinfo
  {author} {\bibfnamefont {S.~L.}\ \bibnamefont {Shapiro}},\ and\ \bibinfo
  {author} {\bibfnamefont {S.~A.}\ \bibnamefont {Teukolsky}},\ }\bibfield
  {title} {\bibinfo {title} {{Implementing an apparent horizon finder in
  three-dimensions}},\ }\href {https://doi.org/10.1103/PhysRevD.54.4849}
  {\bibfield  {journal} {\bibinfo  {journal} {Phys. Rev. D}\ }\textbf {\bibinfo
  {volume} {54}},\ \bibinfo {pages} {4849} (\bibinfo {year} {1996})},\ \Eprint
  {https://arxiv.org/abs/gr-qc/9606010} {arXiv:gr-qc/9606010} \BibitemShut
  {NoStop}%
\bibitem [{\citenamefont {Anninos}\ \emph {et~al.}(1998)\citenamefont
  {Anninos}, \citenamefont {Camarda}, \citenamefont {Libson}, \citenamefont
  {Masso}, \citenamefont {Seidel},\ and\ \citenamefont {Suen}}]{gr-qc/9609059}%
  \BibitemOpen
  \bibfield  {author} {\bibinfo {author} {\bibfnamefont {P.}~\bibnamefont
  {Anninos}}, \bibinfo {author} {\bibfnamefont {K.}~\bibnamefont {Camarda}},
  \bibinfo {author} {\bibfnamefont {J.}~\bibnamefont {Libson}}, \bibinfo
  {author} {\bibfnamefont {J.}~\bibnamefont {Masso}}, \bibinfo {author}
  {\bibfnamefont {E.}~\bibnamefont {Seidel}},\ and\ \bibinfo {author}
  {\bibfnamefont {W.-M.}\ \bibnamefont {Suen}},\ }\bibfield  {title} {\bibinfo
  {title} {{Finding apparent horizons in dynamic 3-D numerical space-times}},\
  }\href {https://doi.org/10.1103/PhysRevD.58.024003} {\bibfield  {journal}
  {\bibinfo  {journal} {Phys. Rev. D}\ }\textbf {\bibinfo {volume} {58}},\
  \bibinfo {pages} {024003} (\bibinfo {year} {1998})},\ \Eprint
  {https://arxiv.org/abs/gr-qc/9609059} {arXiv:gr-qc/9609059} \BibitemShut
  {NoStop}%
\bibitem [{\citenamefont {Gundlach}(1998)}]{gr-qc/9707050}%
  \BibitemOpen
  \bibfield  {author} {\bibinfo {author} {\bibfnamefont {C.}~\bibnamefont
  {Gundlach}},\ }\bibfield  {title} {\bibinfo {title} {{Pseudospectral apparent
  horizon finders: An Efficient new algorithm}},\ }\href
  {https://doi.org/10.1103/PhysRevD.57.863} {\bibfield  {journal} {\bibinfo
  {journal} {Phys. Rev. D}\ }\textbf {\bibinfo {volume} {57}},\ \bibinfo
  {pages} {863} (\bibinfo {year} {1998})},\ \Eprint
  {https://arxiv.org/abs/gr-qc/9707050} {arXiv:gr-qc/9707050} \BibitemShut
  {NoStop}%
\bibitem [{\citenamefont {Shoemaker}\ \emph {et~al.}(2000)\citenamefont
  {Shoemaker}, \citenamefont {Huq},\ and\ \citenamefont
  {Matzner}}]{gr-qc/0004062}%
  \BibitemOpen
  \bibfield  {author} {\bibinfo {author} {\bibfnamefont {D.~M.}\ \bibnamefont
  {Shoemaker}}, \bibinfo {author} {\bibfnamefont {M.~F.}\ \bibnamefont {Huq}},\
  and\ \bibinfo {author} {\bibfnamefont {R.~A.}\ \bibnamefont {Matzner}},\
  }\bibfield  {title} {\bibinfo {title} {{Generic tracking of multiple apparent
  horizons with level flow}},\ }\href
  {https://doi.org/10.1103/PhysRevD.62.124005} {\bibfield  {journal} {\bibinfo
  {journal} {Phys. Rev. D}\ }\textbf {\bibinfo {volume} {62}},\ \bibinfo
  {pages} {124005} (\bibinfo {year} {2000})},\ \Eprint
  {https://arxiv.org/abs/gr-qc/0004062} {arXiv:gr-qc/0004062} \BibitemShut
  {NoStop}%
\bibitem [{\citenamefont {Thornburg}(2004)}]{gr-qc/0306056}%
  \BibitemOpen
  \bibfield  {author} {\bibinfo {author} {\bibfnamefont {J.}~\bibnamefont
  {Thornburg}},\ }\bibfield  {title} {\bibinfo {title} {{A Fast apparent
  horizon finder for three-dimensional Cartesian grids in numerical
  relativity}},\ }\href {https://doi.org/10.1088/0264-9381/21/2/026} {\bibfield
   {journal} {\bibinfo  {journal} {Class. Quant. Grav.}\ }\textbf {\bibinfo
  {volume} {21}},\ \bibinfo {pages} {743} (\bibinfo {year} {2004})},\ \Eprint
  {https://arxiv.org/abs/gr-qc/0306056} {arXiv:gr-qc/0306056} \BibitemShut
  {NoStop}%
\bibitem [{\citenamefont {Pook-Kolb}\ \emph
  {et~al.}(2020{\natexlab{b}})\citenamefont {Pook-Kolb}, \citenamefont
  {Birnholtz}, \citenamefont {Jaramillo}, \citenamefont {Krishnan},\ and\
  \citenamefont {Schnetter}}]{2006.03939}%
  \BibitemOpen
  \bibfield  {author} {\bibinfo {author} {\bibfnamefont {D.}~\bibnamefont
  {Pook-Kolb}}, \bibinfo {author} {\bibfnamefont {O.}~\bibnamefont
  {Birnholtz}}, \bibinfo {author} {\bibfnamefont {J.~L.}\ \bibnamefont
  {Jaramillo}}, \bibinfo {author} {\bibfnamefont {B.}~\bibnamefont
  {Krishnan}},\ and\ \bibinfo {author} {\bibfnamefont {E.}~\bibnamefont
  {Schnetter}},\ }\bibfield  {title} {\bibinfo {title} {{Horizons in a binary
  black hole merger I: Geometry and area increase}},\ }\href@noop {} {\
  (\bibinfo {year} {2020}{\natexlab{b}})},\ \Eprint
  {https://arxiv.org/abs/2006.03939} {arXiv:2006.03939 [gr-qc]} \BibitemShut
  {NoStop}%
\bibitem [{\citenamefont {Ashtekar}\ and\ \citenamefont
  {Krishnan}(2003)}]{gr-qc/0308033}%
  \BibitemOpen
  \bibfield  {author} {\bibinfo {author} {\bibfnamefont {A.}~\bibnamefont
  {Ashtekar}}\ and\ \bibinfo {author} {\bibfnamefont {B.}~\bibnamefont
  {Krishnan}},\ }\bibfield  {title} {\bibinfo {title} {{Dynamical horizons and
  their properties}},\ }\href {https://doi.org/10.1103/PhysRevD.68.104030}
  {\bibfield  {journal} {\bibinfo  {journal} {Phys. Rev. D}\ }\textbf {\bibinfo
  {volume} {68}},\ \bibinfo {pages} {104030} (\bibinfo {year} {2003})},\
  \Eprint {https://arxiv.org/abs/gr-qc/0308033} {arXiv:gr-qc/0308033}
  \BibitemShut {NoStop}%
\bibitem [{\citenamefont {Ashtekar}\ and\ \citenamefont
  {Krishnan}(2002)}]{gr-qc/0207080}%
  \BibitemOpen
  \bibfield  {author} {\bibinfo {author} {\bibfnamefont {A.}~\bibnamefont
  {Ashtekar}}\ and\ \bibinfo {author} {\bibfnamefont {B.}~\bibnamefont
  {Krishnan}},\ }\bibfield  {title} {\bibinfo {title} {{Dynamical horizons:
  Energy, angular momentum, fluxes and balance laws}},\ }\href
  {https://doi.org/10.1103/PhysRevLett.89.261101} {\bibfield  {journal}
  {\bibinfo  {journal} {Phys. Rev. Lett.}\ }\textbf {\bibinfo {volume} {89}},\
  \bibinfo {pages} {261101} (\bibinfo {year} {2002})},\ \Eprint
  {https://arxiv.org/abs/gr-qc/0207080} {arXiv:gr-qc/0207080} \BibitemShut
  {NoStop}%
\bibitem [{\citenamefont {Ashtekar}\ and\ \citenamefont
  {Krishnan}(2004)}]{gr-qc/0407042}%
  \BibitemOpen
  \bibfield  {author} {\bibinfo {author} {\bibfnamefont {A.}~\bibnamefont
  {Ashtekar}}\ and\ \bibinfo {author} {\bibfnamefont {B.}~\bibnamefont
  {Krishnan}},\ }\bibfield  {title} {\bibinfo {title} {{Isolated and dynamical
  horizons and their applications}},\ }\href
  {https://doi.org/10.12942/lrr-2004-10} {\bibfield  {journal} {\bibinfo
  {journal} {Living Rev. Rel.}\ }\textbf {\bibinfo {volume} {7}},\ \bibinfo
  {pages} {10} (\bibinfo {year} {2004})},\ \Eprint
  {https://arxiv.org/abs/gr-qc/0407042} {arXiv:gr-qc/0407042} \BibitemShut
  {NoStop}%
\bibitem [{\citenamefont {Ashtekar}\ \emph
  {et~al.}(2000{\natexlab{a}})\citenamefont {Ashtekar}, \citenamefont
  {Fairhurst},\ and\ \citenamefont {Krishnan}}]{gr-qc/0005083}%
  \BibitemOpen
  \bibfield  {author} {\bibinfo {author} {\bibfnamefont {A.}~\bibnamefont
  {Ashtekar}}, \bibinfo {author} {\bibfnamefont {S.}~\bibnamefont
  {Fairhurst}},\ and\ \bibinfo {author} {\bibfnamefont {B.}~\bibnamefont
  {Krishnan}},\ }\bibfield  {title} {\bibinfo {title} {{Isolated horizons:
  Hamiltonian evolution and the first law}},\ }\href
  {https://doi.org/10.1103/PhysRevD.62.104025} {\bibfield  {journal} {\bibinfo
  {journal} {Phys. Rev. D}\ }\textbf {\bibinfo {volume} {62}},\ \bibinfo
  {pages} {104025} (\bibinfo {year} {2000}{\natexlab{a}})},\ \Eprint
  {https://arxiv.org/abs/gr-qc/0005083} {arXiv:gr-qc/0005083} \BibitemShut
  {NoStop}%
\bibitem [{\citenamefont {Ashtekar}\ \emph
  {et~al.}(2000{\natexlab{b}})\citenamefont {Ashtekar}, \citenamefont {Beetle},
  \citenamefont {Dreyer}, \citenamefont {Fairhurst}, \citenamefont {Krishnan},
  \citenamefont {Lewandowski},\ and\ \citenamefont
  {Wisniewski}}]{gr-qc/0006006}%
  \BibitemOpen
  \bibfield  {author} {\bibinfo {author} {\bibfnamefont {A.}~\bibnamefont
  {Ashtekar}}, \bibinfo {author} {\bibfnamefont {C.}~\bibnamefont {Beetle}},
  \bibinfo {author} {\bibfnamefont {O.}~\bibnamefont {Dreyer}}, \bibinfo
  {author} {\bibfnamefont {S.}~\bibnamefont {Fairhurst}}, \bibinfo {author}
  {\bibfnamefont {B.}~\bibnamefont {Krishnan}}, \bibinfo {author}
  {\bibfnamefont {J.}~\bibnamefont {Lewandowski}},\ and\ \bibinfo {author}
  {\bibfnamefont {J.}~\bibnamefont {Wisniewski}},\ }\bibfield  {title}
  {\bibinfo {title} {{Isolated horizons and their applications}},\ }\href
  {https://doi.org/10.1103/PhysRevLett.85.3564} {\bibfield  {journal} {\bibinfo
   {journal} {Phys. Rev. Lett.}\ }\textbf {\bibinfo {volume} {85}},\ \bibinfo
  {pages} {3564} (\bibinfo {year} {2000}{\natexlab{b}})},\ \Eprint
  {https://arxiv.org/abs/gr-qc/0006006} {arXiv:gr-qc/0006006} \BibitemShut
  {NoStop}%
\bibitem [{\citenamefont {Ashtekar}\ \emph {et~al.}(2002)\citenamefont
  {Ashtekar}, \citenamefont {Beetle},\ and\ \citenamefont
  {Lewandowski}}]{gr-qc/0111067}%
  \BibitemOpen
  \bibfield  {author} {\bibinfo {author} {\bibfnamefont {A.}~\bibnamefont
  {Ashtekar}}, \bibinfo {author} {\bibfnamefont {C.}~\bibnamefont {Beetle}},\
  and\ \bibinfo {author} {\bibfnamefont {J.}~\bibnamefont {Lewandowski}},\
  }\bibfield  {title} {\bibinfo {title} {{Geometry of generic isolated
  horizons}},\ }\href {https://doi.org/10.1088/0264-9381/19/6/311} {\bibfield
  {journal} {\bibinfo  {journal} {Class. Quant. Grav.}\ }\textbf {\bibinfo
  {volume} {19}},\ \bibinfo {pages} {1195} (\bibinfo {year} {2002})},\ \Eprint
  {https://arxiv.org/abs/gr-qc/0111067} {arXiv:gr-qc/0111067} \BibitemShut
  {NoStop}%
\bibitem [{\citenamefont {Courant}\ and\ \citenamefont
  {Hilbert}(1989)}]{Courant1_skip}%
  \BibitemOpen
  \bibfield  {author} {\bibinfo {author} {\bibfnamefont {R.}~\bibnamefont
  {Courant}}\ and\ \bibinfo {author} {\bibfnamefont {D.}~\bibnamefont
  {Hilbert}},\ }\href {https://doi.org/10.1002/9783527617210} {\emph {\bibinfo
  {title} {Methods of Mathematical Physics}}}\ (\bibinfo  {publisher} {John
  Wiley \& Sons, Ltd},\ \bibinfo {year} {1989})\BibitemShut {NoStop}%
\bibitem [{\citenamefont {Baumgarte}\ and\ \citenamefont
  {Shapiro}(2010)}]{Baumgarte1_skip}%
  \BibitemOpen
  \bibfield  {author} {\bibinfo {author} {\bibfnamefont {T.~W.}\ \bibnamefont
  {Baumgarte}}\ and\ \bibinfo {author} {\bibfnamefont {S.~L.}\ \bibnamefont
  {Shapiro}},\ }\href {https://doi.org/10.1017/CBO9781139193344} {\emph
  {\bibinfo {title} {Numerical Relativity: Solving Einstein's Equations on the
  Computer}}}\ (\bibinfo  {publisher} {Cambridge University Press},\ \bibinfo
  {year} {2010})\BibitemShut {NoStop}%
\bibitem [{\citenamefont {Penrose}\ and\ \citenamefont
  {Rindler}(1984)}]{Penrose1_skip}%
  \BibitemOpen
  \bibfield  {author} {\bibinfo {author} {\bibfnamefont {R.}~\bibnamefont
  {Penrose}}\ and\ \bibinfo {author} {\bibfnamefont {W.}~\bibnamefont
  {Rindler}},\ }\href {https://doi.org/10.1017/CBO9780511564048} {\emph
  {\bibinfo {title} {Spinors and Space-Time}}}\ (\bibinfo  {publisher}
  {Cambridge University Press},\ \bibinfo {year} {1984})\BibitemShut {NoStop}%
\bibitem [{\citenamefont {Owen}\ \emph {et~al.}(2011)\citenamefont {Owen} \emph
  {et~al.}}]{1012.4869}%
  \BibitemOpen
  \bibfield  {author} {\bibinfo {author} {\bibfnamefont {R.}~\bibnamefont
  {Owen}} \emph {et~al.},\ }\bibfield  {title} {\bibinfo {title}
  {{Frame-Dragging Vortexes and Tidal Tendexes Attached to Colliding Black
  Holes: Visualizing the Curvature of Spacetime}},\ }\href
  {https://doi.org/10.1103/PhysRevLett.106.151101} {\bibfield  {journal}
  {\bibinfo  {journal} {Phys. Rev. Lett.}\ }\textbf {\bibinfo {volume} {106}},\
  \bibinfo {pages} {151101} (\bibinfo {year} {2011})},\ \Eprint
  {https://arxiv.org/abs/1012.4869} {arXiv:1012.4869 [gr-qc]} \BibitemShut
  {NoStop}%
\bibitem [{\citenamefont {Wald}(1984)}]{Wald1_skip}%
  \BibitemOpen
  \bibfield  {author} {\bibinfo {author} {\bibfnamefont {R.~M.}\ \bibnamefont
  {Wald}},\ }\href {https://doi.org/10.7208/chicago/9780226870373.001.0001}
  {\emph {\bibinfo {title} {{General Relativity}}}}\ (\bibinfo  {publisher}
  {Chicago Univ. Pr.},\ \bibinfo {address} {Chicago, USA},\ \bibinfo {year}
  {1984})\BibitemShut {NoStop}%
\bibitem [{\citenamefont {Hawking}\ and\ \citenamefont
  {Hartle}(1972)}]{Hawking1_skip}%
  \BibitemOpen
  \bibfield  {author} {\bibinfo {author} {\bibfnamefont {S.~W.}\ \bibnamefont
  {Hawking}}\ and\ \bibinfo {author} {\bibfnamefont {J.~B.}\ \bibnamefont
  {Hartle}},\ }\bibfield  {title} {\bibinfo {title} {{Energy and angular
  momentum flow into a black hole}},\ }\href
  {https://doi.org/10.1007/BF01645515} {\bibfield  {journal} {\bibinfo
  {journal} {Commun. Math. Phys.}\ }\textbf {\bibinfo {volume} {27}},\ \bibinfo
  {pages} {283} (\bibinfo {year} {1972})}\BibitemShut {NoStop}%
\bibitem [{\citenamefont {Kinnersley}(1969)}]{Kinnersley1_skip}%
  \BibitemOpen
  \bibfield  {author} {\bibinfo {author} {\bibfnamefont {W.}~\bibnamefont
  {Kinnersley}},\ }\bibfield  {title} {\bibinfo {title} {{Type D Vacuum
  Metrics}},\ }\href {https://doi.org/10.1063/1.1664958} {\bibfield  {journal}
  {\bibinfo  {journal} {J. Math. Phys.}\ }\textbf {\bibinfo {volume} {10}},\
  \bibinfo {pages} {1195} (\bibinfo {year} {1969})}\BibitemShut {NoStop}%
\bibitem [{\citenamefont {Leaver}(1985)}]{Leaver1_skip}%
  \BibitemOpen
  \bibfield  {author} {\bibinfo {author} {\bibfnamefont {E.~W.}\ \bibnamefont
  {Leaver}},\ }\bibfield  {title} {\bibinfo {title} {An analytic representation
  for the quasi-normal modes of kerr black holes},\ }\href
  {https://doi.org/10.1098/rspa.1985.0119} {\bibfield  {journal} {\bibinfo
  {journal} {Proc. R. Soc. Lond. A}\ }\textbf {\bibinfo {volume} {402}},\
  \bibinfo {pages} {285–298} (\bibinfo {year} {1985})}\BibitemShut {NoStop}%
\bibitem [{\citenamefont {Berti}\ \emph {et~al.}(2009)\citenamefont {Berti},
  \citenamefont {Cardoso},\ and\ \citenamefont {Starinets}}]{0905.2975}%
  \BibitemOpen
  \bibfield  {author} {\bibinfo {author} {\bibfnamefont {E.}~\bibnamefont
  {Berti}}, \bibinfo {author} {\bibfnamefont {V.}~\bibnamefont {Cardoso}},\
  and\ \bibinfo {author} {\bibfnamefont {A.~O.}\ \bibnamefont {Starinets}},\
  }\bibfield  {title} {\bibinfo {title} {{Quasinormal modes of black holes and
  black branes}},\ }\href {https://doi.org/10.1088/0264-9381/26/16/163001}
  {\bibfield  {journal} {\bibinfo  {journal} {Class. Quant. Grav.}\ }\textbf
  {\bibinfo {volume} {26}},\ \bibinfo {pages} {163001} (\bibinfo {year}
  {2009})},\ \Eprint {https://arxiv.org/abs/0905.2975} {arXiv:0905.2975
  [gr-qc]} \BibitemShut {NoStop}%
\bibitem [{\citenamefont {Isi}\ and\ \citenamefont {Farr}(2021)}]{2107.05609}%
  \BibitemOpen
  \bibfield  {author} {\bibinfo {author} {\bibfnamefont {M.}~\bibnamefont
  {Isi}}\ and\ \bibinfo {author} {\bibfnamefont {W.~M.}\ \bibnamefont {Farr}},\
  }\bibfield  {title} {\bibinfo {title} {{Analyzing black-hole ringdowns}},\
  }\href@noop {} {\  (\bibinfo {year} {2021})},\ \Eprint
  {https://arxiv.org/abs/2107.05609} {arXiv:2107.05609 [gr-qc]} \BibitemShut
  {NoStop}%
\bibitem [{\citenamefont {Whiting}(1989)}]{Whiting1_skip}%
  \BibitemOpen
  \bibfield  {author} {\bibinfo {author} {\bibfnamefont {B.~F.}\ \bibnamefont
  {Whiting}},\ }\bibfield  {title} {\bibinfo {title} {Mode stability of the
  kerr black hole},\ }\href {https://doi.org/10.1063/1.528308} {\bibfield
  {journal} {\bibinfo  {journal} {J. Math. Phys.}\ }\textbf {\bibinfo {volume}
  {30}},\ \bibinfo {pages} {1301} (\bibinfo {year} {1989})}\BibitemShut
  {NoStop}%
\bibitem [{\citenamefont {Shlapentokh-Rothman}(2015)}]{1302.6902}%
  \BibitemOpen
  \bibfield  {author} {\bibinfo {author} {\bibfnamefont {Y.}~\bibnamefont
  {Shlapentokh-Rothman}},\ }\bibfield  {title} {\bibinfo {title} {{Quantitative
  Mode Stability for the Wave Equation on the Kerr Spacetime}},\ }\href
  {https://doi.org/10.1007/s00023-014-0315-7} {\bibfield  {journal} {\bibinfo
  {journal} {Annales Henri Poincare}\ }\textbf {\bibinfo {volume} {16}},\
  \bibinfo {pages} {289} (\bibinfo {year} {2015})},\ \Eprint
  {https://arxiv.org/abs/1302.6902} {arXiv:1302.6902 [gr-qc]} \BibitemShut
  {NoStop}%
\bibitem [{\citenamefont {Teixeira~da Costa}(2020)}]{1910.02854}%
  \BibitemOpen
  \bibfield  {author} {\bibinfo {author} {\bibfnamefont {R.}~\bibnamefont
  {Teixeira~da Costa}},\ }\bibfield  {title} {\bibinfo {title} {{Mode stability
  for the Teukolsky equation on extremal and subextremal Kerr spacetimes}},\
  }\href {https://doi.org/10.1007/s00220-020-03796-z} {\bibfield  {journal}
  {\bibinfo  {journal} {Commun. Math. Phys.}\ }\textbf {\bibinfo {volume}
  {378}},\ \bibinfo {pages} {705} (\bibinfo {year} {2020})},\ \Eprint
  {https://arxiv.org/abs/1910.02854} {arXiv:1910.02854 [gr-qc]} \BibitemShut
  {NoStop}%
\bibitem [{\citenamefont {Casals}\ and\ \citenamefont
  {da~Costa}(2021)}]{2105.13329}%
  \BibitemOpen
  \bibfield  {author} {\bibinfo {author} {\bibfnamefont {M.}~\bibnamefont
  {Casals}}\ and\ \bibinfo {author} {\bibfnamefont {R.~T.}\ \bibnamefont
  {da~Costa}},\ }\bibfield  {title} {\bibinfo {title} {{Hidden spectral
  symmetries and mode stability of subextremal Kerr(-dS) black holes}},\
  }\href@noop {} {\  (\bibinfo {year} {2021})},\ \Eprint
  {https://arxiv.org/abs/2105.13329} {arXiv:2105.13329 [gr-qc]} \BibitemShut
  {NoStop}%
\bibitem [{\citenamefont {Stein}(2019)}]{1908.10377}%
  \BibitemOpen
  \bibfield  {author} {\bibinfo {author} {\bibfnamefont {L.~C.}\ \bibnamefont
  {Stein}},\ }\bibfield  {title} {\bibinfo {title} {{qnm: A Python package for
  calculating Kerr quasinormal modes, separation constants, and
  spherical-spheroidal mixing coefficients}},\ }\href
  {https://doi.org/10.21105/joss.01683} {\bibfield  {journal} {\bibinfo
  {journal} {J. Open Source Softw.}\ }\textbf {\bibinfo {volume} {4}},\
  \bibinfo {pages} {1683} (\bibinfo {year} {2019})},\ \Eprint
  {https://arxiv.org/abs/1908.10377} {arXiv:1908.10377 [gr-qc]} \BibitemShut
  {NoStop}%
\bibitem [{\citenamefont {Goldberg}\ \emph {et~al.}(1967)\citenamefont
  {Goldberg}, \citenamefont {MacFarlane}, \citenamefont {Newman}, \citenamefont
  {Rohrlich},\ and\ \citenamefont {Sudarshan}}]{Goldberg1_skip}%
  \BibitemOpen
  \bibfield  {author} {\bibinfo {author} {\bibfnamefont {J.~N.}\ \bibnamefont
  {Goldberg}}, \bibinfo {author} {\bibfnamefont {A.~J.}\ \bibnamefont
  {MacFarlane}}, \bibinfo {author} {\bibfnamefont {E.~T.}\ \bibnamefont
  {Newman}}, \bibinfo {author} {\bibfnamefont {F.}~\bibnamefont {Rohrlich}},\
  and\ \bibinfo {author} {\bibfnamefont {E.~C.~G.}\ \bibnamefont {Sudarshan}},\
  }\bibfield  {title} {\bibinfo {title} {{Spin s spherical harmonics and
  edth}},\ }\href {https://doi.org/10.1063/1.1705135} {\bibfield  {journal}
  {\bibinfo  {journal} {J. Math. Phys.}\ }\textbf {\bibinfo {volume} {8}},\
  \bibinfo {pages} {2155} (\bibinfo {year} {1967})}\BibitemShut {NoStop}%
\bibitem [{\citenamefont {Hartle}\ and\ \citenamefont
  {Wilkins}(1974)}]{Hartle2_skip}%
  \BibitemOpen
  \bibfield  {author} {\bibinfo {author} {\bibfnamefont {J.~B.}\ \bibnamefont
  {Hartle}}\ and\ \bibinfo {author} {\bibfnamefont {D.~C.}\ \bibnamefont
  {Wilkins}},\ }\bibfield  {title} {\bibinfo {title} {{Analytic properties of
  the Teukolsky equation}},\ }\href {https://doi.org/10.1007/BF01651548}
  {\bibfield  {journal} {\bibinfo  {journal} {Commun. Math. Phys.}\ }\textbf
  {\bibinfo {volume} {38}},\ \bibinfo {pages} {47–63} (\bibinfo {year}
  {1974})}\BibitemShut {NoStop}%
\bibitem [{\citenamefont {Breuer}\ \emph {et~al.}(1977)\citenamefont {Breuer},
  \citenamefont {Ryan},\ and\ \citenamefont {Waller}}]{Breuer1_skip}%
  \BibitemOpen
  \bibfield  {author} {\bibinfo {author} {\bibfnamefont {R.~A.}\ \bibnamefont
  {Breuer}}, \bibinfo {author} {\bibfnamefont {J.}~\bibnamefont {Ryan},
  \bibfnamefont {M.~P.}},\ and\ \bibinfo {author} {\bibfnamefont
  {S.}~\bibnamefont {Waller}},\ }\bibfield  {title} {\bibinfo {title} {{Some
  Properties of Spin-Weighted Spheroidal Harmonics}},\ }\href
  {https://doi.org/10.1098/rspa.1977.0187} {\bibfield  {journal} {\bibinfo
  {journal} {Proc. R. Soc. Lond. A.}\ }\textbf {\bibinfo {volume} {358}},\
  \bibinfo {pages} {71} (\bibinfo {year} {1977})}\BibitemShut {NoStop}%
\bibitem [{\citenamefont {Bhagwat}\ \emph {et~al.}(2020)\citenamefont
  {Bhagwat}, \citenamefont {Forteza}, \citenamefont {Pani},\ and\ \citenamefont
  {Ferrari}}]{1910.08708}%
  \BibitemOpen
  \bibfield  {author} {\bibinfo {author} {\bibfnamefont {S.}~\bibnamefont
  {Bhagwat}}, \bibinfo {author} {\bibfnamefont {X.~J.}\ \bibnamefont
  {Forteza}}, \bibinfo {author} {\bibfnamefont {P.}~\bibnamefont {Pani}},\ and\
  \bibinfo {author} {\bibfnamefont {V.}~\bibnamefont {Ferrari}},\ }\bibfield
  {title} {\bibinfo {title} {{Ringdown overtones, black hole spectroscopy, and
  no-hair theorem tests}},\ }\href
  {https://doi.org/10.1103/PhysRevD.101.044033} {\bibfield  {journal} {\bibinfo
   {journal} {Phys. Rev. D}\ }\textbf {\bibinfo {volume} {101}},\ \bibinfo
  {pages} {044033} (\bibinfo {year} {2020})},\ \Eprint
  {https://arxiv.org/abs/1910.08708} {arXiv:1910.08708 [gr-qc]} \BibitemShut
  {NoStop}%
\bibitem [{\citenamefont {Dhani}(2021)}]{2010.08602}%
  \BibitemOpen
  \bibfield  {author} {\bibinfo {author} {\bibfnamefont {A.}~\bibnamefont
  {Dhani}},\ }\bibfield  {title} {\bibinfo {title} {{Importance of mirror modes
  in binary black hole ringdown waveform}},\ }\href
  {https://doi.org/10.1103/PhysRevD.103.104048} {\bibfield  {journal} {\bibinfo
   {journal} {Phys. Rev. D}\ }\textbf {\bibinfo {volume} {103}},\ \bibinfo
  {pages} {104048} (\bibinfo {year} {2021})},\ \Eprint
  {https://arxiv.org/abs/2010.08602} {arXiv:2010.08602 [gr-qc]} \BibitemShut
  {NoStop}%
\bibitem [{spe()}]{spec_skip}%
  \BibitemOpen
  \href@noop {} {}\bibinfo {howpublished}
  {\url{http://www.black-holes.org/SpEC.html}}\BibitemShut {NoStop}%
\bibitem [{\citenamefont {Lindblom}\ \emph {et~al.}(2006)\citenamefont
  {Lindblom}, \citenamefont {Scheel}, \citenamefont {Kidder}, \citenamefont
  {Owen},\ and\ \citenamefont {Rinne}}]{gr-qc/0512093}%
  \BibitemOpen
  \bibfield  {author} {\bibinfo {author} {\bibfnamefont {L.}~\bibnamefont
  {Lindblom}}, \bibinfo {author} {\bibfnamefont {M.~A.}\ \bibnamefont
  {Scheel}}, \bibinfo {author} {\bibfnamefont {L.~E.}\ \bibnamefont {Kidder}},
  \bibinfo {author} {\bibfnamefont {R.}~\bibnamefont {Owen}},\ and\ \bibinfo
  {author} {\bibfnamefont {O.}~\bibnamefont {Rinne}},\ }\bibfield  {title}
  {\bibinfo {title} {{A New generalized harmonic evolution system}},\ }\href
  {https://doi.org/10.1088/0264-9381/23/16/S09} {\bibfield  {journal} {\bibinfo
   {journal} {Class. Quant. Grav.}\ }\textbf {\bibinfo {volume} {23}},\
  \bibinfo {pages} {S447} (\bibinfo {year} {2006})},\ \Eprint
  {https://arxiv.org/abs/gr-qc/0512093} {arXiv:gr-qc/0512093} \BibitemShut
  {NoStop}%
\bibitem [{\citenamefont {Lovelace}\ \emph {et~al.}(2008)\citenamefont
  {Lovelace}, \citenamefont {Owen}, \citenamefont {Pfeiffer},\ and\
  \citenamefont {Chu}}]{0805.4192}%
  \BibitemOpen
  \bibfield  {author} {\bibinfo {author} {\bibfnamefont {G.}~\bibnamefont
  {Lovelace}}, \bibinfo {author} {\bibfnamefont {R.}~\bibnamefont {Owen}},
  \bibinfo {author} {\bibfnamefont {H.~P.}\ \bibnamefont {Pfeiffer}},\ and\
  \bibinfo {author} {\bibfnamefont {T.}~\bibnamefont {Chu}},\ }\bibfield
  {title} {\bibinfo {title} {{Binary-black-hole initial data with
  nearly-extremal spins}},\ }\href {https://doi.org/10.1103/PhysRevD.78.084017}
  {\bibfield  {journal} {\bibinfo  {journal} {Phys. Rev. D}\ }\textbf {\bibinfo
  {volume} {78}},\ \bibinfo {pages} {084017} (\bibinfo {year} {2008})},\
  \Eprint {https://arxiv.org/abs/0805.4192} {arXiv:0805.4192 [gr-qc]}
  \BibitemShut {NoStop}%
\bibitem [{\citenamefont {Szilagyi}\ \emph {et~al.}(2009)\citenamefont
  {Szilagyi}, \citenamefont {Lindblom},\ and\ \citenamefont
  {Scheel}}]{0909.3557}%
  \BibitemOpen
  \bibfield  {author} {\bibinfo {author} {\bibfnamefont {B.}~\bibnamefont
  {Szilagyi}}, \bibinfo {author} {\bibfnamefont {L.}~\bibnamefont {Lindblom}},\
  and\ \bibinfo {author} {\bibfnamefont {M.~A.}\ \bibnamefont {Scheel}},\
  }\bibfield  {title} {\bibinfo {title} {{Simulations of Binary Black Hole
  Mergers Using Spectral Methods}},\ }\href
  {https://doi.org/10.1103/PhysRevD.80.124010} {\bibfield  {journal} {\bibinfo
  {journal} {Phys. Rev. D}\ }\textbf {\bibinfo {volume} {80}},\ \bibinfo
  {pages} {124010} (\bibinfo {year} {2009})},\ \Eprint
  {https://arxiv.org/abs/0909.3557} {arXiv:0909.3557 [gr-qc]} \BibitemShut
  {NoStop}%
\bibitem [{\citenamefont {Pretorius}(2005)}]{gr-qc/0407110}%
  \BibitemOpen
  \bibfield  {author} {\bibinfo {author} {\bibfnamefont {F.}~\bibnamefont
  {Pretorius}},\ }\bibfield  {title} {\bibinfo {title} {{Numerical relativity
  using a generalized harmonic decomposition}},\ }\href
  {https://doi.org/10.1088/0264-9381/22/2/014} {\bibfield  {journal} {\bibinfo
  {journal} {Class. Quant. Grav.}\ }\textbf {\bibinfo {volume} {22}},\ \bibinfo
  {pages} {425} (\bibinfo {year} {2005})},\ \Eprint
  {https://arxiv.org/abs/gr-qc/0407110} {arXiv:gr-qc/0407110} \BibitemShut
  {NoStop}%
\bibitem [{\citenamefont {Hemberger}\ \emph {et~al.}(2013)\citenamefont
  {Hemberger}, \citenamefont {Scheel}, \citenamefont {Kidder}, \citenamefont
  {Szil\'agyi}, \citenamefont {Lovelace}, \citenamefont {Taylor},\ and\
  \citenamefont {Teukolsky}}]{1211.6079}%
  \BibitemOpen
  \bibfield  {author} {\bibinfo {author} {\bibfnamefont {D.~A.}\ \bibnamefont
  {Hemberger}}, \bibinfo {author} {\bibfnamefont {M.~A.}\ \bibnamefont
  {Scheel}}, \bibinfo {author} {\bibfnamefont {L.~E.}\ \bibnamefont {Kidder}},
  \bibinfo {author} {\bibfnamefont {B.}~\bibnamefont {Szil\'agyi}}, \bibinfo
  {author} {\bibfnamefont {G.}~\bibnamefont {Lovelace}}, \bibinfo {author}
  {\bibfnamefont {N.~W.}\ \bibnamefont {Taylor}},\ and\ \bibinfo {author}
  {\bibfnamefont {S.~A.}\ \bibnamefont {Teukolsky}},\ }\bibfield  {title}
  {\bibinfo {title} {{Dynamical Excision Boundaries in Spectral Evolutions of
  Binary Black Hole Spacetimes}},\ }\href
  {https://doi.org/10.1088/0264-9381/30/11/115001} {\bibfield  {journal}
  {\bibinfo  {journal} {Class. Quant. Grav.}\ }\textbf {\bibinfo {volume}
  {30}},\ \bibinfo {pages} {115001} (\bibinfo {year} {2013})},\ \Eprint
  {https://arxiv.org/abs/1211.6079} {arXiv:1211.6079 [gr-qc]} \BibitemShut
  {NoStop}%
\bibitem [{\citenamefont {Scheel}\ \emph {et~al.}(2015)\citenamefont {Scheel},
  \citenamefont {Giesler}, \citenamefont {Hemberger}, \citenamefont {Lovelace},
  \citenamefont {Kuper}, \citenamefont {Boyle}, \citenamefont {Szil\'agyi},\
  and\ \citenamefont {Kidder}}]{1412.1803}%
  \BibitemOpen
  \bibfield  {author} {\bibinfo {author} {\bibfnamefont {M.~A.}\ \bibnamefont
  {Scheel}}, \bibinfo {author} {\bibfnamefont {M.}~\bibnamefont {Giesler}},
  \bibinfo {author} {\bibfnamefont {D.~A.}\ \bibnamefont {Hemberger}}, \bibinfo
  {author} {\bibfnamefont {G.}~\bibnamefont {Lovelace}}, \bibinfo {author}
  {\bibfnamefont {K.}~\bibnamefont {Kuper}}, \bibinfo {author} {\bibfnamefont
  {M.}~\bibnamefont {Boyle}}, \bibinfo {author} {\bibfnamefont
  {B.}~\bibnamefont {Szil\'agyi}},\ and\ \bibinfo {author} {\bibfnamefont
  {L.~E.}\ \bibnamefont {Kidder}},\ }\bibfield  {title} {\bibinfo {title}
  {{Improved methods for simulating nearly extremal binary black holes}},\
  }\href {https://doi.org/10.1088/0264-9381/32/10/105009} {\bibfield  {journal}
  {\bibinfo  {journal} {Class. Quant. Grav.}\ }\textbf {\bibinfo {volume}
  {32}},\ \bibinfo {pages} {105009} (\bibinfo {year} {2015})},\ \Eprint
  {https://arxiv.org/abs/1412.1803} {arXiv:1412.1803 [gr-qc]} \BibitemShut
  {NoStop}%
\bibitem [{\citenamefont {Rinne}\ \emph {et~al.}(2007)\citenamefont {Rinne},
  \citenamefont {Lindblom},\ and\ \citenamefont {Scheel}}]{0704.0782}%
  \BibitemOpen
  \bibfield  {author} {\bibinfo {author} {\bibfnamefont {O.}~\bibnamefont
  {Rinne}}, \bibinfo {author} {\bibfnamefont {L.}~\bibnamefont {Lindblom}},\
  and\ \bibinfo {author} {\bibfnamefont {M.~A.}\ \bibnamefont {Scheel}},\
  }\bibfield  {title} {\bibinfo {title} {{Testing outer boundary treatments for
  the Einstein equations}},\ }\href
  {https://doi.org/10.1088/0264-9381/24/16/006} {\bibfield  {journal} {\bibinfo
   {journal} {Class. Quant. Grav.}\ }\textbf {\bibinfo {volume} {24}},\
  \bibinfo {pages} {4053} (\bibinfo {year} {2007})},\ \Eprint
  {https://arxiv.org/abs/0704.0782} {arXiv:0704.0782 [gr-qc]} \BibitemShut
  {NoStop}%
\bibitem [{\citenamefont {Scheel}\ \emph {et~al.}(2006)\citenamefont {Scheel},
  \citenamefont {Pfeiffer}, \citenamefont {Lindblom}, \citenamefont {Kidder},
  \citenamefont {Rinne},\ and\ \citenamefont {Teukolsky}}]{gr-qc/0607056}%
  \BibitemOpen
  \bibfield  {author} {\bibinfo {author} {\bibfnamefont {M.~A.}\ \bibnamefont
  {Scheel}}, \bibinfo {author} {\bibfnamefont {H.~P.}\ \bibnamefont
  {Pfeiffer}}, \bibinfo {author} {\bibfnamefont {L.}~\bibnamefont {Lindblom}},
  \bibinfo {author} {\bibfnamefont {L.~E.}\ \bibnamefont {Kidder}}, \bibinfo
  {author} {\bibfnamefont {O.}~\bibnamefont {Rinne}},\ and\ \bibinfo {author}
  {\bibfnamefont {S.~A.}\ \bibnamefont {Teukolsky}},\ }\bibfield  {title}
  {\bibinfo {title} {{Solving Einstein's equations with dual coordinate
  frames}},\ }\href {https://doi.org/10.1103/PhysRevD.74.104006} {\bibfield
  {journal} {\bibinfo  {journal} {Phys. Rev. D}\ }\textbf {\bibinfo {volume}
  {74}},\ \bibinfo {pages} {104006} (\bibinfo {year} {2006})},\ \Eprint
  {https://arxiv.org/abs/gr-qc/0607056} {arXiv:gr-qc/0607056} \BibitemShut
  {NoStop}%
\bibitem [{\citenamefont {Buchman}\ \emph {et~al.}(2012)\citenamefont
  {Buchman}, \citenamefont {Pfeiffer}, \citenamefont {Scheel},\ and\
  \citenamefont {Szilagyi}}]{1206.3015}%
  \BibitemOpen
  \bibfield  {author} {\bibinfo {author} {\bibfnamefont {L.~T.}\ \bibnamefont
  {Buchman}}, \bibinfo {author} {\bibfnamefont {H.~P.}\ \bibnamefont
  {Pfeiffer}}, \bibinfo {author} {\bibfnamefont {M.~A.}\ \bibnamefont
  {Scheel}},\ and\ \bibinfo {author} {\bibfnamefont {B.}~\bibnamefont
  {Szilagyi}},\ }\bibfield  {title} {\bibinfo {title} {{Simulations of
  non-equal mass black hole binaries with spectral methods}},\ }\href
  {https://doi.org/10.1103/PhysRevD.86.084033} {\bibfield  {journal} {\bibinfo
  {journal} {Phys. Rev. D}\ }\textbf {\bibinfo {volume} {86}},\ \bibinfo
  {pages} {084033} (\bibinfo {year} {2012})},\ \Eprint
  {https://arxiv.org/abs/1206.3015} {arXiv:1206.3015 [gr-qc]} \BibitemShut
  {NoStop}%
\bibitem [{\citenamefont {Lovelace}\ \emph {et~al.}(2011)\citenamefont
  {Lovelace}, \citenamefont {Scheel},\ and\ \citenamefont
  {Szilagyi}}]{1010.2777}%
  \BibitemOpen
  \bibfield  {author} {\bibinfo {author} {\bibfnamefont {G.}~\bibnamefont
  {Lovelace}}, \bibinfo {author} {\bibfnamefont {M.~A.}\ \bibnamefont
  {Scheel}},\ and\ \bibinfo {author} {\bibfnamefont {B.}~\bibnamefont
  {Szilagyi}},\ }\bibfield  {title} {\bibinfo {title} {{Simulating merging
  binary black holes with nearly extremal spins}},\ }\href
  {https://doi.org/10.1103/PhysRevD.83.024010} {\bibfield  {journal} {\bibinfo
  {journal} {Phys. Rev. D}\ }\textbf {\bibinfo {volume} {83}},\ \bibinfo
  {pages} {024010} (\bibinfo {year} {2011})},\ \Eprint
  {https://arxiv.org/abs/1010.2777} {arXiv:1010.2777 [gr-qc]} \BibitemShut
  {NoStop}%
\bibitem [{\citenamefont {Szil\'agyi}(2014)}]{1405.3693}%
  \BibitemOpen
  \bibfield  {author} {\bibinfo {author} {\bibfnamefont {B.}~\bibnamefont
  {Szil\'agyi}},\ }\bibfield  {title} {\bibinfo {title} {{Key Elements of
  Robustness in Binary Black Hole Evolutions using Spectral Methods}},\ }\href
  {https://doi.org/10.1142/S0218271814300146} {\bibfield  {journal} {\bibinfo
  {journal} {Int. J. Mod. Phys. D}\ }\textbf {\bibinfo {volume} {23}},\
  \bibinfo {pages} {1430014} (\bibinfo {year} {2014})},\ \Eprint
  {https://arxiv.org/abs/1405.3693} {arXiv:1405.3693 [gr-qc]} \BibitemShut
  {NoStop}%
\bibitem [{\citenamefont {Buonanno}\ \emph {et~al.}(2011)\citenamefont
  {Buonanno}, \citenamefont {Kidder}, \citenamefont {Mroue}, \citenamefont
  {Pfeiffer},\ and\ \citenamefont {Taracchini}}]{1012.1549}%
  \BibitemOpen
  \bibfield  {author} {\bibinfo {author} {\bibfnamefont {A.}~\bibnamefont
  {Buonanno}}, \bibinfo {author} {\bibfnamefont {L.~E.}\ \bibnamefont
  {Kidder}}, \bibinfo {author} {\bibfnamefont {A.~H.}\ \bibnamefont {Mroue}},
  \bibinfo {author} {\bibfnamefont {H.~P.}\ \bibnamefont {Pfeiffer}},\ and\
  \bibinfo {author} {\bibfnamefont {A.}~\bibnamefont {Taracchini}},\ }\bibfield
   {title} {\bibinfo {title} {{Reducing orbital eccentricity of precessing
  black-hole binaries}},\ }\href {https://doi.org/10.1103/PhysRevD.83.104034}
  {\bibfield  {journal} {\bibinfo  {journal} {Phys. Rev. D}\ }\textbf {\bibinfo
  {volume} {83}},\ \bibinfo {pages} {104034} (\bibinfo {year} {2011})},\
  \Eprint {https://arxiv.org/abs/1012.1549} {arXiv:1012.1549 [gr-qc]}
  \BibitemShut {NoStop}%
\bibitem [{\citenamefont {Boyle}\ \emph {et~al.}(2019)\citenamefont {Boyle}
  \emph {et~al.}}]{1904.04831}%
  \BibitemOpen
  \bibfield  {author} {\bibinfo {author} {\bibfnamefont {M.}~\bibnamefont
  {Boyle}} \emph {et~al.},\ }\bibfield  {title} {\bibinfo {title} {{The SXS
  Collaboration catalog of binary black hole simulations}},\ }\href
  {https://doi.org/10.1088/1361-6382/ab34e2} {\bibfield  {journal} {\bibinfo
  {journal} {Class. Quant. Grav.}\ }\textbf {\bibinfo {volume} {36}},\ \bibinfo
  {pages} {195006} (\bibinfo {year} {2019})},\ \Eprint
  {https://arxiv.org/abs/1904.04831} {arXiv:1904.04831 [gr-qc]} \BibitemShut
  {NoStop}%
\bibitem [{\citenamefont {Varma}\ and\ \citenamefont
  {Scheel}(2018)}]{1808.07490}%
  \BibitemOpen
  \bibfield  {author} {\bibinfo {author} {\bibfnamefont {V.}~\bibnamefont
  {Varma}}\ and\ \bibinfo {author} {\bibfnamefont {M.~A.}\ \bibnamefont
  {Scheel}},\ }\bibfield  {title} {\bibinfo {title} {{Constructing a boosted,
  spinning black hole in the damped harmonic gauge}},\ }\href
  {https://doi.org/10.1103/PhysRevD.98.084032} {\bibfield  {journal} {\bibinfo
  {journal} {Phys. Rev. D}\ }\textbf {\bibinfo {volume} {98}},\ \bibinfo
  {pages} {084032} (\bibinfo {year} {2018})},\ \Eprint
  {https://arxiv.org/abs/1808.07490} {arXiv:1808.07490 [gr-qc]} \BibitemShut
  {NoStop}%
\bibitem [{\citenamefont {Iozzo}\ \emph
  {et~al.}(2021{\natexlab{a}})\citenamefont {Iozzo} \emph
  {et~al.}}]{2104.07052}%
  \BibitemOpen
  \bibfield  {author} {\bibinfo {author} {\bibfnamefont {D.~A.~B.}\
  \bibnamefont {Iozzo}} \emph {et~al.},\ }\bibfield  {title} {\bibinfo {title}
  {{Comparing Remnant Properties from Horizon Data and Asymptotic Data in
  Numerical Relativity}},\ }\href {https://doi.org/10.1103/PhysRevD.103.124029}
  {\bibfield  {journal} {\bibinfo  {journal} {Phys. Rev. D}\ }\textbf {\bibinfo
  {volume} {103}},\ \bibinfo {pages} {124029} (\bibinfo {year}
  {2021}{\natexlab{a}})},\ \Eprint {https://arxiv.org/abs/2104.07052}
  {arXiv:2104.07052 [gr-qc]} \BibitemShut {NoStop}%
\bibitem [{\citenamefont {Boyle}\ and\ \citenamefont
  {Mroue}(2009)}]{0905.3177}%
  \BibitemOpen
  \bibfield  {author} {\bibinfo {author} {\bibfnamefont {M.}~\bibnamefont
  {Boyle}}\ and\ \bibinfo {author} {\bibfnamefont {A.~H.}\ \bibnamefont
  {Mroue}},\ }\bibfield  {title} {\bibinfo {title} {{Extrapolating
  gravitational-wave data from numerical simulations}},\ }\href
  {https://doi.org/10.1103/PhysRevD.80.124045} {\bibfield  {journal} {\bibinfo
  {journal} {Phys. Rev. D}\ }\textbf {\bibinfo {volume} {80}},\ \bibinfo
  {pages} {124045} (\bibinfo {year} {2009})},\ \Eprint
  {https://arxiv.org/abs/0905.3177} {arXiv:0905.3177 [gr-qc]} \BibitemShut
  {NoStop}%
\bibitem [{\citenamefont {Boyle}(2013)}]{1302.2919}%
  \BibitemOpen
  \bibfield  {author} {\bibinfo {author} {\bibfnamefont {M.}~\bibnamefont
  {Boyle}},\ }\bibfield  {title} {\bibinfo {title} {{Angular velocity of
  gravitational radiation from precessing binaries and the corotating frame}},\
  }\href {https://doi.org/10.1103/PhysRevD.87.104006} {\bibfield  {journal}
  {\bibinfo  {journal} {Phys. Rev. D}\ }\textbf {\bibinfo {volume} {87}},\
  \bibinfo {pages} {104006} (\bibinfo {year} {2013})},\ \Eprint
  {https://arxiv.org/abs/1302.2919} {arXiv:1302.2919 [gr-qc]} \BibitemShut
  {NoStop}%
\bibitem [{\citenamefont {Boyle}\ \emph {et~al.}(2014)\citenamefont {Boyle},
  \citenamefont {Kidder}, \citenamefont {Ossokine},\ and\ \citenamefont
  {Pfeiffer}}]{1409.4431}%
  \BibitemOpen
  \bibfield  {author} {\bibinfo {author} {\bibfnamefont {M.}~\bibnamefont
  {Boyle}}, \bibinfo {author} {\bibfnamefont {L.~E.}\ \bibnamefont {Kidder}},
  \bibinfo {author} {\bibfnamefont {S.}~\bibnamefont {Ossokine}},\ and\
  \bibinfo {author} {\bibfnamefont {H.~P.}\ \bibnamefont {Pfeiffer}},\
  }\bibfield  {title} {\bibinfo {title} {{Gravitational-wave modes from
  precessing black-hole binaries}},\ }\href@noop {} {\  (\bibinfo {year}
  {2014})},\ \Eprint {https://arxiv.org/abs/1409.4431} {arXiv:1409.4431
  [gr-qc]} \BibitemShut {NoStop}%
\bibitem [{\citenamefont {Boyle}(2016)}]{1509.00862}%
  \BibitemOpen
  \bibfield  {author} {\bibinfo {author} {\bibfnamefont {M.}~\bibnamefont
  {Boyle}},\ }\bibfield  {title} {\bibinfo {title} {{Transformations of
  asymptotic gravitational-wave data}},\ }\href
  {https://doi.org/10.1103/PhysRevD.93.084031} {\bibfield  {journal} {\bibinfo
  {journal} {Phys. Rev. D}\ }\textbf {\bibinfo {volume} {93}},\ \bibinfo
  {pages} {084031} (\bibinfo {year} {2016})},\ \Eprint
  {https://arxiv.org/abs/1509.00862} {arXiv:1509.00862 [gr-qc]} \BibitemShut
  {NoStop}%
\bibitem [{\citenamefont {Iozzo}\ \emph
  {et~al.}(2021{\natexlab{b}})\citenamefont {Iozzo}, \citenamefont {Boyle},
  \citenamefont {Deppe}, \citenamefont {Moxon}, \citenamefont {Scheel},
  \citenamefont {Kidder}, \citenamefont {Pfeiffer},\ and\ \citenamefont
  {Teukolsky}}]{2010.15200}%
  \BibitemOpen
  \bibfield  {author} {\bibinfo {author} {\bibfnamefont {D.~A.~B.}\
  \bibnamefont {Iozzo}}, \bibinfo {author} {\bibfnamefont {M.}~\bibnamefont
  {Boyle}}, \bibinfo {author} {\bibfnamefont {N.}~\bibnamefont {Deppe}},
  \bibinfo {author} {\bibfnamefont {J.}~\bibnamefont {Moxon}}, \bibinfo
  {author} {\bibfnamefont {M.~A.}\ \bibnamefont {Scheel}}, \bibinfo {author}
  {\bibfnamefont {L.~E.}\ \bibnamefont {Kidder}}, \bibinfo {author}
  {\bibfnamefont {H.~P.}\ \bibnamefont {Pfeiffer}},\ and\ \bibinfo {author}
  {\bibfnamefont {S.~A.}\ \bibnamefont {Teukolsky}},\ }\bibfield  {title}
  {\bibinfo {title} {{Extending gravitational wave extraction using Weyl
  characteristic fields}},\ }\href
  {https://doi.org/10.1103/PhysRevD.103.024039} {\bibfield  {journal} {\bibinfo
   {journal} {Phys. Rev. D}\ }\textbf {\bibinfo {volume} {103}},\ \bibinfo
  {pages} {024039} (\bibinfo {year} {2021}{\natexlab{b}})},\ \Eprint
  {https://arxiv.org/abs/2010.15200} {arXiv:2010.15200 [gr-qc]} \BibitemShut
  {NoStop}%
\bibitem [{\citenamefont {Hartle}(1974)}]{Hartle1_skip}%
  \BibitemOpen
  \bibfield  {author} {\bibinfo {author} {\bibfnamefont {J.~B.}\ \bibnamefont
  {Hartle}},\ }\bibfield  {title} {\bibinfo {title} {{Tidal shapes and shifts
  on rotating black holes}},\ }\href {https://doi.org/10.1103/PhysRevD.9.2749}
  {\bibfield  {journal} {\bibinfo  {journal} {Phys. Rev. D}\ }\textbf {\bibinfo
  {volume} {9}},\ \bibinfo {pages} {2749} (\bibinfo {year} {1974})}\BibitemShut
  {NoStop}%
\bibitem [{\citenamefont {Berti}\ and\ \citenamefont
  {Klein}(2014)}]{1408.1860}%
  \BibitemOpen
  \bibfield  {author} {\bibinfo {author} {\bibfnamefont {E.}~\bibnamefont
  {Berti}}\ and\ \bibinfo {author} {\bibfnamefont {A.}~\bibnamefont {Klein}},\
  }\bibfield  {title} {\bibinfo {title} {{Mixing of spherical and spheroidal
  modes in perturbed Kerr black holes}},\ }\href
  {https://doi.org/10.1103/PhysRevD.90.064012} {\bibfield  {journal} {\bibinfo
  {journal} {Phys. Rev. D}\ }\textbf {\bibinfo {volume} {90}},\ \bibinfo
  {pages} {064012} (\bibinfo {year} {2014})},\ \Eprint
  {https://arxiv.org/abs/1408.1860} {arXiv:1408.1860 [gr-qc]} \BibitemShut
  {NoStop}%
\bibitem [{\citenamefont {Booth}\ and\ \citenamefont
  {Fairhurst}(2007)}]{gr-qc/0610032}%
  \BibitemOpen
  \bibfield  {author} {\bibinfo {author} {\bibfnamefont {I.}~\bibnamefont
  {Booth}}\ and\ \bibinfo {author} {\bibfnamefont {S.}~\bibnamefont
  {Fairhurst}},\ }\bibfield  {title} {\bibinfo {title} {{Isolated, slowly
  evolving, and dynamical trapping horizons: Geometry and mechanics from
  surface deformations}},\ }\href {https://doi.org/10.1103/PhysRevD.75.084019}
  {\bibfield  {journal} {\bibinfo  {journal} {Phys. Rev. D}\ }\textbf {\bibinfo
  {volume} {75}},\ \bibinfo {pages} {084019} (\bibinfo {year} {2007})},\
  \Eprint {https://arxiv.org/abs/gr-qc/0610032} {arXiv:gr-qc/0610032}
  \BibitemShut {NoStop}%
\bibitem [{\citenamefont {Good}\ and\ \citenamefont {Ong}(2015)}]{1412.5432}%
  \BibitemOpen
  \bibfield  {author} {\bibinfo {author} {\bibfnamefont {M.~R.~R.}\
  \bibnamefont {Good}}\ and\ \bibinfo {author} {\bibfnamefont {Y.~C.}\
  \bibnamefont {Ong}},\ }\bibfield  {title} {\bibinfo {title} {{Are black holes
  springlike?}},\ }\href {https://doi.org/10.1103/PhysRevD.91.044031}
  {\bibfield  {journal} {\bibinfo  {journal} {Phys. Rev. D}\ }\textbf {\bibinfo
  {volume} {91}},\ \bibinfo {pages} {044031} (\bibinfo {year} {2015})},\
  \Eprint {https://arxiv.org/abs/1412.5432} {arXiv:1412.5432 [gr-qc]}
  \BibitemShut {NoStop}%
\end{thebibliography}%
\end{document}